\documentclass[a4paper,11pt]{article}

\pdfoutput=1
\usepackage{comment}
\usepackage{fixltx2e}
\usepackage[table,xcdraw]{xcolor}
\usepackage{jcappub} 
\usepackage{float} 
\usepackage[scr=boondox]{mathalfa}
\usepackage{bm}
\usepackage{stmaryrd}
\usepackage{bbold}
\usepackage[normalem]{ulem}
\usepackage{amsmath}
\usepackage{bbm}
\usepackage{geometry}
\usepackage{cancel}
\usepackage{xcolor}

\def\be{\begin{equation}}
\def\ee{\end{equation}}
\def\bea{\begin{eqnarray}}
\def\eea{\end{eqnarray}}
\newcommand{\vs}{\nonumber\\}
\def\ba#1\ea{\begin{align}#1\end{align}}
\def\bg#1\eg{\begin{gather}#1\end{gather}}

\newcommand{\s}{\sigma}

\newcommand{\refeq}[1]{Eq.~(\ref{eq:#1})}

\newcommand{\reffig}[1]{Fig.~\ref{fig:#1}}

\newcommand{\refsec}[1]{Sec.~\ref{sec:#1}}

\newcommand{\unit}{{\mathbb{1}}}

\renewcommand{\vec}{\textbf}

\newcommand*{\df}  {\delta}

\newcommand*{\non}  {\nonumber}
\newcommand*{\lb}  {\left(}
\newcommand*{\rb}  {\right)}

\newcommand{\vk} {\vec{k}}

\def\dirac{\delta_{\rm D}}

\def\L{\Lambda}

\def\O{\mathcal{O}}

\renewcommand{\d}{\delta}
\newcommand{\G}{\mathcal{G}}

\newcommand{\ea}{\end{aligned}\]}

\newcommand{\eq}[1]{\begin{align}#1\end{align}}
\newcommand{\eeq}[1]{\begin{equation}#1\end{equation}}

\newcommand{\nn}{\nonumber}

\renewcommand{\emph}[1]{\textit{#1}}

\renewcommand{\d}{\delta}

\def\L{\Lambda}

\def\L{\Lambda}

\def\O{\mathcal{O}}

\newcommand{\perm}[1]{ \expandafter\ifstrempty\expandafter{#1} {\mbox{perm.}} {\mbox{$#1$ perm.}} }

\makeatletter
\newlength{\apb@width}
\newcommand{\autoparbox}[2][c]{\settowidth{\apb@width}{#2}\parbox[#1]{\apb@width}{#2}}

\makeatother

\DeclareMathOperator{\tr}{tr}

\def\dirac{\delta_{\rm D}}

\renewcommand{\comment}[1]{}

\usepackage[T1]{fontenc}
\usepackage{graphbox}
\usepackage{subfig}
\usepackage{booktabs}
\subheader{TUM-HEP-1566/25\\
RBI-ThPhys-2025-30}
\title{
Two-loop renormalization and running of galaxy bias
}

\usepackage{amsmath}
\author[a]{Thomas Bakx,}
\author[b]{Mathias Garny,}
\author[c,d,e]{Henrique Rubira,}
\author[d,e,f]{Zvonimir Vlah}
\affiliation[a]{Institute for Theoretical Physics, Utrecht University, Princetonplein 5, 3584 CC, Utrecht, The Netherlands}
\affiliation[b]{Physik Department T31, School of Natural Sciences, Technische Universit\"at M\"unchen,\\
James-Franck-Stra{\ss}e 1, D-85748 Garching, Germany}

\affiliation[c]{University Observatory, Faculty of Physics, Ludwig-Maximilians-Universit\"at, Scheinerstr. 1, D-81679 München, Germany}
\affiliation[d]{Kavli Institute for Cosmology Cambridge, Madingley Road, Cambridge CB3 0HA, UK}
\affiliation[e]{Centre for Theoretical Cosmology, Department of Applied Mathematics and Theoretical Physics
University of Cambridge, Wilberforce Road, Cambridge, CB3 0WA, UK}
\affiliation[f]{Division of Theoretical Physics, Ru\dj er Bo\u skovi\' c Institute, 10000 Zagreb, Croatia,}%

\abstract{
We systematically extend the framework of galaxy bias renormalization to two-loop order.
For the minimal complete basis of 29 deterministic bias operators up to fifth order in the density field and at leading order in gradient expansion we explicitly work out one- and two-loop renormalization.
The latter is provided in terms of double-hard limits of bias kernels, which we find to depend on only one function of the ratio of the loop momenta. 
After including stochasticity in terms of composite operator renormalization, we apply the framework to the two-loop power spectrum of biased tracers and provide a simple result suitable for numerical evaluation.
In addition, we work out one- and two-loop renormalization group equations (RGE) for deterministic bias coefficients related to bias operators constructed from a smoothed density field, generalizing previous works. 
We identify a linear combination of bias operators with enhanced UV sensitivity, related to a positive eigenvalue of the RGE. Finally, we present an analogy with the RGE as used in quantum field theory, suggesting that a resummation of large logarithms as employed in the latter may also yield useful applications in the study of large-scale galaxy bias.
} 


\begin{document}

\maketitle
\flushbottom

\section{Introduction}\label{sec:intro}

Precise measurements of the large-scale structure (LSS) of the Universe by current and next-generation galaxy surveys demand robust and efficient theoretical modeling of galaxy clustering~\cite{BOSS:2016wmc,DESI:2024jis,Euclid:2023tog}. 
A key element is the large-scale bias expansion, being a systematic mapping of the density contrast $\delta_g({\bf x})$ of a biased tracer population observed in a galaxy survey to the underlying matter density distribution within the mildly non-linear regime~\cite{Kaiser:1984sw, Fry:1992vr, McDonald:2009dh, Chan:2012jj, Mirbabayi:2014zca, Senatore:2014eva, egge_oneloopbisp}, see~\cite{Desjacques:2016bnm} for a review. In real space and within its Eulerian formulation, on which we focus in this work, it can be written in the form
\ba \label{eq:biasexp}
\delta_g (\vec x) = \sum_{\mathcal{O}} b_{a} \mathcal{O}_a (\vec x) + \text{h.d.} + \text{stoch.} \,,
\ea
where $\mathcal{O}_a (\vec x)$ are universal `operators' and $b_a$ bias coefficients specific to the considered galaxy population. The operators can be classified according to the minimal order in perturbation theory in terms of the initial density fluctuations at which they contribute, and are constrained by the assumed symmetries, related to the cosmological and equivalence principles. In addition, the operators are characterized by a gradient expansion, with higher derivative operators being suppressed by powers of the wavenumber on large scales in Fourier space, and complemented by stochastic terms, both of which are left implicit in Eq.\,\eqref{eq:biasexp} for simplicity. At leading order in  gradient expansion, as considered in this work, the number of non-redundant operators starting at orders $1-5$ in perturbation theory is $1, 2, 4, 8, 14$, respectively, being 29 in total~\cite{Schmidt:2020tao, Donath:2023sav, Ansari:2025nsf}.

The operators are well-defined when smoothing the underlying linear matter density field on an arbitrary smoothing scale $\L$. Since physical quantities such as $N$-point correlation functions of $\delta_g$ cannot depend on it, it is desirable to control the dependence on $\L$. One method is to consider renormalized bias operators $[{\cal O}_a]$ as done by McDonald~\cite{McDonald:2006mx} and Assassi, Baumann, Green and Zaldarriaga~\cite{Assassi:2014fva} (see also \cite{McDonald:2009dh,Schmidt:2012ys,Baldauf:2012hs,Chan:2012jj,Saito:2014qha,Angulo:2015eqa,egge_oneloopbisp}). The $[{\cal O}_a]$ are defined by a renormalization condition, constructed such that the $\L$-dependence of arbitrary $N$-point correlation functions of the renormalized operators cancels out (to a certain desired accuracy in the gradient expansion). This implies that the corresponding renormalized bias coefficients $b_{[a]}$ are independent of $\L$ as well, allowing for a clear physical interpretation. The renormalization condition considered in~\cite{Assassi:2014fva} ensures that the $b_{[a]}$ match their separate Universe expectation~\cite{Lazeyras:2015lgp, Baldauf:2015vio, Li:2015jsz}. One-loop renormalization for the 15 leading-gradient Eulerian bias operators up to fourth order has been worked out in~\cite{DAmico:2022ukl}, see also~\cite{Assassi:2014fva,egge_oneloopbisp,philcox_oneloopbisp} and~\cite{Abidi:2018eyd} for a discussion of two-loop contributions.

An alternative approach is to use the $\L$-dependent bias operators ${\cal O}_a$, and compensate for it by choosing $\L$-dependent bias coefficients $b_a(\L)$ such that the $\L$-dependence cancels in physical observables. This has the advantage of being more directly related to field-level approaches \cite{Jasche_2013, Taruya:2018jtk, Elsner:2019rql, nguyen_fli} for extracting bias parameters~\cite{Schmittfull:2014tca,Lazeyras:2017hxw,Abidi:2018eyd,Schmittfull:2018yuk,Peron:2025lgh}, and provides additional information on the scale-dependence. In this case, it is mandatory to control the evolution of the bias when varying $\L$, which can be conveniently formulated in terms of the renormalization group (RG). For deterministic leading-gradient bias up to third order, this was done in~\cite{Rubira:2023vzw} employing the Wilson-Polchinski approach~\cite{wilson:1971,polchinski:1984,Carroll:2013oxa} (see also~\cite{McDonald:2006hf,Matarrese:2007wc,Blas:2015qsi,Floerchinger:2016hja,Cabass:2019lqx}), and generalized to stochastic terms~\cite{Rubira:2024tea} as well as primordial non-Gaussianity in~\cite{Nikolis:2024kbx}. We also highlight another complementary approach based on multi-point propagators~\cite{egge_oneloopbisp}, allowing for key insights on e.g.~the interplay of bias renormalization and time-evolution as well as additional redundancies of bias operators that arise when computing $N$-point functions to a certain loop order.

In this work we systematically extend the concept of bias renormalization to two-loop order. We consider both the method of renormalized bias operators $[{\cal O}_a]$ as well as the renormalization group techniques, and the relation among them. We furthermore go beyond previous work by explicitly computing one- and two-loop renormalization for all 29 leading-gradient deterministic bias operators up to fifth order. Additionally, we consider stochastic bias, being given by contact terms arising in the renormalization of products of two (or more) bias operators, known as composite operator renormalization.
While bias renormalization works at the operator level and is applicable to any $N$-point function or field-level analysis, we apply it to the two-loop power spectrum and provide an explicit, manageable expression at leading order in gradient expansion [see Eq.\,\eqref{eq:pren1L2L}]. We then proceed to the alternative RG technique. Since the Wilson-Polchinski approach is not easily generalizeable beyond one-loop, we consider an alternative, more direct derivation of renormalization group equations (RGEs) using the $\L$-independence of galaxy $N$-point functions. We compute one- and two-loop `$\beta$-functions', generalizing one-loop results of \cite{Rubira:2023vzw} from third to fifth-order operators. 
We finally discuss solutions of the RGE, and the relation to the resummation of `large logarithms' known from quantum field theory (QFT).

The structure of this work is as follows: in \refsec{fifthorder} we describe the fifth-order bias basis and in \refsec{singledoublelimits} we discuss the single and double-hard limits of bias kernels, that are widely used throughout this work. In \refsec{biasrenorm} we derive the renormalized bias operators of \cite{McDonald:2006mx,Assassi:2014fva} at fifth-order and two loops, including also stochastic contributions. We calculate the RGEs at fifth-order and two loops in \refsec{rg}. In \refsec{connection_RG_renorm} we provide an alternative derivation of the RGEs using the renormalized bias operators of \refsec{biasrenorm}, making the connection between both frameworks manifest. Solutions for the now extended RGEs are analyzed in \refsec{solutions}. We discuss resummation of higher-loop terms in \refsec{resum} and conclude in \refsec{conclusions}. The appendices contain recursion relations used to construct the bias operators (Appendix~\ref{app:recursivekernel}), explicit tabulated results for coefficients entering bias renormalization and RGEs (Appendices~\ref{app:det} and~\ref{sec:stoch}), as well as further details for stochastic terms (Appendix~\ref{app:hierarchynoise}).

\newpage
\section{The fifth-order bias expansion}\label{sec:fifthorder}
In this Section, we review the fifth-order bias operator basis and also set the notation.
We write for the matter overdensity $\delta$ and velocity divergence field $\theta$ in Fourier space,  using the Einstein--de~Sitter approximation,
\ba
\delta (\vec k)= \sum_{n=1}^\infty \int_{q_1\cdots q_n} (2\pi)^3 \dirac(\vec k - \vec q_{1 n}) 
D^n F^{(n)} (\vec q_1, \ldots , \vec q_n) \df_L(\vec q_1) \ldots  \df_L(\vec q_n)\,,
\\
\theta (\vec k)= \sum_{n=1}^\infty \int_{q_1\cdots q_n} (2\pi)^3 \dirac(\vec k - \vec q_{1 n})
D^n G^{(n)} (\vec q_1, \ldots , \vec q_n) \df_L(\vec q_1) \ldots  \df_L(\vec q_n)\,,
\ea
where $\dirac(\vec k - \vec q_{1 n})\equiv \dirac(\vec k-\sum_{i=1}^n\vec q_i)$ is the Dirac delta, $F^{(n)}$ and $G^{(n)}$ are the standard Eulerian matter kernels~\cite{Bernardeau:2001qr}, $\int_{q_1q_2\cdots}\equiv \int d^3q_1/(2\pi)^3\int d^3q_2/(2\pi)^3\cdots$, $\df_L$ is the linear density field and $D=D(z)$ is the growth factor. We omit the time dependence of the operators and bias coefficients throughout this work.
We can also construct kernels $K_{a}$ for the fields  ${\cal O}_a$ of biased tracers defined in \refeq{biasexp},
\ba \label{eq:K_kernels}
{\cal O}_{a} (\vec k) &= \sum_{n=1}^\infty \int_{q_1\cdots q_n}  (2\pi)^3 \dirac(\vec k - \vec q_{1n}) D^n K_{a}^{(n)} (\vec q_1 \ldots , \vec q_n) \df_L(\vec q_1) \ldots  \df_L(\vec q_n)\,.
\ea

We use in this work the basis of bias operators defined in \cite{Mirbabayi:2014zca} (see also \cite{Desjacques:2016bnm}) constructed from the building blocks $\Pi^{[m]}_{ij}$ for $m\geq 1$. The superscript indicates that $\Pi^{[m]}_{ij}$ contributes at order $n\geq m$ in the perturbative expansion in powers of the linear density field. They are defined recursively, with recursion relation in Fourier given by
\begin{equation} \label{eq:Pin}
    \begin{aligned}
        \Pi^{[n]}_{ij}(\vec k) = \frac{1}{(n-1)!} \Bigg[ \bigg( \frac{\partial}{\partial \ln D} &- (n-1) \bigg) \Pi^{[n-1]}_{ij}(\vec k) \\
&- (2\pi)^3 \dirac(\vec k - \vec q_{12}) \frac{\vec q_1 . \vec q_2}{q_1^2} \theta(\vec q_1) \Pi^{[n-1]}_{ij}(\vec q_2)
\Bigg]\,.
    \end{aligned}
\end{equation}
For $\Pi^{[1]}_{ij}$ we have in terms of the matter density field $\df$
\eq{
\Pi^{[1]}_{ij}(\vec k) = 
\frac{k_ik_j}{k^2}
\df(\vec k)\,,
}
with $k\equiv |\vec k|$. We provide more details about the construction of $\Pi^{[n]}_{ij}$ in Appendix~\ref{app:recursivekernel}. A complete basis of deterministic bias operators ${\cal O}_a$  at leading order in gradient expansion for biased tracer fields that are scalars under spatial rotations (such as the galaxy density field) can be constructed by  considering all possible scalar contractions of products of $\Pi_{ij}^{[n]}$. At up to fifth order in the linear density field, we have
\eq{
\label{eq:bias_basis}
{\rm (1)} \qquad 
    &{\rm tr}\big[ \Pi^{[1]} \big], \non\\
{\rm (2)} \qquad 
    &{\rm tr}\big[ \big( \Pi^{[1]} \big)^2 \big], 
    ~~ \Big( {\rm tr}\big[ \Pi^{[1]} \big] \Big)^2, \non\\
{\rm (3)} \qquad 
   &{\rm tr}\big[ \big( \Pi^{[1]} \big)^3 \big],
   ~~{\rm tr}\big[ \big( \Pi^{[1]} \big)^2 \big] {\rm tr}\big[ \Pi^{[1]} \big],
   ~~ \Big( {\rm tr}\big[ \Pi^{[1]} \big] \Big)^3,
   ~~{\rm tr}\big[\Pi^{[1]} \Pi^{[2]} \big], \non\\
{\rm (4)} \qquad 
   &{\rm tr}\big[ \big( \Pi^{[1]} \big)^2 \big] \big({\rm tr}\big[ \Pi^{[1]} \big]\big)^2, 
   ~~{\rm tr}\big[ \big( \Pi^{[1]} \big)^3 \big] {\rm tr}\big[ \Pi^{[1]} \big],
   ~~\Big( {\rm tr}\big[ \big( \Pi^{[1]} \big)^2 \big] \Big)^2,   
   ~~\Big( {\rm tr}\big[ \Pi^{[1]} \big] \Big)^4, \\
   &{\rm tr}\big[\Pi^{[1]} \Pi^{[3]} \big], 
   ~~{\rm tr}\big[\Pi^{[1]} \Pi^{[1]} \Pi^{[2]} \big],    
   ~~{\rm tr}\big[\Pi^{[2]} \Pi^{[2]} \big],   
   ~~{\rm tr}\big[ \Pi^{[1]} \big] {\rm tr}\big[\Pi^{[1]} \Pi^{[2]} \big], \non\\
 {\rm (5)} \qquad 
    &{\rm tr}\big[ \big( \Pi^{[1]} \big)^3 \big]{\rm tr}\big[ \big( \Pi^{[1]} \big)^2 \big] ,
 ~~{\rm tr}\big[ \big( \Pi^{[1]} \big)^3 \big] \left({\rm tr}\big[ \Pi^{[1]} \big]\right)^2,
 ~~{\rm tr}\big[ \big( \Pi^{[1]} \big)^2 \big] \left({\rm tr}\big[ \Pi^{[1]} \big]\right)^3,\non\\
 &{\rm tr}\big[ \Pi^{[1]} \big]\Big( {\rm tr}\big[ \big( \Pi^{[1]} \big)^2 \big] \Big)^2,   
 ~~\Big( {\rm tr}\big[ \Pi^{[1]} \big] \Big)^5, {\rm tr}\big[ \Pi^{[1]} \big] {\rm tr}\big[\Pi^{[1]} \Pi^{[3]} \big],
 ~~{\rm tr}\big[\Pi^{[1]} \Pi^{[4]} \big] \non\\  
   &{\rm tr}\big[\Pi^{[1]} \Pi^{[1]} \Pi^{[3]} \big],  
~~{\rm tr}\big[\Pi^{[1]} \Pi^{[2]} \Pi^{[2]} \big],    
~~{\rm tr}\big[\Pi^{[2]} \Pi^{[3]} \big],   
~~{\rm tr}\big[ \Pi^{[1]} \big] {\rm tr}\big[\Pi^{[2]} \Pi^{[2]} \big],\non\\ 
&
{\rm tr}\big[ \Pi^{[1]} \big] {\rm tr}\big[\Pi^{[1]} \Pi^{[1]} \Pi^{[2]}\big]\,,
~~{\rm tr}\big[ \Pi^{[1]}  \Pi^{[1]} \big] {\rm tr}\big[\Pi^{[1]} \Pi^{[2]}\big]\,,
~~\left({\rm tr}\big[ \Pi^{[1]}\big]\right)^2 {\rm tr}\big[\Pi^{[1]} \Pi^{[2]} \big]. \non
}
Each of the five sublists contains all the operators that \textit{start} at this order, but all operators starting at order $m=1,\dots 5$ also receive contributions at higher orders $m' \geq m$.  Notice that stochastic contributions and `counter-terms' for the density field also have to be included to account for small-scale physics. We discuss the former below, while the latter are not relevant at leading-gradient order considered in this work.

We checked that this basis is non-redundant at fifth order, containing in total 29 bias operators. Apart from the well-known redundancy of ${\rm tr}\big[\Pi^{[n]}\big]$ for $n>1$, we also used the Cayley-Hamilton relation (see also \cite{egge_oneloopbisp}) in three dimensions (here $\delta^K$ is the Kronecker symbol)
\eeq{\label{eq:CayleyHamilton}
\big( \Pi^{[1]} \big)^3_{ij}
 - {\rm tr}\big[ \Pi^{[1]} \big] \big( \Pi^{[1]} \big)^2_{ij} 
 + \frac{1}{2} \lb \lb {\rm tr}\big[ \Pi^{[1]} \big] \rb^2  -  {\rm tr}\big[ \big( \Pi^{[1]} \big)^2 \big] \rb  \Pi^{[1]} _{ij}
 - \det\big[ \Pi^{[1]} \big] \delta^{\rm K}_{ij} = 0 \, ,
}
to eliminate e.g. ${\rm tr}\big[ \big( \Pi^{[1]} \big)^n \big]$ for $n>3$ or ${\rm tr}\big[ \big( \Pi^{[1]} \big)^3 \Pi^{[2]}\big]$.
We note that the number of independent operators at fifth order agrees with that found for the different Lagrangian \cite{Schmidt:2020tao} and Eulerian basis~\cite{Donath:2023sav} (see also \cite{Ansari:2025nsf}). We also verified that when constructing the basis from only $\Pi_\theta^{[1]} = \partial_{(i} v_{j)}$ and $\Pi^{[1]}$, the basis would be complete up to fourth order, but one would miss three operators at fifth order (in line with~\cite{Donath:2023sav}).

Furthermore, the zeroth-order operator 
\be \label{eq:unit}
\unit(\vk) = (2\pi)^3\dirac(\vk)\,,
\ee
which is a constant in real space, has to be added to the basis when dealing with stochastic contributions \cite{Rubira:2024tea} but is removed from the deterministic component of the bias expansion since $\langle \d_g \rangle = 0$.


\newpage
\section{Single and double-hard limits of bias kernels}\label{sec:singledoublelimits}

The bias expansion up to fifth order is required to compute the power spectrum of a biased tracer at two-loop order. Using~\refeq{biasexp}, we can write it as
\be
\langle \df_g(\vec k) \df_g(\vec k')\rangle=(2\pi)^3\dirac(\vec k+\vec k')\Big(\sum_{{\cal O}_a,{\cal O}_b} b_a b_b P_{ab}(k) + \text{h.d.} + \text{stoch.}\Big)\,,
\ee
where we have kept higher-derivative and stochastic terms implicit. It involves the cross spectra 
\be
\langle {\cal O}_a(\vec k) {\cal O}_b(\vec k')\rangle=(2\pi)^3\dirac(\vec k+\vec k')P_{ab}(k)
\ee
of two bias operators ${\cal O}_a$ and ${\cal O}_b$ from the basis set~\refeq{bias_basis}. Up to two-loop order 
\be
  P_{ab}(k)=D^2P^\text{lin}_{ab}(k)+D^4P^{1\text{L}}_{ab}(k)+D^6P^{2\text{L}}_{ab}(k)\,,
\ee
with 
\bea
  P^\text{lin}_{ab}(k)&=& P^{(11)}_{ab}\,,\nn\\
  P^{1\text{L}}_{ab}(k) &=& P^{(13)}_{ab}+P^{(31)}_{ab}+P^{(22)}_{ab}\,,\nn\\
  P^{2\text{L}}_{ab}(k) &=& P^{(15)}_{ab}+P^{(51)}_{ab}+P^{(24)}_{ab}+P^{(42)}_{ab}+P^{(33),I}_{ab}+P^{(33),II}_{ab}\,,
\eea
where $P^{(nm)}_{ab}(k)=P^{(mn)}_{ba}(k)$ and
\bea\label{eq:twoloopcontribs}
  P^{(11)}_{ab} &=& K^{(1)}_a(k)K^{(1)}_b(k)P^{\text{lin}}(k)\,,\\
  P^{(13)}_{ab} &=& 3\int_p K^{(1)}_a(k)K_b^{(3)}(\vec k,\vec p,-\vec p)P^{\text{lin}}(k)P^{\text{lin}}(p)\,,\nn\\
  P^{(22)}_{ab} &=& 2\int_p K^{(2)}_a(\vec k-\vec p,\vec p)K_b^{(2)}(\vec k-\vec p,\vec p)P^{\text{lin}}(|\vec k-\vec p|)P^{\text{lin}}(p)\,,\nn\\
  P^{(15)}_{ab} &=& 15\int_{pq} K^{(1)}_a(k)K_b^{(5)}(\vec k,\vec p,-\vec p,\vec q,-\vec q)P^{\text{lin}}(k)P^{\text{lin}}(p)P^{\text{lin}}(q)\,,\nn\\
  P^{(24)}_{ab} &=& 12\int_{pq} K^{(2)}_a(\vec k-\vec p,\vec p)K_b^{(4)}(\vec k-\vec p,\vec p,\vec q,-\vec q)P^{\text{lin}}(|\vec k-\vec p|)P^{\text{lin}}(p)P^{\text{lin}}(q)\,,\nn\\
  P^{(33),I}_{ab} &=& 9\int_{pq} K^{(3)}_a(\vec k,\vec p,-\vec p)K_b^{(3)}(\vec k,\vec q,-\vec q)P^{\text{lin}}(k)P^{\text{lin}}(p)P^{\text{lin}}(q)\,,\nn\\
  P^{(33),II}_{ab}\hspace{-0.2cm} &=&  6\int_{pq}\hspace{-0.2cm} K^{(3)}_a(\vec k-\vec p-\vec q,\vec p,\vec q)K_b^{(3)}(\vec k-\vec p-\vec q,\vec p,\vec q)P^{\text{lin}}(|\vec k-\vec p-\vec q|)P^{\text{lin}}(p)P^{\text{lin}}(q).\nn
\eea
Here $P^{\text{lin}}(k)$ is the linear matter power spectrum, and the linear bias kernels $K^{(1)}_a(k)\equiv K^{(1)}_a(\vec k)$ are $1$ for ${\cal O}_a={\rm tr}\big[\Pi^{[1]}\big]=\delta$ and $0$ otherwise. Throughout this work, we assume that the linear density field $\d_{L}$, and therefore the linear power spectra $P^{\text{lin}}$, are always smoothed with a sharp-$k$ filter\footnote{One could in principle alleviate this sharp-$k$ assumption by 
considering other filter functions, replacing $\Theta(\Lambda-p)$ by $W_\Lambda(p)$ with $W_\Lambda(p)\simeq 1$ for $p\lesssim\Lambda$ and $W_\Lambda(p)\to 0$ for $p/\Lambda\to\infty$. Most of our results can be easily generalized to this case by replacing the UV cutoff  $p<\Lambda$ ($p,q<\Lambda$) in one (two) loop integrals by a weighting with the filter function $W_\Lambda(p)$ ($W_\Lambda(p)W_\Lambda(q)$). This applies in particular to single- and double hard limits discussed below, one- and two-loop bias renormalization discussed in Sec.~\ref{sec:biasrenorm} [specifically the main result Eq.~\eqref{eq:Oren1L2L}], as well as the one-loop renormalization group equations from Sec.~\ref{sec:rg} [in particular Eq.~\eqref{eq:RG1L} retains the same form, with the quantity $\sigma_\Lambda$ defined in Eq.~\eqref{eq:sigmaLambda} generalized to $\int_p W_\Lambda(p)P^\text{lin}(p)$]. Generalizing the two-loop renormalization group equations to a generic filter function would make their form slightly more involved. We also note that, within the Wilson-Polchinski approach~\cite{Rubira:2024tea}, a sharp-$k$ filter is singled out as being the simplest choice.
} at a cutoff scale $\L$ 
\be \label{eq:filter}
\delta_L^\Lambda({\bm p})\equiv \Theta(\Lambda-|{\bm p}|)\delta_L({\bm p})\,.
\ee
Therefore, the loop integrals are always regularized by the cutoff in its internal propagators. We omit the $\L$ superscript hereafter, but emphasize the cutoff used inside the loop integrals for a generic $n$-variable integrand ${\cal F}({\bm p}_1,\dots {\bm p}_n)$  by writing 
\be
\int_{p_1, \dots p_n < \Lambda}{\cal F}({\bm p}_1,\dots {\bm p}_n) \equiv \int_{p_1\dots p_n} {\cal F}({\bm p}_1,\dots {\bm p}_n)\Theta(\Lambda-|{\bm p}_1|) \dots \Theta(\Lambda-|{\bm p}_n|).
\ee

The UV-sensitivity of the loop integrals can be accounted for by using appropriate renormalized bias operators, see Sec.\,\ref{sec:biasrenorm}. Here we already briefly describe the correspondence between the various possible UV limits of the loop integrands and the bias terms by which they are renormalized. As is well known, at one-loop the hard limit $p\gg k$ of $P^{(13)}$ can be absorbed into a shift of the linear bias $b_{\d}$ entering the galaxy power spectrum along with $P^{(11)}$, while the hard limit of $P^{(22)}$ is renormalized by a $k$-independent stochastic term at leading order in the gradient expansion.
 Notice that there is also a contribution to a $\nabla^2 \d$ term as well as stochastic terms suppressed by $k^2$, but throughout this work we neglect higher-derivative terms focusing on the leading-in-derivative operators.

At two-loop order, we have to consider both the \textit{single-hard} and \textit{double-hard} limits.
 For the expressions \refeq{twoloopcontribs} that means
the integration region for which (parametrically) either $p\gg k,q$ or $q\gg k,p$ (single-hard), as well as the case $p,q\gg k$ (double-hard). The contributions within these regions are defined by replacing the kernels by their Taylor-expanded limits when either one of the loop wavenumbers becomes large, or both of them with a fixed ratio $r=p/q$, respectively. The double-hard limits of $P^{(15)}$ and $P^{(33),I}$ can again be renormalized by shifting the linear bias entering the tracer power spectrum along with $P^{(11)}$, while those of $P^{(24)}$ and $P^{(33),II}$ can be absorbed in stochastic terms. The single-hard limits of $P^{(15)}$ and $P^{(33),I}$ [see \refeq{twoloopcontribs} above] can be absorbed in a shift in deterministic bias operators up to third order entering the tracer power spectrum via $P^{(13)}$, and the single-hard limit $q\gg p,k$ of $P^{(24)}$ into a shift in deterministic bias at up to second order entering via $P^{(22)}$. On the other hand, the single hard limit $p\gg q,k$ of $P^{(24)}$ as well as all single-hard limits of $P^{(33),II}$ are accounted for by stochastic terms \cite{Rubira:2024tea}. As we will see, no mixed stochastic--deterministic bias terms are required to renormalize the \textit{power spectrum} at any loop order (these enter only when considering higher $N$-point functions, see e.g.~\cite{Rubira:2024tea}).

To obtain explicit results for bias renormalization at two-loop order, we compute in a first step the relevant limits of the bias kernels $K$, considering single- and double-hard regions, as well as deterministic and stochastic terms, respectively. The deterministic part is related to the limits of a single kernel, while the stochastic terms are, for the power spectrum, associated to limits of a product of two kernels. Apart from bias renormalization, these limits also determine the RGE coefficients.

\subsection{Deterministic terms: single-hard limit}

Loop contributions that are renormalized by deterministic bias operators in the single-hard limit $p\to\infty$  correspond to those containing `${\bm p},-{\bm p}$' configurations of wavenumber arguments inside a single kernel $K_b$ of an operator $\O_b$. The general structure of the kernel at $(n+2)$th order is $K_b^{(n+2)}(\vec k_1,\dots,\vec k_n,\vec  p,-\vec p)$, where the $\vec k_i$ stand for external scales or other loop wavenumbers that are all being kept fixed in the single-hard limit. Averaging over the direction of $\hat{\bf p}\equiv\vec p/p$ and then taking the limit $p\to\infty$ yields a function of the $\vec k_1,\dots,\vec k_n$ that can again be decomposed into a linear combination of bias operators ${\cal O}_a$, specifically into a sum of the lower-order kernels $K_a^{(n)}(\vec k_1,\dots,\vec k_n)$,
\be \label{eq:biaskernelexpansionsinglehard}
  K_b^{(n+2)}({\bm k}_1,\dots,{\bm k}_n,{\bm p},-{\bm p})^{p\to\infty}_{\text{av}_{\hat{\bf p}}} = c^{(n+2)}_{ba}K_a^{(n)}({\bm k}_1,\dots,{\bm k}_n)\,,
\ee
where a sum over all contributions from deterministic bias operators $\O_a$ \emph{up to} order $n$ is implicit with $c_{ba}$ being a coefficient matrix.
We say in that case that the operator $\O_b$ \emph{generates} the operator $\O_a$ in the single-hard limit, or that the operator $\O_b$ \emph{is renormalized} by $\O_a$.
Concretely, this means 
\begin{eqnarray}\label{eq:singlehard}
  K^{(3)}_b({\bm k},{\bm p},-{\bm p})^{p\to\infty}_{\text{av}_{\hat{\bf p}}} &=&  c_{ba}^{(3)}K^{(1)}_a({\bm k})\,,\nn\\
  K^{(4)}_b({\bm k}_1,{\bm k}_2,{\bm p},-{\bm p})^{p\to\infty}_{\text{av}_{\hat{\bf p}}} &=&  c_{ba}^{(4)}K^{(2)}_a({\bm k}_1,{\bm k}_2)\,,\nn\\
  K^{(5)}_b({\bm k}_1,{\bm k}_2,{\bm k}_3,{\bm p},-{\bm p})^{p\to\infty}_{\text{av}_{\hat{\bf p}}} &=&  c_{ba}^{(5)}K^{(3)}_a({\bm k}_1,{\bm k}_2,{\bm k}_3)\,.
\end{eqnarray}
 Explicit results for $c^{(3)}_{ba}, c^{(4)}_{ba}, c^{(5)}_{ba}$ are given in Tables \eqref{eq:doublehardcoeff}, \eqref{eq:singlehardcoeffK4} and \eqref{eq:singlehardcoeffK5}, respectively. Note that for $c^{(3)}_{ba}$ the index $b$ ranges over all $\O_b$ up to third order, while the index $a$ contains only the linear bias operator $a=\delta$. For $c^{(4)}_{ba}$ the index $b$ ($a$) runs over fourth (second) order bias operators, and for $c^{(5)}_{ba}$ over fifth (third) order, respectively. 

When Taylor expanding in powers of $1/p^2$,~\refeq{singlehard} corresponds to the leading terms. Keeping the first subleading term would yield contributions that could be accounted for by higher-gradient bias operators, which we do not consider here.
We also note that it is important to keep the wavevectors $\vec k_i$ in~\refeq{singlehard} independent from each other to be able to uniquely determine the $c^{(n+2)}_{ba}$ coefficients.

We stress that~\refeq{singlehard} is applicable to general $N$-point functions, relevant for e.g. the one-loop bispectrum or the one-loop trispectrum for $K^{(4)}$ and $K^{(5)}$, respectively. Here we apply it to the
single-hard limits of the one- and two-loop contributions to the power spectrum $P_{ab}(k)$ that are related to deterministic bias renormalization, finding
\bea\label{eq:singleharddet}
  P^{(13)}_{ab}(k)\Big|^\text{single-hard}_\text{det.} &=& 3P^{(11)}_{ac}(k)c^{(3)}_{bc}\int_{p<\Lambda} P^{\text{lin}}(p)\,,\nn\\
  P^{(15)}_{ab}(k)\Big|^\text{single-hard}_\text{det.} &=& 10P^{(13)}_{ac}(k)c^{(5)}_{bc}\int_{p<\Lambda} P^{\text{lin}}(p)\,,\nn\\
  P_{ab}^{(33),I}(k)\Big|^\text{single-hard}_\text{det} &=& 3\left( P_{ac}^{(31)}(k)c^{(3)}_{bc}+c^{(3)}_{ac} P_{cb}^{(13)}(k)\right)\int_{p<\Lambda} P^{\text{lin}}(p)\,,\nn\\
  P_{ab}^{(24)}(k)\Big|^\text{single-hard}_\text{det} &=& 6P_{ac}^{(22)}(k)c^{(4)}_{bc} \int_{p<\Lambda} P^{\text{lin}}(p)\,.
\eea
Here a sum over $c_{bc}$ coefficients is implied. For the two-loop contributions, we highlight that each summand factorizes into a one-loop integral (either $P^{(13)}$, $P^{(31)}$ or $P^{(22)}$ with loop wavenumber denoted by $q$ below) and a hard loop factor given by $\int_{p<\Lambda} P^{\text{lin}}(p)$. We note that the wavenumber $q$ is unconstrained and thus can be soft, $q \ll k$, and one might worry about the residual IR divergencies known to exist in isolated $P^{(22)}$ and $P^{(13)}$ terms. However, these cancel out in the sum for the full two-loop power spectrum, as required by Galilean invariance, which is guaranteed by relations among the $c^{(n)}_{ab}$ coefficients for various $n$ as we show below [see Eq.\,\eqref{eq:ctos_relation} and Eq.\,\eqref{eq:singlehardrencorrespondence}]. Moreover, compared to the full two-loop expressions, we can bring the single-hard limit in the form above by renaming $p\leftrightarrow q$ if necessary, such as e.g. for $P^{(24)}$. For the hard region, it is understood that $p$ is restricted to values $p\gg q, k$. Nevertheless, the loop wavenumber $q$ inside the one-loop factor can also become hard (i.e.~$p\gg q\gg k$), and we deal with this `overlap' contribution with the double-hard limit $p,q\gg k$ below. Note that we included a factor $2$ for $P^{(15)}$, accounting for the two equivalent possibilities of taking one of the two loops to be hard. The result~\refeq{singlehard} is instrumental for taking into account the UV-sensitivity arising from the single-hard region when considering bias renormalization in Sec.\,\ref{sec:biasrenorm}.

\subsection{Deterministic terms: double-hard limit}

\begin{figure}[t]
	\centering
	\includegraphics[width = 0.6\textwidth]{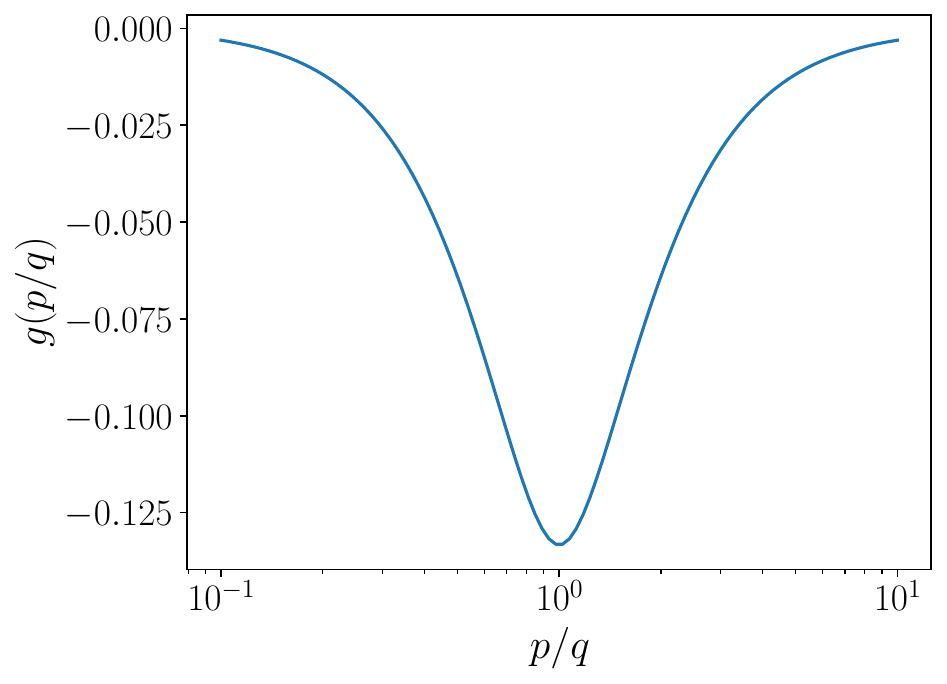}
	\caption{The universal function $g(p/q)$ that appears in the double-hard limit \refeq{K5doublehard} of all bias operators up to fifth order [see~\refeq{doublehardfuncg}]. The function $g(p/q)$ enters in the two-loop renormalization. It is symmetric around $p/q = 1$, as $g(r) = g(1/r)$, and has its extremum at $r=1$, where $g(1) = -2/15$. }
	\label{fig:gfunc}
\end{figure}

Here we consider the double-hard limit $p,q\gg k$, with ratio $r=p/q$ fixed.
Contributions to the two-loop power spectrum that correspond to renormalization of deterministic bias terms at two loop order arise from $P^{(15)}$ and $P^{(33),I}$.
For the former we require the double hard limit of the fifth-order bias kernels.
Specifically, this limit can be seen as the special $n=1$ case of the more general relation
\be\label{eq:kernelexpansiondoublehard}
  K_b^{(n+4)}({\bm k}_1,\dots,{\bm k}_n,{\bm q},-{\bm q},{\bm p},-{\bm p})^{p,q\to\infty }_{\text{av}_{\hat{\bf p},\hat{\bf q}}} \Big|_{r\equiv p/q=\text{fixed}}= d_{ba}^{(n+4)}(r)\, K_a^{(n)}({\bm k}_1,\dots,{\bm k}_n)\,,
\ee
where a sum over $a$ is implied, and the coefficients $d_{ba}^{(n+4)}(r)$ are functions of the (fixed) ratio $r=p/q$. This can be viewed as the extension  of the single-hard limit~\refeq{biaskernelexpansionsinglehard} to the double hard case. For the two-loop power spectrum we only need the case $n=1$,
\begin{eqnarray}\label{eq:K5doublehardgeneral}
  K^{(5)}_b({\bm k},{\bm p},-{\bm p},{\bm q},-{\bm q})_{\text{av}_{\hat{\bf p},\hat{\bf q}}}^{p,q\to\infty}\Big|_{p/q=\text{fixed}} &=& d^{(5)}_{ba}(p/q)\, K_a^{(1)}({\bm k})\,,
\end{eqnarray}
for which the sum over $a$ collapses to the single contribution $a=\delta$ from the linear bias operator $\O_a=\delta$. By explicit computation, we find that for all operators $\O_b$ up to fifth order, the double-hard limit can be decomposed into a sum of an $r$-independent part and a part proportional to a \textit{common} function $g(p/q)$, that is, independently of the operator $b$ under consideration,
\begin{eqnarray}\label{eq:K5doublehard}
   d^{(5)}_{b\delta}(p/q) &=& d^{(5)}_{b}+\tilde d^{(5)}_{b}\times g(p/q) \,,
\end{eqnarray}
where $g(p/q) \equiv g(r)$ is given by\footnote{Note that for $b = \delta$, we have $d_b^{(5)} = \tilde{d}_b^{(5)} = 0$, so that the double-hard limit starts at $O(k^2)$ as required by momentum conservation. This $O(k^2)$ contribution already appeared in e.g. \cite{blas_nl,Baldauf:2015aha,bernardeau_2loop}.} 
\begin{eqnarray}\label{eq:doublehardfuncg}
\boxed{
g(r) = \frac{15 r^8 - 40 r^6 + 18 r^4- 40 r^2 + 15}{240r^4}-\frac{(r^2-1)^4(r^2+1)}{64r^5}\ln\left(\frac{r+1}{r-1}\right)^2.
}
\end{eqnarray}
Our result for the numerical coefficients $d^{(5)}_{b}$ and $\tilde d^{(5)}_{b}$ is given in Table~\eqref{eq:doublehardcoeff}.
We display the function $g(p/q)$ in \reffig{gfunc}.
Note that $g(r)=g(1/r)$ due to the symmetry in $p\leftrightarrow q$ in the original kernel. This function will be relevant for the next Sections, since it determines the two-loop contribution to bias renormalization and the RGE of the linear bias parameter. We note that it is supported for $p/q=O(1)$, and vanishes in the hierarchical limit $p\ll q$ as $g\propto (p/q)^2$, and equivalently for $q\ll p$. This means that it captures the double-hard two-loop region, without overlap with single-hard contributions. This point will be discussed in more detail later on. 

On the other hand, we find that the $r$-independent contribution proportional to $d^{(5)}_b$ in~\refeq{K5doublehard} is related to sequentially taking the single-hard limits with respect to $p$, and then $q$. This means iterating~\refeq{biaskernelexpansionsinglehard} twice is equivalent to first taking the double-hard limit and then expanding for $r\ll 1$. Specifically, this equivalence implies the relation 
\be\label{eq:doublehardvssequentialsinglehard}
  d_{ba}^{(n+4)}(r=0) = c^{(n+4)}_{bc}c^{(n+2)}_{ca}\,,
\ee
between double- and single-hard limits. 
This expression says that when the ratio of the momenta is large ($r \to 0$), the contribution from the double-hard limit of an operator ${\cal O}_b$ (expanded at order $n+4$) to another operator ${\cal O}_a$ (expanded at order $n$) can be decomposed into the sum of products of two single-hard pieces. One from the single-hard contribution of ${\cal O}_b$ to another operator ${\cal O}_c$ (summed over $c$, and expanded at order $n+2$) and the subsequent contribution of ${\cal O}_c$ to ${\cal O}_a$.
This relation will be used for deriving two-loop RGEs below, and also for understanding the bias renormalization at two loops. We checked explicitly that this relation holds for $n+4=5$ and all bias operators up to fifth order. It is plausible to hold in general since the kernels are rational functions without any features such as e.g.~branch cuts, so that taking first the limit $q\to \infty$ and then $p\to\infty$ agrees with taking the limit $p,q\to\infty$ with $p/q$ fixed, and afterwards expanding for $p/q\to 0$.

Using~\refeq{K5doublehardgeneral} as well as~\refeq{biaskernelexpansionsinglehard}, the double-hard contributions at two loop which are related to the renormalization of deterministic bias operators are given by
\begin{eqnarray}\label{eq:doubleharddet}
  P^{(15)}_{ab}(k)\Big|^\text{double-hard}_\text{det.} &=& 15P^{(11)}_{a\delta}(k)\left(d^{(5)}_{b}\left(\int_{p<\Lambda} P^{\text{lin}}(p)\right)^2 \right.
  \\
  &&\hspace{2cm} \left.+ \,\tilde d^{(5)}_{b}\int_{p,q<\Lambda} P^{\text{lin}}(p)P^{\text{lin}}(q)g(p/q)\right)\,, \nonumber\\
  P^{(33),I}_{ab}(k)\Big|^\text{double-hard}_\text{det.} &=& 9c^{(3)}_{ac}P^{(11)}_{cd}(k)c^{(3)}_{bd}\left(\int_{p<\Lambda} P^{\text{lin}}(p)\right)^2\,.
\end{eqnarray}

\subsection{Stochastic terms: single-hard limit}\label{sec:shstoch}

We turn to single-hard loop contributions that are renormalized by stochastic terms in the bias expansion. In the case of the power spectrum, these correspond to loop integrals where the loop wavenumber appears in {\it both} kernels $K_a$ and $K_b$ in the loop integrand.\footnote{One can of course generalize that to taking limits involving loops of more than two operators. The limits involving more than two operators lead to non-Gaussian noise terms \cite{Rubira:2024tea}, but those do not appear in the two-point function.} This occurs in the one-loop term $P^{(22)}$, as well as for $P^{(33),II}$ and in one of the single-hard limits of $P^{(24)}$ at two-loop order. For $P^{(22)}$ the relevant limit is 
\ba \label{eq:K2sh}
\left(K^{(2)}_a({\bm k}-{\bm p},{\bm p})K^{(2)}_b({\bm k}-{\bm p},{\bm p})\right)^{p\to\infty} &= n_{ab}^{(22)} =n^{(2)}_{a}n^{(2)}_{b} \,,
\ea
with some numerical coefficients $n_{ab}^{(22)}$. For these kernel one can check that the limit of the product is equal to the product of the limits of each kernel, i.e. $n^{(22)}_{ab}=n^{(2)}_{a}n^{(2)}_{b}$ where $K^{(2)}_a({\bm k}-{\bm p},{\bm p})^{p\to\infty}=n_{a}^{(2)}$. The $n_{a}^{(2)}$ are given in the first column of Table~\eqref{eq:doublehardcoeffnoise}. We note that this limit does not require to average over angles. The corresponding single-hard limit of $P^{(22)}$ is thus given by
\begin{eqnarray}\label{eq:oneloophardlimit22}
  P^{(22)}_{ab}(k)\Big|^\text{single-hard}_\text{stoch.} &=& 2\,n^{(22)}_{ab}\int_{p<\Lambda} P^{\text{lin}}(p)^2 \,.
\end{eqnarray}
At two-loops, stochastic terms appear for $P^{(24)}$ and $P^{(33),II}$. 
The single-hard limits of the relevant kernels are 
\bea
  \frac12\Big( K_a^{(3)}({\bm k}-{\bm p}-{\bm  q},{\bm p},{\bm q})K_b^{(3)}({\bm k}-{\bm p}-{\bm  q},{\bm p},{\bm q}) + {}  &&\\
  K_a^{(3)}({\bm k}-{\bm p}+{\bm  q},{\bm p},-{\bm q})K_b^{(3)}({\bm k}-{\bm p}+{\bm  q},{\bm p},-{\bm q}) \Big)^{p\to\infty}_{\text{av}_{\hat{\bf p}}} &=& n_{ab}^{(33)}+\left(\frac{{\bm k}\cdot{\bm q}}{q^2}\right)^2\tilde n_{ab}^{(33)}\,,\nn\\
  K_a^{(2)}({\bm k}-{\bm p},{\bm p})K_b^{(4)}({\bm k}-{\bm p},{\bm p},{\bm q},-{\bm  q})^{p\to\infty}_{\text{av}_{\hat{\bf p}}} &=& n_{ab}^{(24)}+\left(\frac{{\bm k}\cdot{\bm q}}{q^2}\right)^2 \tilde n_{ab}^{(24)}\,, \nonumber
\eea
with numerical coefficients $n_{ab}^{(33,24)}$ and $\tilde n_{ab}^{(33,24)}$. Here we also symmetrized the kernels entering $P^{(33),II}$ under ${\bm q}\to-{\bm q}$, since this makes a cancellation of terms linear in ${\bm k}\cdot{\bm q}/q^2$ manifest.
We also define $n_{ab}^{(42)}\equiv n_{ba}^{(24)}$ and $\tilde n_{ab}^{(42)}\equiv \tilde n_{ba}^{(24)}$ for the kernels entering $P^{(42)}$. 
The wavenumber $q$ is unconstrained and can for example also be soft, $q\ll k$. We thus expect that terms proportional to $({\bm k}\cdot{\bm q}/q^2)^2$ cancel due to Galilean invariance. This is indeed the case in the sum $P^{(24)}_{ab}+P^{(42)}_{ab}+P^{(33),II}_{ab}$.
More precisely, we checked that
\be
  3\times 6\times \tilde n_{ab}^{(33)} + 12\times \tilde n_{ab}^{(24)} + 12\times \tilde n_{ab}^{(42)}=0\,.
\ee
Here  the factors $6$ and $12$ account for the prefactors contained in $P^{(33),II}_{ab}$ and $P^{(24)}_{ab}$ (as well as $P^{(42)}_{ab}$), respectively. The extra factor $3$ in the first term is best understood by considering the diagrammatic representation of $P_{ab}^{(33),II}$, where one can choose any one of the three lines connecting the two vertices to remain soft.

The resulting stochastic contribution from the single-hard limit of the sum of all two-loop terms of the power spectrum is 
\be\label{eq:P2Lstochsinglehard}
  P^\text{2L}_{ab}(k)\Big|^\text{single-hard}_\text{stoch.} = n_{ab}\left(\int_{p<\Lambda}P^{\text{lin}}(p)^2\right)\times\left(\int_{q<\Lambda} P^{\text{lin}}(q)\right)\,,
\ee
where $p\gg k,q$ is hard while the relative size of $q$ and $k$ is arbitrary and we make explicit that we are considering just the single-hard stochastic part. The numerical coefficients are
\be \label{eq:nab}
  n_{ab}= 3\times 6\times  n_{ab}^{(33)} + 12\times  n_{ab}^{(24)} + 12\times n_{ab}^{(42)}\,.
\ee
They are reported in Table~\eqref{eq:singlehardcoeffnoisetwoloop} for all combinations of bias operators $\O_a,\O_b$ (up to fourth order, since those are the relevant operators for the two-loop power spectra). 
We highlight that Eq.\,\eqref{eq:P2Lstochsinglehard} is independent of $k$, even though it could have in principle featured a non-trivial dependence on $k/q$ in the single-hard limit, such as for the single-hard deterministic terms, see Eq.\,\eqref{eq:singleharddet}. This is not a coincidence, but related to the structure of stochastic bias renormalization for the power spectrum, as we will see below.

\subsection{Stochastic terms: double-hard limit}\label{sec:noisedouble}

Finally, we consider the double-hard limit for two-loop contributions related to stochastic bias. Those appear in $P^{(24)}$ and $P^{(33),II}$. For the relevant limits of the corresponding kernels we find
\begin{eqnarray}\label{eq:K34doublehard}
  K^{(3)}_b({\bm k}-{\bm p}-{\bm q},{\bm p},{\bm q})^{p,q\to\infty} &=& n_b^{(3)} + \tilde n_b^{(3)}\times f({\bm p},{\bm q})\,,  \\
  K^{(4)}_b({\bm k}-{\bm p},{\bm p},{\bm q},-{\bm q})_{\text{av}_{\hat{\bf p},\hat{\bf q}}}^{p,q\to\infty} &=& n_b^{(4)} + \tilde n_b^{(4)}\times g(p/q) \,,\label{eq:K4doublehard}
\end{eqnarray}
with numerical coefficients $n_b^{(3)}$, $\tilde n_b^{(3)}$ and $n_b^{(4)}$, $\tilde n_b^{(4)}$ given in Table~\eqref{eq:doublehardcoeffnoise}, the function $g(r)$ was defined in \refeq{doublehardfuncg} and
\begin{eqnarray}\label{eq:doublehardfuncf}
f({\bm p},{\bm q}) &=& \frac{\left[({\bm p}+{\bm q})\cdot{\bm p}\right]\, \left[({\bm p}+{\bm q})\cdot{\bm q}\right]\, \left[{\bm p}\cdot{\bm q}\right]}{p^2q^2({\bm p}+{\bm q})^2}\,,
\end{eqnarray}
again independently of the operator $b$.  The double-hard limit of term $P^{(33),II}$ contains \refeq{K34doublehard} squared and the double-hard limit of $P^{(24)}$ \refeq{K4doublehard}  multiplied by a single-hard limit calculated in \refeq{K2sh}. 
Note that the limit of $K^{(3)}_b$ is taken without averaging over the directions of $\vec p$ and $\vec q$, as appropriate for $P^{(33),II}$. In contrast, for $K^{(4)}_b$ it is possible to average over directions due to the structure of $P^{(24)}$ and since $K^{(2)}_a(\vec k-\vec p,\vec p)^{p\to\infty}=n^{(2)}_a$ is independent of the direction of $\vec p$, see Sec.\,\ref{sec:shstoch}. Altogether, we find 
\begin{eqnarray}\label{eq:doublehardnoise}
  P^{(24)}_{ab}(k)\Big|^\text{double-hard}_{\rm stoch.} &=& 12\,n^{(2)}_{a}\left(n^{(4)}_{b}\left(\int_{p<\Lambda} P^{\text{lin}}(p)^2\right)\left(\int_{p<\Lambda} P^{\text{lin}}(p)\right) \right. \nonumber\\ 
  &&\hspace{4cm} \left. +\, \tilde n^{(4)}_{b}\int_{p,q<\Lambda} P^{\text{lin}}(p)^2P^{\text{lin}}(q)g(p/q)\right)\,,\nonumber\\
  P^{(33),II}_{ab}(k)\Big|^\text{double-hard}_{\rm stoch.} &=& 6\,n^{(3)}_{a}n^{(3)}_{b}\int_{p<\Lambda} P^{\text{lin}}(p)P^{\text{lin}}(q)P^{\text{lin}}(|{\bm p}+{\bm q}|) \nonumber\\ 
  && {} + 6\,(n^{(3)}_{a}\tilde n^{(3)}_{b}+\tilde n^{(3)}_{a}n^{(3)}_{b})\int_{p,q<\Lambda} P^{\text{lin}}(p)P^{\text{lin}}(q)P^{\text{lin}}(|{\bm p}+{\bm q}|)f({\bm p},{\bm q})\nonumber\\
  && {} + 6\,\tilde n^{(3)}_{a}\tilde n^{(3)}_{b}\int_{p,q<\Lambda} P^{\text{lin}}(p)P^{\text{lin}}(q)P^{\text{lin}}(|{\bm p}+{\bm q}|)f({\bm p},{\bm q})^2\,. 
\end{eqnarray}
We note that an analogous relation as~\refeq{doublehardvssequentialsinglehard} between double- and single-hard limits exists also for the stochastic contributions, see Appendix~\ref{app:hierarchynoise}.

\newpage
\section{Bias renormalization}\label{sec:biasrenorm}

Here we extend the treatment of renormalized bias presented in~\cite{McDonald:2006mx,Assassi:2014fva} to two-loop order, as well as to fifth order in the bias operators. We derive explicit expressions for renormalized bias operators, applicable to any $N$-point functions. Furthermore we discuss the extension of bias renormalization to stochastic terms, and apply our results to the two-loop power spectrum. As we make clear in~\refsec{connection_RG_renorm}, this renormalization scheme closely relates to the bias RG~\cite{Rubira:2023vzw}. 

We start by defining the renormalized bias operator
\be\label{eq:biasren}
  [{\cal O}_a]\equiv{\cal O}_a+ Z_{ab}{\cal O}_b\,,
\ee
where the ${\cal O}_a$ are the (usual) deterministic bias operators, and a sum over $b$ is implied (here by exception including also the constant operator $\unit=1$, see below). The coefficients $Z_{ab}$ are fixed by  {\it renormalization conditions}. Following~\cite{Assassi:2014fva}, we consider the conditions\footnote{Here and in the remainder of this work, we omit the wavenumber argument on the operator ${\cal O}_a$ (or its renormalized equivalent) when discussing the \textit{single-operator} renormalization condition. In this case, the operator argument also approaches zero by momentum conservation. }
\be\label{eq:rencond}
\boxed{
  \langle [{\cal O}_a] \delta_L({\bm k}_1)\cdots\delta_L({\bm k}_n)\rangle\big|_{{\bm k}_i\to 0} = \langle {\cal O}_a \delta_L({\bm k}_1)\cdots\delta_L({\bm k}_n)\rangle_\text{tree}\big|_{{\bm k}_i\to 0}\,,
}
\ee
to be satisfied for any value of $n\geq 0$. 
The left-hand side is the full correlation function, while the right-hand side contains only the tree-level contribution\footnote{It is sufficient to consider only connected diagrams, since the condition for disconnected ones is then automatically satisfied. To see why, consider for a moment some (arbitrary) disconnected contribution, of the form
$\langle{\cal O}_a\delta_L({\bm k}_1)\cdots\delta_L({\bm k}_n)\rangle|_\text{dis}=\langle{\cal O}_a\delta_L({\bm k}_1)\cdots\delta_L({\bm k}_m)\rangle^c\times \langle\delta_L({\bm k}_{m+1})\cdots\delta_L({\bm k}_n)\rangle$, where the first factor is connected (as indicated by the superscript), and the second can be either connected or again be a product of several disconnected pieces. If the renormalization condition holds for the connected pieces, then it also holds for the disconnected piece since the second factor contains only linear fields, and hence equals its tree-level expectation. Therefore~\refeq{rencond} is satisfied for the complete correlation function if it is satisfied for the connected parts. We thus only consider connected correlation functions below, and omit any superscript `c' for brevity.}.
Intuitively, this condition fixes the renormalized bias parameters of $n^{\text{th}}$-order to  match their tree-level measurement via the $(n+1)$-point function, e.g., fixing the linear bias parameter to match its value measured by the large-scale limit of the tracer-matter cross power spectrum. 

The tracer density field can be equivalently expanded in terms of bare or renormalized bias operators, with corresponding bias coefficients (and omitting stochastic terms for the moment, as well as higher derivative contributions), 
\be
  \delta_g =  b_a{\cal O}_a \equiv 
   b_{[a]}\, [{\cal O}_a]\,.
\ee
Using~\refeq{biasren}, the bare and renormalized bias are related as
\be\label{eq:biastrafo}
  b_a = b_{[a]} + b_{[c]} Z_{ca}\,.
\ee
In matrix-vector notation, $b=[b](1+Z)$,
where $b~ ([b])$ is a vector of all (renormalized) bias coefficients, and $(1+Z)_{ab}=\delta^K_{ab}+Z_{ab}$ with the Kronecker-delta. Thus, $[b]=b(1+Z)^{-1}$, where $(1+Z)^{-1}$ is the inverse matrix.

The case $n=0$ with no external legs is somewhat special, since it is the only one where the unit operator $\unit$ contributes to connected diagrams. This is the only operator for which $\langle{\cal O}\rangle_\text{tree}$ is non-zero at tree level, while other operators may possess an average value starting at loop level only. The renormalization condition evaluated with zero external legs thus yields (for all $a\not= \unit$)
\be\label{eq:zerocase}
  Z_{a\unit}=-\langle{\cal O}_a\rangle-Z_{ab}\langle{\cal O}_b\rangle\,,
\ee
where it is understood that the sum over $b$ now excludes the unit operator (which is our usual convention)\footnote{Note also that the unit operator has only a zeroth order piece, i.e. $\unit^{(n)} = 0$ for all $n>0$. Thus, there is no ambiguity when decomposing hard limits of kernels such as \refeq{singlehard}.}.
We can therefore rewrite~\refeq{biasren} in the form~\cite{Assassi:2014fva}  (for all $a\not= \unit$)
\be\label{eq:biasren2}
  [{\cal O}_a]={\cal O}_a-\langle{\cal O}_a\rangle+ Z_{ab}({\cal O}_b-\langle{\cal O}_b\rangle)\,.
\ee
On the other hand, for the unit operator, we have trivially (in position space)
\be\label{eq:unitop}
  [\unit]=\unit=1\,.
\ee
In practice, when inserting the renormalized bias into power-spectra, the average values yield contributions proportional to delta functions $\dirac({\bm k})$, and can thus be ignored at any $k>0$. A similar argument holds for the bispectrum, etc. We thus omit the averages in the following for simplicity, noting that they can always be easily restored by requiring that the renormalized bias operator has vanishing average.
We can expand~\refeq{rencond} in loop orders to obtain the $Z_{ab}$. At zero loops, one trivially has $Z_{ab}^\text{tree}=0$. Thus,
\be
  Z_{ab} = Z_{ab}^\text{1L}+Z_{ab}^\text{2L}+\dots\,.
\ee

\subsection{One-loop renormalization conditions} \label{sec:oneloop_renorm}

The renormalization conditions~\refeq{rencond} evaluated at one-loop read
\be
0 = \langle {\cal O}_a \delta_L({\bm k}_1)\cdots\delta_L({\bm k}_n)\rangle_\text{1L}\big|_{{\bm k}_i\to 0} + Z_{ab}^\text{1L}\langle {\cal O}_b \delta_L({\bm k}_1)\cdots\delta_L({\bm k}_n)\rangle_\text{tree}\big|_{{\bm k}_i\to 0}\,.
\ee
As mentioned above we only consider connected diagrams, and with at least $n\geq 1$, for which the average values of operators in~\refeq{biasren2} drop out. Furthermore, as also discussed above, the sum over $b$ excludes the trivial $\unit$ operator.
The tree-level and one-loop correlation function can be computed as 
\bea\label{eq:corrtree1L}
\langle{\cal O}_a\delta_L({\bm k}_1)\cdots\delta_L({\bm k}_n)\rangle'_\text{tree} &=& n! \,K_a^{(n)}({\bm k}_1,\dots,{\bm k}_n)P^{\text{lin}}(k_1)\cdots P^{\text{lin}}(k_n)\,,\nn\\
\langle{\cal O}_a\delta_L({\bm k}_1)\cdots\delta_L({\bm k}_n)\rangle'_{\rm 1L} &=& \\
&&\hspace{-3cm} \frac{(n+2)!}{2}\int_{p<\Lambda}K_a^{(n+2)}({\bm k}_1,\dots,{\bm k}_n,{\bm p},-{\bm p})P^{\text{lin}}(p)\Theta(\Lambda-p)P^{\text{lin}}(k_1)\cdots P^{\text{lin}}(k_n)\,, \nn
\eea
where the prefactors arise from all possible Wick contractions among linear fields and the prime denotes the omission of $(2\pi)^3 \dirac({\bm k}+{\bm k}_{1n})$, where ${\bm k}$ is the wavenumber attached to ${\cal O}_a$. The limit ${\bm k}_i\to 0$ corresponds to taking the (single-)hard limit of the loop, which can be computed using the expansion of the bias kernels from~\refeq{biaskernelexpansionsinglehard}. Altogether, this yields 
\be\label{eq:Z1L}
\boxed{
	Z_{ab}^\text{1L} = -\sigma_\Lambda^2 s_{ab}^{\rm 1L}\,,
}
\ee
with
\be\label{eq:ctos_relation}
  s_{ab}^{\rm 1L} \equiv \frac{(n+2)(n+1)}{2}c^{(n+2)}_{ab}\,,
\ee
where the $c^{(n+2)}_{ab}$ coefficients are related to the single-hard limits of the bias kernels $K^{(n+2)}_a$ [see~\refeq{biaskernelexpansionsinglehard} and Tables \eqref{eq:doublehardcoeff}, \eqref{eq:singlehardcoeffK4} and \eqref{eq:singlehardcoeffK5} for explicit expressions for $n+2=3,4,5$], and
\be\label{eq:sigmaLambda}
\sigma_\Lambda^2=\int_{p<\Lambda} P^{\text{lin}}(p)=\int_{0}^\Lambda dp \frac{p^2P^{\text{lin}}(p)}{2\pi^2}=\int_{0}^\Lambda d\Lambda' \frac{d\sigma^2_{\Lambda'}}{d\Lambda'}\,.
\ee
Our results from~\refeq{singlehardcoeffK5} thus generalize the bias renormalization from~\cite{McDonald:2006mx,Assassi:2014fva} up to fifth order for $a$ and third order for $b$. See \cite{DAmico:2022ukl} for the fourth-order operator renormalization and \cite{Assassi:2015fma,Patrone:2023cqe, Nikolis:2024kbx} for primordial non-Gaussianities.

Since bias renormalization holds at the operator level, $Z_{ab}$ cannot depend on $n$.
Moreover, this statement has to be true order by order in the loop expansion.
The one-loop result~\refeq{Z1L} thus implies that the coefficients $s_{ab}^{\rm 1L}$ are universal, which means the right-hand side of~\refeq{ctos_relation} has to be independent of $n$. Therefore, the single-hard limits~\refeq{biaskernelexpansionsinglehard} for different values of $n$ have to be related. We explicitly checked that our results for $c^{(n)}_{ab}$ at $n=3,4,5$ satisfy these relations, for example $10c^{(5)}_{ab}=6c^{(4)}_{ab}$. Note that $c^{(4)}_{ab}$, being related to the single-hard limit of the $K^{(4)}$ kernel, can capture only bias operators $\O_a$ up to fourth order and $\O_b$ up to second order. When restricting $c^{(5)}_{ab}$ to all of those cases, we indeed find $10c^{(5)}_{ab}=6c^{(4)}_{ab}$
using Tables \eqref{eq:singlehardcoeffK4} and \eqref{eq:singlehardcoeffK5}. Similarly we checked $6c^{(4)}_{ab}=3c^{(3)}_{ab}$ for all $\O_a$ up to third order and $\O_b=\delta$ using Tables \eqref{eq:doublehardcoeff} and \eqref{eq:singlehardcoeffK4}.

\subsection{Two-loop renormalization conditions}  \label{sec:twoloop_renorm}

The renormalization conditions~\refeq{rencond} evaluated at two-loop order are
\ba
  0 =\,& \langle {\cal O}_a \delta_L({\bm k}_1)\cdots\delta_L({\bm k}_n)\rangle_\text{2L}\big|_{{\bm k}_i\to 0} + Z_{ab}^\text{1L}\langle {\cal O}_b \delta_L({\bm k}_1)\cdots\delta_L({\bm k}_n)\rangle_\text{1L}\big|_{{\bm k}_i\to 0}
  \vs
  &+ Z_{ab}^\text{2L}\langle {\cal O}_b \delta_L({\bm k}_1)\cdots\delta_L({\bm k}_n)\rangle_\text{tree}\big|_{{\bm k}_i\to 0}\,.
\ea
Using
\bea\label{eq:corr2L}
  \langle{\cal O}_a\delta_L({\bm k}_1)\cdots\delta_L({\bm k}_n)\rangle'_{\text{2L}} &=& \frac{(n+4)!}{2^3}\int_{p,q<\Lambda}  K_a^{(n+4)}({\bm k}_1,\dots,{\bm k}_n,{\bm p},-{\bm p},{\bm q},-{\bm q})\nn\\
  && \hspace{-1cm} {} \times P^{\text{lin}}(p)\Theta(\Lambda-p)P^{\text{lin}}(q)\Theta(\Lambda-q)P^{\text{lin}}(k_1)\cdots P^{\text{lin}}(k_n),
\eea
as well as the double-hard limit expansion~\refeq{kernelexpansiondoublehard} of the bias kernels yields
\be\label{eq:Z2L}
  \boxed{
    Z_{ab}^\text{2L} = - Z_{ac}^\text{1L}\sigma_\Lambda^2 s_{cb}^{\rm 1L}
  -\frac12 \int_{p,q<\Lambda} s_{ab}^\text{2L}(p/q)P^{\text{lin}}(p)P^{\text{lin}}(q)\,,
  }
\ee
noticing the first term can be written as $Z_{ac}^\text{1L}\sigma_\Lambda^2 s_{cb}^{\rm 1L} = \sigma_\Lambda^4 s_{ac}^{\rm 1L} s_{cb}^{\rm 1L}$, making clear it is a one-loop contribution iterated twice. 
Here we defined
\be\label{eq:twoloopbetafunc}
  s^\text{2L}_{ab}(p/q) \equiv \frac{(n+4)!}{4n!}d_{ab}^{(n+4)}(p/q)\,,
\ee
being related to the expansion coefficients $d_{ab}^{(n+4)}(r)$ of the bias kernels $K^{(n+4)}_a$ in terms of $K^{(n)}_b$ in the double-hard limit, see~\refeq{kernelexpansiondoublehard}. These coefficients are functions of the ratio $r=p/q$, and are the two-loop generalization of the one-loop coefficients $s_{ab}^{\rm 1L}$ from~\refeq{ctos_relation}. The symmetry under $\vec p\leftrightarrow\vec q$ implies $s^\text{2L}_{ab}(r)= s^\text{2L}_{ab}(1/r)$. Using~\refeq{K5doublehard}, they are explicitly given by
\be\label{eq:twoloopbetafuncfordelta}
s^\text{2L}_{a\delta}(q/p)=30d^{(5)}_a+30\tilde d^{(5)}_a g(q/p)\,,
\ee
for ${\cal O}_b=\delta$. The coefficients $d^{(5)}_a$ and $\tilde d^{(5)}_a$ for all $\O_a$ up to fifth order are given in Table~\eqref{eq:doublehardcoeff}, and the function $g(r)$ in~\refeq{doublehardfuncg}. 

We expect consistency conditions similar as for the one-loop renormalization to hold, i.e. the double-hard limits of kernels at any order $n>5$ should be related to those at order $5$ such that the coefficients in~\refeq{twoloopbetafunc} are independent of $n$. This means e.g. that we expect that $90d^{(6)}_{ba}(q/p)=30d^{(5)}_{ba}(q/p)$ for all operators ${\cal O}_b$ up to order $n\leq 5$. Testing this would require to compute the double-hard limit of the sixth order kernel, which is beyond the scope of this work.

We note that the commutation-of-limits relation~\refeq{doublehardvssequentialsinglehard} implies  that when $q/p \to 0$ 
\be\label{eq:commutationoflimitsrelation}
s^\text{2L}_{ab}(0)  = s_{ac}^{\rm 1L}s_{cb}^{\rm 1L}\,.
\ee
Using also that $- Z_{ac}^\text{1L}\sigma_\Lambda^2 s_{cb}^{\rm 1L}=(\sigma_\Lambda^2)^2s_{ac}^{\rm 1L}s_{cb}^{\rm 1L}$ this implies that the contributions to $Z_{ab}^\text{2L}$ in the limit of hierarchical loop wavenumbers $p\gg q$ (or equivalently $q\gg p$) factorize into a product of two one-loop contributions, see~\refeq{Z2L} and~\refeq{Z1L}. This is physically reasonable, since configurations with $p\gg q\gg k$ can either be described by sequentially taking two single-hard limits, or by taking the double hard limit $p,q\gg k$ and then expanding for $p\gg q$.  This relation enters in the derivation of two-loop RGEs in Sec.~\ref{sec:rg}.

\subsection{Renormalized deterministic bias at two loop order}

Inserting the $Z$-factors~\refeq{Z1L} and~\refeq{Z2L} yields expressions for the renormalized deterministic bias operators $[\O_a]$ to be used when inserted in either one- or two-loop $N$-point correlation functions, 
\begin{equation}
\boxed{
\begin{aligned}
\label{eq:Oren1L2L}
[{\cal O}_a]\Big|_\text{1L}&=\, {\cal O}_a-\s^2_\L s_{ab}^{\rm 1L}{\cal O}_b\,,\\ 
 [ {\cal O}_a ]\Big|_\text{2L}&=\, {\cal O}_a-\s^2_\L s_{ab}^{\rm 1L}[{\cal O}_b]\Big|_\text{1L}
 -\frac12 \left(\int_{p,q<\Lambda} s_{ab}^\text{2L}(q/p)P^{\text{lin}}(p)P^{\text{lin}}(q)\right){\cal O}_b\,.
\end{aligned}  
}
\end{equation}
This is one of the main results in this work, together with the explicit expressions for the numerical coefficients $s_{ab}^{\rm 1L}$ and for $s_{ab}^\text{2L}(r)$ for bias operators up to fifth order. 
At one-loop, the subtraction term precisely cancels the (deterministic) hard (UV) part.
At two-loop, the last term takes care of the renormalization of deterministic bias from the double-hard limit, and the middle term of the single-hard limit. The fact that the one-loop renormalized bias appears in that term is related to the first term on the right-hand side of~\refeq{Z2L}, and takes care that the subtractions of single-hard and double-hard limits are not over-counted in the region where both limits overlap.

We stress that~\refeq{Oren1L2L} is valid when inserting the bias operators in arbitrary $N$-point functions at up to two-loop order, and thus also at field level. Below we apply it to the two-loop power spectrum. Before that, we discuss an extension of bias renormalization accounting also for stochastic bias.

\subsection{Stochastic bias renormalization}  \label{sec:noise_renorm}
To treat the stochastic contributions to the bias expansion, bias renormalization needs to be extended to products of bias operators at locations ${\bm x}$ and ${\bm y}$.
The renormalization of such composite operators needs to be complemented by contact terms that are supported only when both locations coincide.
We denote the renormalized product by $[{\cal O}_a({\bm x}){\cal O}_b({\bm y})]$, given by 
\be\label{eq:OPE}
  [{\cal O}_a({\bm x}){\cal O}_b({\bm y})]
  = [{\cal O}_a({\bm x})]\, [{\cal O}_b({\bm y})]
  + Z_{abc}\, [{\cal O}_c({\bm X})]\,\dirac({\bm x}-{\bm y})
  + \dots\,,
\ee
where the ellipsis stand for higher-gradient corrections, involving derivatives of the Dirac--delta, and ${\bm X}=({\bm x}+{\bm y})/2$.
This structure is analogous to that of the stochastic terms within the RG approach~\cite{Rubira:2024tea}, which are treated at the field level via higher-order-in-the-current terms in the partition function. In QFT language, it corresponds to an operator product expansion. 
In Fourier space, it reads
\be
  [{\cal O}_a({\bm k}){\cal O}_b({\bm k}')]
  = [{\cal O}_a({\bm k})]\, [{\cal O}_b({\bm k}')]
  + Z_{abc}\, [{\cal O}_c({\bm k}+{\bm k}')]
  + \dots\,,
\ee
with higher-gradient contributions indicated by the ellipsis suppressed by powers of $({\bm k}-{\bm k}')^2$.
Note that here the sum over index $c$ includes the unit operator $\unit$, and we introduce the notation
\be
  \Delta N_{ab} \equiv Z_{ab\unit}\,,
\ee
for that particular contribution.

To obtain the $Z_{abc}$ coefficients, we have to generalize the renormalization condition~\refeq{rencond}. An obvious choice is to require that
\begin{equation}\label{eq:rencondnoise}
\boxed{
\begin{aligned}
  \langle [{\cal O}_a({\bm k}){\cal O}_b({\bm k}')]\delta_L({\bm k}_1)\cdots\delta_L({\bm k}_n)\rangle|_{{\bm k},{\bm k}',{\bm k}_i\to 0} & \\
  &\hspace{-2.0cm}=\langle [{\cal O}_a({\bm k})]\,[{\cal O}_b({\bm k}')]\delta_L({\bm k}_1)\cdots\delta_L({\bm k}_n)\rangle_\text{tree}|_{{\bm k},{\bm k}',{\bm k}_i\to 0}\,,
  \end{aligned}
}
\end{equation}
where we explicitly note that \textit{all} wavenumbers are sent to zero.  Once more the choice of the renormalization conditions in the IR limits is such that the loop-contributions vanish in that limit and the stochastic contributions are defined in terms of tree-level $N$-point functions.\footnote{Note that the $k,k' \to 0$ limit agrees with the conditions imposed in \cite{Rubira:2024tea}, in which the support of the currents is chosen in the IR. }
At one-loop this yields the conditions
\ba\label{eq:rencondnoise1L}
  0 &= 
  \langle [{\cal O}_a({\bm k})]\,[{\cal O}_b({\bm k}')]\delta_L({\bm k}_1)\cdots\delta_L({\bm k}_n)\rangle_\text{1L}|_{{\bm k},{\bm k}',{\bm k}_i\to 0} \vs
  &\hspace{4cm} + Z_{abc}^\text{1L}\langle [{\cal O}_c({\bm k}+{\bm k}')]\delta_L({\bm k}_1)\cdots\delta_L({\bm k}_n)\rangle_\text{tree}|_{{\bm k},{\bm k}',{\bm k}_i\to 0} \,.
\ea
Applying the renormalization condition for $n=0$ and using that $[\unit(\vec k+\vec k')]=(2\pi)^3\dirac(\vec k+\vec k')$ as well as $\langle[\O_c]\rangle=0$ for all non-trivial renormalized bias operators yields 
\be\label{eq:Nab1L}
  \Delta N_{ab}^\text{1L} \equiv Z^\text{1L}_{ab\unit} = -P^\text{1L}_{[a][b]}(0)\,,
\ee
where $P^\text{1L}_{[a][b]}(k)$ is the one-loop power spectrum with two renormalized operators inserted. 
This argument can be simply extended to two-loop order,
\be\label{eq:Nab2L}
  \Delta N_{ab}^\text{2L} \equiv Z^\text{2L}_{ab\unit} = -P^\text{2L}_{[a][b]}(0)\,,
\ee
and in fact holds at any loop order, i.e. $\Delta N_{ab}= -P_{[a][b]}(0)$. We present the calculation of $\Delta N_{ab}^\text{1L}$ and $\Delta N_{ab}^\text{2L}$ below, in \refsec{appicationto2L}. We denote the power spectra for fully renormalized composite bias operators $[{\cal O}_a(\vec k){\cal O}_b(\vec k')]$ by
\be
  \langle [{\cal O}_a({\bm k}){\cal O}_b({\bm k}')]\rangle \equiv (2\pi)^3\dirac({\bm k}+{\bm k}')P_{[ab]}(k) \,.
\ee
Thus, for the power spectrum, composite operator renormalization [with renormalization condition given by~\refeq{rencondnoise}] at leading order in the derivative expansion simply amounts to subtracting the $k$-independent shot noise term such that the fully renormalized tracer power spectrum $P_{[ab]}$ approaches zero in the $k\to 0$ limit,
\be\label{eq:noiseren}
  P_{[ab]}(k) = P_{[a][b]}(k)+\Delta N_{ab}\,.
\ee
This result holds at any loop order. 

Explicit expressions for $Z_{abc}$ for non-trivial operators $c\neq \unit$ can be derived for $n>0$ at one-loop  from~\refeq{rencondnoise1L}, and similarly at two loop order. We note that those mixed stochastic-deterministic contributions ($Z_{abc}$ for $c\neq \unit$) are required for renormalization of the bispectrum and higher $N$-point correlators, along with a generalization of Eq.~\eqref{eq:OPE} to products of three or in general $N$ operators. However, neither $Z_{abc}$ for $c\neq \unit$ nor composite operator renormalization for a product of three or more bias operators is needed for renormalizing the power spectrum at any loop order, and thus not considered in this work.

\subsection{Application to the two-loop power spectrum} \label{sec:appicationto2L}

We apply bias renormalization at two-loop order to the power spectrum, including deterministic and stochastic contributions at leading order in the derivative expansion. 
We also consider the power spectrum of the individually renormalized bias operators $[\O_a]$ and $[\O_b]$, $P_{[a][b]}(k)$, and the one for the unrenormalized bias $\O_a$ and $\O_b$, which we denote by $P_{ab}(k)$. 
The tracer power spectrum can be equivalently expressed in terms of any of these spectra (leaving higher-derivative terms implicit) as 
\bea\label{eq:Pggrewriting}
  P_{gg}(k) &=& \sum_{a,b}b_ab_bP_{ab}(k)+\text{cst.}+\text{h.d.}\nn\\
   &=& \sum_{a,b}b_{[a]}b_{[b]}P_{[a][b]}(k)+\text{cst.}+\text{h.d.}\nn\\
   &=& \sum_{a,b}b_{[a]}b_{[b]}P_{[ab]}(k)+[\text{cst.}]+\text{h.d.}\,,
\eea
where $b_a$ and cst. (not to be confused with the unit operator $\mathbb{1}$ from \refeq{unitop}) are the `bare' deterministic and constant stochastic bias coefficients, respectively, and $b_{[a]}=b_c[(1+Z)^{-1}]_{ca}$ as well as $[\text{cst.}]\equiv \text{cst.}-b_{[a]}b_{[b]}\Delta N_{ab}$ the renormalized ones.

We note that at linear level,
\be
  P^\text{lin}_{[ab]}(k)=P^\text{lin}_{[a][b]}(k)=P^\text{lin}_{ab}(k)\equiv P^{(11)}_{ab}(k)=\delta^K_{a\delta}\delta^K_{b\delta}P^{\text{lin}}(k)\,,
\ee
given by the usual linear spectrum $P^{\text{lin}}(k)$ for $a=b=\delta$, and zero otherwise.\footnote{Note that, in real space and with the basis constructed from \refeq{Pin} the velocity divergence $\theta$ is not included in the bias basis, such that $\delta$ is the only bias operator starting at linear order. 
}
To derive explicit expressions at one- and two-loop level we proceed in two steps:
\begin{enumerate}
\item As a first step, we consider the power spectra $P^\text{1L/2L}_{[a][b]}(k)$ of the individually renormalized bias operators derived in~\refeq{Oren1L2L}. This corresponds to taking renormalization of deterministic bias into account. 
\item In the second step, we include also stochastic bias renormalization, i.e. go from $P^\text{1L/2L}_{[a][b]}(k)$ to the fully renormalized power spectrum $P^\text{1L/2L}_{[ab]}(k)$ for the renormalized composite operator $[{\cal O}_a(\vec k){\cal O}_b(\vec k')]$ using~\refeq{noiseren}. 
\end{enumerate}
Using~\refeq{Oren1L2L}, the first step yields 
\bea
  \label{eq:P1Lrennonnoise}
  P^\text{1L}_{[a][b]} &=&  P^\text{1L}_{ab} - \s^2_\L\left(s_{ac}^{\rm 1L}P^\text{lin}_{cb}+(a\leftrightarrow b)\right)\,,\\
 \label{eq:P2Lrennonnoise}
 P^\text{2L}_{[a][b]} &=& P^\text{2L}_{ab} -\s^2_\L \bigg( s_{ac}^{\rm 1L}P^\text{1L}_{[c][b]}+(a\leftrightarrow b)\bigg) \\
	&&~ \hspace{0.5cm} {} - \Bigg(\frac12 \int_{p,q<\Lambda} \bigg(s_{ac}^\text{2L}(q/p) P^\text{lin}_{cb}+(a\leftrightarrow b)\bigg) P^{\text{lin}}(p)P^{\text{lin}}(q) + (\s^2_\L)^2s_{ac}^{\rm 1L}P^\text{lin}_{cd}\bar s_{db}^{\rm 1L}\Bigg)\,,\nn
\eea
where we introduced the transposed matrix 
\be\label{eq:sbardef}
\bar s_{db}^{\rm 1L}\equiv s_{bd}^{\rm 1L}.
\ee
Using~\refeq{ctos_relation} and~\refeq{twoloopbetafunc} we find the important result that the various subtraction terms arising from the deterministic bias renormalization relation~\refeq{Oren1L2L} \emph{precisely} give rise to the deterministic single- and double-hard limits of the one- and two-loop power spectrum derived in~\refeq{singleharddet} and~\refeq{doubleharddet}, respectively, 
\bea\label{eq:singlehardrencorrespondence}
	\s^2_\L\left(s_{ac}^{\rm 1L}P^\text{lin}_{cb}+(a\leftrightarrow b)\right) &=&   P^\text{1L}_{ab}\Big|^\text{hard}_\text{det.}
	\,,\nn\\
    \s^2_\L \bigg( s_{ac}^{\rm 1L}P^\text{1L}_{[c][b]}+(a\leftrightarrow b)\bigg) 
    &=&  P^\text{2L}_{ab}\Big|^\text{single-hard}_\text{det.}
		- P^\text{2L}_{ab}\Big|^\text{single-hard}_\text{det.}\Bigg|^\text{hard}  \,,
\eea
and
\bea\label{eq:doublehardrencorrespondence}
	\frac12 \int_{p,q<\Lambda} \bigg(s_{ac}^\text{2L}(q/p) P^\text{lin}_{cb}+(a\leftrightarrow b)\bigg) P^{\text{lin}}(p)P^{\text{lin}}(q) + (\s^2_\L)^2s_{ac}^{\rm 1L}P^\text{lin}_{cd}\bar s_{db}^{\rm 1L}
    &=&  P^\text{2L}_{ab}\Big|^\text{double-hard}_\text{det.} 
    \,.\nn\\
\eea
Thus, as expected, bias renormalization removes the leading UV sensitive parts of the loop integrals. We stress that the relation between the single-hard coefficients $c_{ab}^{(n)}=2s_{ab}^\text{1L}/(n(n-1))$ for $n=3,4,5$ implied by~\refeq{ctos_relation} is crucial for obtaining this result, particularly for seeing that the $P^{(13)}$ and $P^{(22)}$ contributions in the single-hard limit of the two-loop power spectrum can be combined into $P^\text{1L}$. Furthermore, note that the single-hard two-loop subtraction term involves the one-loop power spectrum $P^\text{1L}_{[c][d]}$ of the {\it renormalized} deterministic bias operators $[\O_c]$ and $[\O_d]$. Thus, as already hinted at above, bias renormalization automatically takes care of subtracting the overlap of single- and double-hard limits, such that no double-counting arises. This can be seen by the fact that the single-hard deterministic subtraction term in the second line in Eq.\,\eqref{eq:singlehardrencorrespondence} is itself a difference between the two expressions given on the right-hand side. 

In the second step we use~\refeq{Nab1L} and~\refeq{Nab2L} as well as~\refeq{noiseren}. At one-loop order, using~\refeq{P1Lrennonnoise}
and assuming that the linear power spectrum vanishes for $k=0$, $P^{\text{lin}}(0)=0$, we have 
\be 
  \Delta N_{ab}^\text{1L} = -P^\text{1L}_{ [a][b] }(0)
  = -P^\text{1L}_{ ab }(0)
  = -P^\text{1L}_{ ab }(k)\Big|^\text{single-hard}_\text{stoch.}\,,
\ee
with the latter given by the stochastic contribution to the single-hard limit, see~\refeq{oneloophardlimit22}. At two-loop,
\be 
  \Delta N_{ab}^\text{2L} = -P^\text{2L}_{ [a][b] }(0) = -P^\text{2L}_{ ab }(0) - \s^2_\L\big(s_{ac}^{\rm 1L}\Delta N_{cb}^\text{1L}+(a\leftrightarrow b)\big)\,.
\ee
Here the first term is given by the stochastic contribution to the double-hard limit,
\be
  P^\text{2L}_{ ab }(0) = P^\text{2L}_{ ab }(k)\Big|^\text{double-hard}_\text{stoch.}\,,
\ee
see~\refeq{doublehardnoise} for the explicit result, while the second term can be combined with the deterministic single-hard subtraction term~\refeq{P2Lrennonnoise} within the fully renormalized power spectrum~\refeq{noiseren}. 

Altogether, the fully renormalized power spectrum at one and two loops is given by
\begin{equation}\label{eq:pren1L2L}
\boxed{
\begin{aligned}
  P^\text{1L}_{[ab]}(k) &= P^\text{1L}_{ab}(k)-P^\text{1L}_{ab}(k)\Big|^\text{single-hard}_\text{det.+stoch.}\\
  &= P^\text{1L}_{ab}(k)-\s^2_\L\big(s_{ac}^{\rm 1L}P^\text{lin}_{cb}(k)+(a\leftrightarrow b)\big)-P^\text{1L}_{ab}(0)\,,\\
  P^\text{2L}_{[ab]}(k) &= P^\text{2L}_{ab}(k) 
  -\s^2_\L\big(s_{ac}^{\rm 1L}P^\text{1L}_{[cb]}(k)+(a\leftrightarrow b)\big)-P^\text{2L}_{ab}(k)\Big|^\text{double-hard}_\text{det.+stoch.}\\
  &= P^\text{2L}_{ab}(k) 
  -\s^2_\L\big(s_{ac}^{\rm 1L}P^\text{1L}_{[cb]}(k)+(a\leftrightarrow b)\big)-P^\text{2L}_{ab}(k)\Big|^\text{double-hard}_\text{det.}-P^\text{2L}_{ab}(0)\,.
\end{aligned}
}
\end{equation}
The one-loop expression agrees with the standard result, for which the two subtraction terms in the second line take care of the UV sensitivity of $P^{(13)}$ and $P^{(22)}$, respectively, associated to deterministic and stochastic bias renormalization. The expressions in \refeq{pren1L2L} can be computed numerically using for example the methodology outlined in \cite{bakx_cobra}. We highlight several features of the two-loop result:
\begin{itemize}
\item The subtraction term taking care of removing the UV sensitivity associated to the single-hard limit of the bare two-loop power spectrum $P^\text{2L}_{ab}(k)$ contains itself the fully {\it renormalized} one-loop power spectrum $P^\text{1L}_{[cb]}(k)$ as well as the numerical one-loop bias renormalization coefficients $s_{ac}^{\rm 1L}$ [see Table~\eqref{eq:singlehardcoeffK5} for explicit expressions up to fifth order bias, noting that $s_{ab}^{\rm 1L}=10c^{(5)}_{ab}$].
\item The further subtraction term takes care of the double-hard limit at two-loop order. It is given by the sum of deterministic and stochastic contributions, and explicit results for both are given in~\refeq{doubleharddet} and~\refeq{doublehardnoise} for up to fifth order bias.
\item No mixed stochastic/deterministic terms (corresponding to $Z_{abc}$ with nontrivial $c\neq\unit$) are required for the power spectrum at any loop order. They would appear when inserting the composite renormalized bias operator in three- or higher $N$-point functions only. 
\item Stochastic bias renormalization simply amounts to subtracting the $k=0$ limit for the power spectrum at leading order in the derivative expansion, as naively expected. This holds at any loop order.
\end{itemize}
Lastly, while all fifth-order kernels from \refeq{bias_basis} are mutually independent for general momenta ${\bm q}_1, \dots {\bm q}_5$, these fifth-order kernels only enter the two-loop power spectrum through the $P^{(15)}$ diagram in \refeq{twoloopcontribs}. Similarly, fourth-order kernels enter only in $P^{(24)}$ in \refeq{twoloopcontribs}.
The special configurations of wavenumber arguments of the type `${\bm p},-{\bm p}$' in these diagrams as well as the integration over the angles of loop wavenumbers implies that some of the fourth- and fifth-order ${\cal O}_a$ operators become redundant in the two-loop power spectrum and need not be considered (see~\cite{egge_oneloopbisp} for a general discussion, and detailed treatment of fourth order bias applied to the one-loop bispectrum).
They are shown in gray in Tables \eqref{eq:doublehardcoeff} and \eqref{eq:singlehardcoeffK5}.
As a first example, this applies to the fifth-order operators that are built out of contractions of five instances of $\Pi_{ij}^{[1]}$.
We refer to them collectively as $(\Pi^{[1]})^5$ operators below.
Their fifth-order kernels are built out of contractions of the \textit{linear} operator $(k_i k_j / k^2)\delta^{(1)}$ only.
Hence, the $P^{(15)}$ integrand from \refeq{twoloopcontribs} does not depend on the magnitude of the external wavenumber $k$, and after integrating over angles of the loop momenta, it must be a constant function of $k$. In other words, these operators only contribute to the renormalization of the linear bias coefficient [since $P^{(15)}$ becomes proportional to $P^{\text{lin}}(k)$]\footnote{As a corollary, this same argument applied at one loop in the power spectrum shows that only ${\rm tr}\big[\Pi^{[1]} \Pi^{[2]} \big]$ needs to be included at third order, which is expressable in terms of more commonly used choices such as $\Gamma_3$ \cite{Desjacques:2016bnm}.}.

This argument can be extended and made more systematic by considering the renormalized power spectrum Eq.~\eqref{eq:Pggrewriting}, composed of the renormalized operators from Eq.~\eqref{eq:Oren1L2L}. For the present discussion it is most convenient to consider $P_{[a][b]}$ from the middle line in Eq.~\eqref{eq:Pggrewriting}, noting that all lines in that equation are equivalent representations of the tracer power spectrum. The power spectra $P_{[a][b]}$ are given by  replacing the bias kernels in Eq.~\eqref{eq:twoloopcontribs} by those for the renormalized operators from Eq.~\eqref{eq:Oren1L2L}, which involve subtraction terms removing the leading contributions in the hard limits.
For certain bias operators, those subtraction terms have the effect that the renormalized $P_{[a][b]}$ is identically zero. This is the case for the $(\Pi^{[1]})^5$ operators mentioned already above. To give an impression of how this occurs, consider first the slightly generalized fifth-order kernels of the form $K_a^{(5)}({\bm k}_1,{\bm k}_2,{\bm k}_3,{\bm p},-{\bm p})$, which are relevant for $P^{(15)}$ (for ${\bm k}_3=-{\bm k}_2$), as well as the one-loop trispectrum, which we take as an illustrative example for a moment. One-loop renormalization from Eq.~\eqref{eq:Oren1L2L} implies that the difference $K_a^{(5)}({\bm k}_1,{\bm k}_2,{\bm k}_3,{\bm p},-{\bm p})_{\text{av}_{\hat p}}-K_a^{(5)}({\bm k}_1,{\bm k}_2,{\bm k}_3,{\bm p},-{\bm p})_{\text{av}_{\hat p}}^{p\to\infty}$ enters the renormalized one-loop trispectrum, noting that both terms are averaged over the direction of ${\bm p}$, but the limit $p\to\infty$ is taken only in the subtraction term obtained from the one-loop renormalization condition. Since for $(\Pi^{[1]})^5$ operators the kernels depend only on angles between the wavevectors (as argued above), this difference is precisely zero for those operators. Thus, $(\Pi^{[1]})^5$ operators do not contribute to the renormalized one-loop trispectrum. Using instead the two-loop renormalization condition from Eq.~\eqref{eq:Oren1L2L} as appropriate for $P^{(15)}$, yields in a similar way that they also do not contribute to the renormalized two-loop power spectrum.

Consider next the fourth-order bias operators composed of four instances of $\Pi_{ij}^{[1]}$, that we refer to as $(\Pi^{[1]})^4$ operators\footnote{The authors thank Rom\'an  Scoccimarro and Alexander Eggemeier for making them aware of the redundancy of those fourth-order operators in the two-loop power spectrum.}. They enter the two-loop power spectra $P_{[a][b]}$ via the $P^{(24)}$ and $P^{(15)}$ diagrams. In the former, we need to insert the one-loop renormalized bias from Eq.~\eqref{eq:Oren1L2L}. For  $(\Pi^{[1]})^4$ operators, the resulting subtraction term makes $P^{(24)}$ vanish identically, in close analogy to the case discussed above. 
However, a priori those operators could contribute to $P^{(15)}$.
Applying the appropriate two-loop renormalization from Eq.~\eqref{eq:Oren1L2L}, reveals that the various subtraction terms arising in that case lead to a precise cancellation as well, such that the renormalized $P^{(15)}$ vanishes identically for $(\Pi^{[1]})^4$ operators. This finding is in line with the discussion of the general structure of operator redundancy in~\cite{egge_oneloopbisp}, implying that $(\Pi^{[1]})^4$ operators could give a non-redundant contribution only starting from the three-loop power spectrum (specifically the $P^{(44)}$ contribution).

Finally, we find that the renormalized two-loop power spectra for all fifth-order operators involving three instances of $\Pi_{ij}^{[1]}$ and one of $\Pi_{ij}^{[2]}$ vanish identically as well. This is due to a cancellation among all of the subtraction terms arising from the two-loop renormalization Eq.~\eqref{eq:Oren1L2L} inserted in $P^{(15)}$. This means in summary that $3+4+5=12$ of the 29 leading-gradient bias operators up to fifth order need not to be considered for the two-loop power spectrum.

We finally note that, for practical application, the bias basis considered in this work should be complemented by higher-derivative corrections. When adopting the commonly used scheme of counting loop orders equally to the orders in the gradient expansion, all bias operators suppressed with two gradients and up to third perturbative order should be added, as well as one with four gradients and a single linear density field. Furthermore, also gradient corrections to the stochastic terms need to be added. A discussion of renormalization at next-to-leading power in the gradient expansion is left to future work.

\newpage
\section{The renormalization group equations}\label{sec:rg}

In this and the following Sections we turn to the alternative RG approach for controlling the dependence of bias operators on the smoothing scale $\L$, counter-balanced by correspondingly $\L$-dependent bias coefficients $b_a(\L)$. While~\cite{Rubira:2023vzw}  considered the Wilson-Polchinski formulation, in which modes are integrated out in shells in Fourier space, we follow an alternative RG approach exploiting more directly the independence of physical observables, i.e.~the galaxy density contrast $\delta_g({\bf x})$ (and its $N$-point functions), on $\L$ as starting point. 
This is advantageous for extending the RG to two loop order, while agreeing with the Wilson-Polchinski results~\cite{Rubira:2023vzw}  at one loop. We also note that the relation to renormalized bias, namely $b_{[a]}=b_a(\L\to 0)$, carries over from~\cite{Rubira:2023vzw} to the one- and two-loop RG we consider here.\footnote{
Within the RG approach, we focus on deterministic bias in this work, noting that an extension of the one-loop RG for stochastic terms from~\cite{Rubira:2024tea}  to two loop would also be possible within the RG approach employed here,  using the formulation from~\refsec{noise_renorm}.}

\subsection{The N-point function renormalization conditions}

To derive two-loop RG equations we 
start from the requirement that the tracer field is independent of the smoothing scale $\Lambda$ (we suppress time arguments in all quantities for brevity),
\be\label{eq:RGraw}
  0=\frac{d}{d\Lambda}\delta_g({\bm x})=\frac{db_a}{d\Lambda}{\cal O}_a({\bm x})+b_a\frac{d{\cal O}_a({\bm x})}{d\Lambda}\,.
\ee
As before, the smoothing is defined such that it is applied to the linear density field with a sharp-$k$ filter in Fourier space, see \refeq{filter}, and the bias operators ${\cal O}_a$ are constructed from this smoothed field as in \refeq{K_kernels} (we emphasize once more that smoothed initial conditions are implicitly assumed throughout this work). 
To derive explicit RG equations we use that~\refeq{RGraw} is an operator equation and thus has to be valid when inserted into arbitrary correlation functions,\footnote{We note that it is sufficient to consider only the \emph{connected} parts of the correlations functions in both summands (see footnote at the beginning of Section \ref{sec:biasrenorm}).}
\be\label{eq:RGcorr}
\boxed{0 = \frac{db_a}{d\Lambda}\langle{\cal O}_a\delta_L({\bm k}_1)\cdots\delta_L({\bm k}_n)\rangle + b_a \frac{d}{d\Lambda}\langle{\cal O}_a\delta_L({\bm k}_1)\cdots\delta_L({\bm k}_n)\rangle\,.}
\ee
Here we used that in the second term the $\Lambda$-derivative can be pulled out of the average, since the only source of $\Lambda$-dependence comes from the smoothed fields entering the ${\cal O}_a$. Below we also assume that $|{\bm k}_i|\ll \Lambda$ are IR wavenumbers 
below the smoothing scale. Of course other correlators could be chosen to derive the RG, but in analogy with \cite{Assassi:2014fva}, the correlation with the linear density field allows for a simple derivation in terms of a single operator kernel limit for the deterministic part, as we see below. Moreover, the smoothing of the initial Gaussian fluctuations resembles the smoothing of the free field considered in the Wilson-Polchinski framework \cite{Carroll:2013oxa, Rubira:2023vzw}.

Next, we treat the peculiar case $n=0$ separately, as in the discussion around \refeq{zerocase}, for which no external lines are attached. There is one special bias operator which contributes to~\refeq{RGcorr} only in that case (when restricting to connected correlation functions), being the trivial operator $\unit$.
Since $\langle \unit \rangle=1$, this yields 
\be
  0 = \frac{db_\unit}{d\Lambda} + \frac{d}{d\Lambda}\left(b_a\langle {\cal O}_a\rangle\right),
\ee
where the unit operator $\unit$ is excluded from the summation over all bias operators $a$ in the second term. We use this convention also below. The solution is simply $b_\unit=-b_a\langle {\cal O}_a\rangle$. The role of this constant bias operator is thus just to ensure that the average of the tracer field vanishes,
\be
\langle\delta_g\rangle =  0 \,.
\ee
We do not need to consider $b_\unit$ in~\refeq{RGcorr} for all non-trivial cases with $n>0$, since $\unit$ does not contribute to connected diagrams. This also means that $b_\unit$ cannot show up in the RG equations for any non-trivial bias coefficients $b_a$.
In summary, we consider~\refeq{RGcorr} for $n>0$ in the following, and take only bias operators into account that are at least linear in the density fields (i.e. exclude the constant bias $b_\unit$ in any summation over bias terms).

\refeq{RGcorr} has to be true order by order in the loop expansion. We can decompose the running into contributions coming from one, two-loop and so on,
\be
  \frac{db_a}{d\Lambda} = \frac{db_a}{d\Lambda}\Big|_{\text{1L}} +  \frac{db_a}{d\Lambda}\Big|_{\text{2L}} + \dots\,.
\ee
In the following Sections, we discuss the one and two-loop results separately.

\subsection{One-loop renormalization group} \label{sec:oneLoopRG}

The one-loop part of \,\refeq{RGcorr} yields 
\be\label{eq:RGcorr1L}
0 = \frac{db_a}{d\Lambda}\Big|_{\text{1L}} \langle{\cal O}_a\delta_L({\bm k}_1)\cdots\delta_L({\bm k}_n)\rangle_\text{tree} + b_a \frac{d}{d\Lambda}\langle{\cal O}_a\delta_L({\bm k}_1)\cdots\delta_L({\bm k}_n)\rangle_{\text{1L}}\,.
\ee
Using \refeq{corrtree1L} and $k_i<\Lambda$ we get 
\be\label{eq:RGkernel1L}
0 = \frac{db_a}{d\Lambda}\Big|_{\text{1L}}  n!\,K_a^{(n)}({\bm k}_1,\dots,{\bm k}_n) + b_b \frac{(n+2)!}{2}K_b^{(n+2)}({\bm k}_1,\dots,{\bm k}_n,{\bm p},-{\bm p})_{\text{av}_p}^{p=\Lambda}\frac{\Lambda^2P^{\text{lin}}(\Lambda)}{2\pi^2}\,.
\ee
Note that we renamed the summation index into $b$ in the second term for later convenience.
The notation means that the $(n+2)$th order kernel is averaged over the direction of ${\bm p}$, and evaluated at $p\equiv |{\bm p}|=\Lambda$.
Following~\cite{Rubira:2023vzw}, the last factor can be written as
\be \label{eq:ds2dL}
  \frac{d\sigma^2_\Lambda}{d\Lambda}\equiv \frac{d}{d\Lambda}\int_{p<\Lambda} P^{\text{lin}}(p) = \frac{\Lambda^2P^{\text{lin}}(\Lambda)}{2\pi^2}\,.
\ee
Expanding the kernel in the limit $|{\bm k}_i| \ll \Lambda$ yields a series of contributions, that correspond to higher and higher orders in the gradient expansion, suppressed by powers of $k_i/\Lambda$. The leading-power contribution is obtained by taking $k_i\to 0$, or equivalently $p\to \infty$. In that limit, we can decompose the $(n+2)$th kernel into the kernels of the bias terms at $n$th order using \refeq{biaskernelexpansionsinglehard}, since they form a complete basis, 
to get after inserting into~\refeq{RGkernel1L}
\be 
0 = \frac{db_a}{d\Lambda}\Big|_{\text{1L}}  n!\,K_a^{(n)}({\bm k}_1,\dots,{\bm k}_n) + b_b \frac{(n+2)!}{2}c^{(n+2)}_{ba} K_a^{(n)}({\bm k}_1,\dots,{\bm k}_n) \frac{d\sigma^2_\Lambda}{d\Lambda}+O\left(\frac{|{\bm k}_i|}{\Lambda}\right)^2\,.
\ee
Using that the bias operators form a minimal basis such that the equation has to satisfied for each coefficient of the $K_a^{(n)}({\bm k}_1,\dots,{\bm k}_n)$,  we obtain the one-loop RG equation at leading order in gradient expansion 
\be\label{eq:RG1L}
\boxed{
 \frac{db_a}{d\Lambda}\Big|_{\text{1L}}  = -b_b s_{ba}^{\rm 1L} \frac{d\sigma^2_\Lambda}{d\Lambda}\,,
}
\ee
where the right-hand side corresponds in QFT language to the one-loop `$\beta$-function', with coefficients defined as in \refeq{ctos_relation}.
This precisely matches the one-loop result derived from the Polchinski shell-approach~\cite{Rubira:2023vzw}. The coefficients of Tables \eqref{eq:singlehardcoeffK4} and \eqref{eq:singlehardcoeffK5} extend the results of \cite{Rubira:2023vzw} to fourth and fifth-order. This allows for a complete derivation of running of the second and third-order bias coefficients, which we discuss in \refsec{solutions}.

The structure of the one-loop RG has been studied in~\cite{Rubira:2023vzw} and we can reproduce some general properties:
\begin{itemize}
  \item For bias operators ${\cal O}_a^{[n]}$ that start at order $n$ in perturbation theory, we can derive the RG equation using correlation functions with at least $n'\geq n$ external legs.
  This could yield a priori contributions from ${\cal O}_a^{[m]}$ with $m\leq n'+2$. For the minimal choice $n'=n$ of external legs, this means we can have contributions at most up to $m\leq n+2$. Thus,
  \be\label{eq:sabcondition}
    s_{ba}^{\rm 1L}=0\quad\text{for}\quad {\cal O}_b={\cal O}_b^{[m]}, {\cal O}_a={\cal O}_a^{[n]}\ \text{with}\ m>n+2\,.
  \ee
  This means for example for the running of $b_{\d}$ at one-loop, the only non-zero contributions can be $s_{\delta\delta}$ and $s_{b\delta}$ for $\O_b={\cal O}_b^{[2]}$ and $\O_b={\cal O}_b^{[3]}$, i.e. at one-loop order only the first, second and third-order biases enter in the RG for $b_{\d}$, in line with~\cite{Rubira:2023vzw} (in fact $s_{\delta\delta}=0$, see below).\footnote{In the Polchinski language, these terms appears as a `shell contraction' of the operator ${\cal O}$, denoted by $\mathcal{S}_{\cal O}^2$  in \cite{Rubira:2023vzw}.}
  \item Momentum conservation implies that the kernels for the density can contribute only at the level of ${\bm k}_i^2$, such that for any bias operator ${\cal O}_a$ at leading order in gradients
  \be\label{eq:sda}
    s_{\delta a}^{\rm 1L} = 0 \,.
  \ee
  The renormalization of $\d$ can then only lead to higher-derivative operators, i.e.~operators resembling the usual `counter-terms' for matter. 
  \item The non-renormalization property of the Galileon operators, which in our basis correspond to the linear combinations 
  \bea
   {\cal G}_2 &=& {\rm tr}\big[ \big( \Pi^{[1]} \big)^2 \big] - {\rm tr}\big[  \Pi^{[1]}  \big]^2\,,\nn\\
   {\cal G}_3 &=& -\frac{1}{2}\left[2\,{\rm tr}\big[ \big( \Pi^{[1]} \big)^3 \big]-3\,{\rm tr}\big[ \big( \Pi^{[1]} \big)^2 \big] {\rm tr}\big[ \Pi^{[1]} \big]+ {\rm tr}\big[  \Pi^{[1]}  \big]^3 \right]\,,
 \eea
 of bias operators, imply that
  \bea\label{eq:Galileonconstraints}
    s_{\tiny{\rm tr}\big[ \big( \Pi^{[1]} \big)^2 \big], a}^{\rm 1L}
    - s_{\tiny{\rm tr}\big[  \Pi^{[1]}  \big]^2, a}^{\rm 1L} &=& 0\,,\nn\\
    2\,s_{\tiny{\rm tr}\big[ \big( \Pi^{[1]} \big)^3 \big], a}^{\rm 1L}
    -3\, s_{\tiny{\rm tr}\big[ \big( \Pi^{[1]} \big)^2 \big] {\rm tr}\big[ \Pi^{[1]} \big], a}^{\rm 1L} 
    + s_{\tiny{\rm tr}\big[  \Pi^{[1]}  \big]^3, a}^{\rm 1L} &=& 0\,.
  \eea
  We checked these relations for all ${\cal O}_a$ up to fifth order. Note that the higher-order Galilean operators ${\cal G}_4$ and ${\cal G}_5$~\cite{Assassi:2014fva} are actually identically zero in $d=3$ dimensions~\cite{egge_oneloopbisp}, which can be seen as a consequence of the Cayley-Hamilton relations Eq.\,\eqref{eq:CayleyHamilton}.
\end{itemize}

\subsection{Two-loop renormalization group}

Considering the two-loop part of \,\refeq{RGcorr}, we have
\ba\label{eq:RGcorr2L}
0 &= \frac{db_a}{d\Lambda}\Big|_{\text{2L}} \langle{\cal O}_a\delta_L({\bm k}_1)\cdots\delta_L({\bm k}_n)\rangle_\text{tree} + \frac{db_a}{d\Lambda}\Big|_{\text{1L}} \langle{\cal O}_a\delta_L({\bm k}_1)\cdots\delta_L({\bm k}_n)\rangle_{\text{1L}} \vs
&+ b_a \frac{d}{d\Lambda}\langle{\cal O}_a\delta_L({\bm k}_1)\cdots\delta_L({\bm k}_n)\rangle_{\text{2L}}\,.
\ea
Apart from the first and last term, which are a direct generalization of the one-loop case, we also have to take into account a mixed term containing the product of the one-loop $\beta$-function and a one-loop diagram.
Using~\refeq{corrtree1L} and \refeq{corr2L}, as well as $k_i<\Lambda$, we get
\ba\label{eq:RGkernel2L}
0 = \frac{db_a}{d\Lambda}&\Big|_{\text{2L}} n!\,K_a^{(n)}({\bm k}_1,\dots,{\bm k}_n) +  \frac{db_c}{d\Lambda}\Big|_{\text{1L}} \frac{(n+2)!}{2}\int_{q< \Lambda}  K_c^{(n+2)}({\bm k}_1,\dots,{\bm k}_n,{\bm q},-{\bm q})P^{\text{lin}}(q)\nn\\
&+  b_b \frac{(n+4)!}{2^3}2\times \frac{d\sigma^2_\Lambda}{d\Lambda}\int_{q< \Lambda} K_b^{(n+4)}({\bm k}_1,\dots,{\bm k}_n,{\bm q},-{\bm q},{\bm p},-{\bm p})_{\text{av}_p}^{p=\Lambda}P^{\text{lin}}(q)\,.
\ea
Note the factor of two coming from taking the $\Lambda$-derivative of the two-loop term (bottom line), acting either on $\Theta(\Lambda-p)$ or on $\Theta(\Lambda-q)$. In addition we renamed the summation index $a\to c$ and $a\to b$ in the second and third terms, respectively.
Inserting the one-loop RG equation~\refeq{RG1L} gives
\ba
0 = \frac{db_a}{d\Lambda}\Big|_{\text{2L}}  K_a^{(n)}({\bm k}_1,\dots,{\bm k}_n) & + b_b \frac{(n+4)!}{4n!}\frac{d\sigma^2_\Lambda}{d\Lambda}\int_{q< \Lambda} \Big[ K_b^{(n+4)}({\bm k}_1,\dots,{\bm k}_n,{\bm q},-{\bm q},{\bm p},-{\bm p})_{\text{av}_p}^{p=\Lambda}\nn\\
&-\frac{2s_{bc}^{\rm 1L}}{(n+4)(n+3)}K_c^{(n+2)}({\bm k}_1,\dots,{\bm k}_n,{\bm q},-{\bm q})\Big]P^{\text{lin}}(q)\,.
\ea
We can use the result in~\refeq{ctos_relation} for the one-loop $\beta$-function giving $2s^{\rm 1L}_{bc}/(n+4)(n+3)=c_{bc}^{(n+4)}$. The resulting sum over $c$ precisely yields the single-hard limit of $K_c^{(n+4)}$, see~\refeq{biaskernelexpansionsinglehard}. Thus we arrive at
\ba\label{eq:RGkernel2Lb}
0 = \frac{db_a}{d\Lambda}\Big|_{\text{2L}} K_a^{(n)}({\bm k}_1,\dots,{\bm k}_n) &+ b_b \frac{(n+4)!}{4n!}\frac{d\sigma^2_\Lambda}{d\Lambda}\int_{q< \Lambda} \Big[ K_b^{(n+4)}({\bm k}_1,\dots,{\bm k}_n,{\bm q},-{\bm q},{\bm p},-{\bm p})_{\text{av}_p}^{p=\Lambda}\nn\\
& - K_b^{(n+4)}({\bm k}_1,\dots,{\bm k}_n,{\bm q},-{\bm q},{\bm p},-{\bm p})_{\text{av}_p}^{p\to\infty}\Big]P^{\text{lin}}(q)\,.
\ea
As before we assume $k_i \ll\Lambda$, but $q$ can take any value up to the smoothing scale. However, if $q$ is `soft', i.e. $q\ll\Lambda$, the two kernels in the square bracket cancel (up to gradient corrections), such that only the `hard' region where $q\sim \Lambda$ contributes effectively. This matches the intuition that the $\beta$-function is dominated by fluctuations of the order of the smoothing scale. Nevertheless, this does not mean that $q$ is precisely equal to $\Lambda$, but just that the integration over $q$ is supported at scales where $q/\Lambda=O(1)$.
This is consistent with the physical picture of the $\beta$-function, but it is challenging to match it to the Polchinski shell description. As is well-known, even though the Polchinski equation is formally exact, it is hard to use it for obtaining results in a systematic loop expansion beyond the one-loop order.
However, the approach we follow here yields a two-loop result for the RG equation without any conceptual problems.

To extract the two-loop $\beta$-function at leading order in the gradient expansion, we consider again the limit $k_i\to 0$, i.e. small external wavenumbers. By the above argument, we can assume that $q/\Lambda = O(1)$ so that $k_i \ll p,q$. Thus, we use \refeq{kernelexpansiondoublehard} to write the first kernel in the square bracket of~\refeq{RGkernel2Lb} in terms of the \emph{double-hard limit}. We then use \refeq{K5doublehard} to write it in terms of $d_b^{(5)}+\tilde d_b^{(5)}g(r)$.
The subtraction term involving the second kernel in the square bracket in~\refeq{RGkernel2Lb} is evaluated in the \emph{sequential single-hard limit}, where $k_i \ll q \ll p$, and we can use \refeq{doublehardvssequentialsinglehard}. Finally, we obtain the two-loop RG equation
\bea\label{eq:RG2Lraw}
\frac{db_a}{d\Lambda}\Big|_{\text{2L}} &=& -   b_b \frac{(n+4)!}{4n!}\frac{d\sigma^2_\Lambda}{d\Lambda}\int_{q<\Lambda} \Big[ d_{ba}^{(n+4)}(q/\Lambda) - c^{(n+4)}_{bc}c^{(n+2)}_{ca} \Big]P^{\text{lin}}(q)\,,
\eea
which after using \refeq{doublehardvssequentialsinglehard} and \refeq{twoloopbetafunc} simplifies to
\be\label{eq:RG2L}
\boxed{
 \frac{db_a}{d\Lambda}\Big|_{\text{2L}} = -  b_b \frac{d\sigma^2_\Lambda}{d\Lambda}\int_{q< \Lambda} \Big[ s^\text{2L}_{ba}(q/\Lambda) - s^\text{2L}_{ba}(0) \Big]P^{\text{lin}}(q)\,.
}
\ee
We observe that the integral over $q$ is indeed dominated by scales of order $\Lambda$, since for $q\ll\Lambda$ the two terms in the square bracket cancel.
Thus, the running is dominated by scales of the order of $\Lambda$, as expected.
Moreover, the two-loop contribution to the $\beta$-function only involves the \emph{double}-hard limit, while the one-loop contribution is related to the single-hard limit. 
Applying this result for the running of the linear bias $b_1\equiv b_\delta$
and using the explicit result for the double hard limit of $K^{(5)}$ from~\refeq{K5doublehard} yields
\be\label{eq:RG2Lb1}
\boxed{
 \frac{db_\delta}{d\Lambda}\Big|_{\text{2L}} = -  30 b_b \tilde d^{(5)}_b \frac{d\sigma^2_\Lambda}{d\Lambda}\int_0^\Lambda dq \frac{q^2P^{\text{lin}}(q)}{2\pi^2} g(q/\Lambda)\,,
}
\ee
where we used
\be\label{eq:twoloopbetafuncfordeltarepeat}
  s^\text{2L}_{b\delta}(q/p)=30d^{(5)}_b+30\tilde d^{(5)}_b g(q/p)\,.
\ee
The coefficients $d^{(5)}_b$ drop out in the difference in the square bracket of~\refeq{RG2L}, i.e. only the $\tilde d^{(5)}_b$ coefficients proportional to $g(r)$ in~\refeq{K5doublehard} enter the running. This also means that the dependence on $\Lambda$ is encapsulated in a single integral, while the relative contributions from the bias operators ${\cal O}_b$ up to fifth order are given by the numerical weights $\tilde d^{(5)}_b$, see Table~\eqref{eq:doublehardcoeff}. We consider the split \refeq{twoloopbetafuncfordeltarepeat} for $\d$, but a generic $s^\text{2L}_{bc}(q/p)$ can always be decomposed into a constant piece $s^\text{2L}_{bc}(0)$  [related to the square of a one-loop contribution via Eq.~\eqref{eq:commutationoflimitsrelation}] and the `genuine' two-loop contribution related to the remainder $s^\text{2L}_{bc}(q/p)-s^\text{2L}_{bc}(0)$.
We discuss this point in more detail in Sec.~\ref{sec:resum}, and the result for the two-loop running in \refsec{twoloopsolution}.

Note that $s^\text{2L}$ satisfies the same relations as $s^\text{1L}$ from momentum conservation, i.e. $s^\text{2L}_{\delta a}=0$, and additionally also the relations Eq.\,\eqref{eq:Galileonconstraints} implied by non-renormalization of ${\cal G}_2$ and ${\cal G}_3$. We checked this for the explicit result Eq.\,\eqref{eq:twoloopbetafuncfordeltarepeat}.

In summary, in this Section we provided an alternative derivation of the RGE for the bias parameters using the $\L$-independence of the galaxy density $\delta_g$ and its $N$-point functions, instead of integrating out `shell modes' as in \cite{Rubira:2023vzw}. The one-loop RGE \refeq{RG1L} was extended to include fifth-order operators in the $b$ indices of $s$ (see Appendix~\ref{app:det}), leading to a full derivation of the one-loop RG for third-order equations in the $a$ index of \refeq{RG1L}. We have also derived for the first time the two-loop RGE \refeq{RG2L}, with all the coefficients for the running of $b_\d$, \refeq{RG2Lb1}, also determined.

\
\newpage
\section{Connection between RGE and renormalized bias} \label{sec:connection_RG_renorm}

In this Section, we show that the (one and two-loop) renormalization group equations from Sec.\,\ref{sec:rg} can also be derived via the renormalized bias parameters discussed in Sec.\,\ref{sec:biasrenorm}, making the connection between both approaches explicit.

Since the renormalized bias coefficients $b_{[c]}$ do (by construction) not depend on the cutoff $\Lambda$, we have from~\refeq{biastrafo}
\ba
  \frac{db_a}{d\Lambda}=b_{[c]}\frac{dZ_{ca}}{d\Lambda}=b_b[(1+Z)^{-1}]_{bc}\frac{dZ_{ca}}{d\Lambda}\equiv b_b\gamma_{ba}\,,
\ea
where $\gamma_{ba}$ is referred to in QFT as the \emph{anomalous dimension matrix}.
Thus, as usual, the anomalous dimension matrix $\gamma$ is directly related to the renormalization constants $Z$,
\be\label{eq:rgconnection}
\boxed{
  \gamma_{ba} = [(1+Z)^{-1}]_{bc}\frac{dZ_{ca}}{d\Lambda}\,.
  }
\ee
Expanding in loops gives  $\gamma_{ba}^{\text{tree}} = 0$, and
\bea
  \gamma_{ba}^\text{1L} &=& \frac{dZ^\text{1L}_{ba}}{d\Lambda}\,,\nn\\
  \gamma_{ba}^\text{2L} &=& \frac{dZ^\text{2L}_{ba}}{d\Lambda}-Z^\text{1L}_{bc} \frac{dZ^\text{1L}_{ca}}{d\Lambda}\,.
\eea
Using the one-loop result for $Z$ from~\refeq{Z1L} yields directly
\be
  \gamma_{ba}^\text{1L} = -\frac{d\sigma_\Lambda^2}{d\Lambda}s_{ba}^{\rm 1L}\,,
\ee
which precisely matches the one-loop RG result~\refeq{RG1L}.
Using the two-loop result~\refeq{Z2L} and the rewritings of coefficients mentioned below that equation yields
\bea
  \gamma_{ba}^\text{2L} &=& \frac{d}{d\Lambda}\Big[(\sigma_\Lambda^2)^2s_{bc}^{\rm 1L}s_{ca}^{\rm 1L}-\frac12 \int_{p,q<\Lambda} s_{ba}^\text{2L}(q/p)P^{\text{lin}}(p)P^{\text{lin}}(q)\Big]-Z^\text{1L}_{bc} \frac{dZ^\text{1L}_{ca}}{d\Lambda}\nn\\
  &=& \frac{d(\sigma_\Lambda^2)^2}{d\Lambda}s_{bc}^{\rm 1L}s_{ca}^{\rm 1L} 
  - \frac{\Lambda^2P^{\text{lin}}( \Lambda)}{2\pi^2}\int_{q<\Lambda}s_{ba}^\text{2L}(q/\Lambda)P^{\text{lin}}(q)
  - \sigma_\Lambda^2\frac{d\sigma_\Lambda^2}{d\Lambda}s_{bc}^{\rm 1L}s_{ca}^{\rm 1L}\nn\\
  &=& \sigma_\Lambda^2\frac{d\sigma_\Lambda^2}{d\Lambda}s_{bc}^{\rm 1L}s_{ca}^{\rm 1L} 
  - \frac{d\sigma_\Lambda^2}{d\Lambda}\int_{q<\Lambda}s_{ba}^\text{2L}(q/\Lambda)P^{\text{lin}}(q) \nn\\
  &=&  
  - \frac{d\sigma_\Lambda^2}{d\Lambda}\int_{q<\Lambda}\left[s_{ba}^\text{2L}(q/\Lambda)-s_{ba}^\text{2L}(0)\right]P^{\text{lin}}(q)  \,,
\eea
where we used \refeq{ds2dL} in the second step and $s_{ba}^\text{2L}(0)=s_{bc}^{\rm 1L}s_{ca}^{\rm 1L}$
in the last step, see~\refeq{commutationoflimitsrelation}.
This precisely agrees with the two-loop RG result~\refeq{RG2L}. Therefore, starting from the connection between the renormalized and bare bias parameters \refeq{biastrafo} we have derived the one and two-loop RG. 

Moreover, it was shown in Sec.~3 of \cite{Rubira:2023vzw} [see Eq.~(3.1) therein] that $b_a(\L) \to b_{[a]}$ in the IR limit $\L \to 0$, with $k/\L$ fixed. Within the formulation of the RG used in this work, this result is equivalent to the statement that $Z_{ca} \to 0 $ when $\L \to 0$  in \refeq{biastrafo} after taking $k \to 0$ in \refeq{rencond}. This implies that $b_a(\L) \to b_{[a]}$ for $\L\to 0$ holds also in the RG approach employed here, and {\it at all loop orders}.
 This is a central point and implies that one could use the non-perturbative separate Universe results for the bias coefficients from~\cite{Lazeyras:2015lgp, Baldauf:2015vio, Li:2015jsz} as initial conditions for the bias RG.

\newpage
\section{Solutions to the RGE}\label{sec:solutions}

In this Section we discuss solutions of the bias RG. 
We sum up the results of the previous two Sections writing down  \refeq{RG1L} and \refeq{RG2L} [see also \refeq{sabcondition}] as
\ba
\frac{d b_{\d}}{d \ln\L} &= -\left[ \sum_{c \in {\cal O}^{[2]}} s_{c\d}^{\rm 1L} b_{c} +  \sum_{c \in {\cal O}^{[3]}}s_{c\d}^{\rm 1L} b_{c}  \right]\frac{d \s^2_\L}{d \ln\L} \vs
&\hspace{3cm}- \sum_{c \in \text{all}}   \frac{d\sigma^2_\Lambda}{d\ln\Lambda}\int_0^\Lambda \frac{d\Lambda'}{\Lambda'}\frac{d\sigma^2_{\Lambda'}}{d\ln\Lambda'} \left[s_{c\delta}^\text{2L}(\Lambda'/\Lambda)-s_{c\delta}^\text{2L}(0)\right]b_c\,,\label{eq:drun}
\\
\frac{d b_{a}}{d \ln\L} &= -\left[ \sum_{c \in {\cal O}^{[2]}} s_{ca}^{\rm 1L} b_{c} +  \sum_{c \in {\cal O}^{[3]}}s_{ca}^{\rm 1L} b_{c} +  \sum_{c \in {\cal O}^{[4]}}s_{ca}^{\rm 1L} b_{c}  \right]\frac{d \s^2_\L}{d \ln\L};\, \quad a \in {\cal O}^{[2]}, \label{eq:2ndrunning}
\\
\frac{d b_{a}}{d \ln\L} &= -\left[ \sum_{c \in {\cal O}^{[2]}} s_{ca}^{\rm 1L} b_{c} +  \sum_{c \in {\cal O}^{[3]}}s_{ca}^{\rm 1L} b_{c} +  \sum_{c \in {\cal O}^{[4]}}s_{ca}^{\rm 1L} b_{c} +  \sum_{c \in {\cal O}^{[5]}}s_{ca}^{\rm 1L} b_{c}  \right]\frac{d \s^2_\L}{d \ln\L}\,; \quad a \in {\cal O}^{[3]}, \label{eq:3rdrunning}
\ea
with 
\be
\Delta^2_\Lambda \equiv d\s^2_\Lambda/d\ln\Lambda=\Lambda^3P^{\text{lin}}(\Lambda)/(2\pi^2)  \,,
\ee
defined as the change of the density variance with smoothing scale as defined in Eq.~\eqref{eq:sigmaLambda}, widely used in \refsec{resum}. The one-loop $\beta$-function matrix $s^\text{1L}_{ab}$ is obtained from the expansion coefficients of  bias kernel in the single-hard limit \refeq{ctos_relation} with explicit values given in Tables \eqref{eq:singlehardcoeffK4} and \eqref{eq:singlehardcoeffK5}, and the two-loop $\beta$-function related to $s_{c\delta}^\text{2L}(q/\Lambda)-s_{c\delta}^\text{2L}(0)=30\tilde d_c^{(5)}g(q/\Lambda)$ with expansion coefficients $\tilde d_c^{(5)}$ in the double-hard limit tabulated in \eqref{eq:doublehardcoeff} and a universal scale-dependent weight function $g(p/q)$ from Eq.\,\eqref{eq:doublehardfuncg}. The notation $c \in {\cal O}^{[m]}$ denotes all operators $c$ that start at order $m$. A few comments are in order.
First, we emphasize once more that by going to fifth-order in perturbation theory in our calculations, we are only able to calculate one-loop contributions to third-order terms and two-loop contributions to the linear bias. The calculation of the running of fourth and fifth-order parameters would require calculating the operators starting from sixth-order in perturbation theory, and is beyond our scope. The same applies to two-loop contributions to \refeq{2ndrunning} and \refeq{3rdrunning}.
Second, if not for the two-loop term in the equations above, one could change variables and solve for $d b(\s^2)/d\s^2$, which admits a solution in terms of  exponential functions, as described in Sec.~2.4 of \cite{Rubira:2023vzw}.

Therefore, this work goes beyond the results of \cite{Rubira:2023vzw} by including terms up to fifth-order and moreover considering two-loop running. Effectively, this means:
\begin{enumerate}
    \item Including the running of third-order bias coefficients.
    \item Including fourth and fifth-order operators. This means fourth-order bias coefficients contributing to the running of second-order bias, i.e.~the last term in \refeq{2ndrunning}, and fourth and fifth-order bias contributing to the running of third-order bias, i.e.~the last two terms in \refeq{3rdrunning};
    \item Calculating the two-loop running of the linear bias, the last term in \refeq{drun}. 
\end{enumerate}
We discuss each of those points separately in the following Sections. In this part, we restrict our analysis to the deterministic part of the bias RG. As discussed in \cite{Rubira:2024tea}, the running of stochastic components does not affect deterministic bias. We use a linear power spectrum according to Planck 2018 $\L$CDM cosmology \cite{Planck:2018vyg} for numerical results shown in this Section (unless mentioned otherwise), but note that one-loop results for $b_a(\L)$ are cosmology-independent when parameterized in terms of the rescaled variable $\sigma^2\equiv \sigma_\L^2$ instead of $\L$.

\subsection{Including the running of third-order bias parameters}

\begin{figure}[t]
    \centering
    \includegraphics[width = 0.7\textwidth]{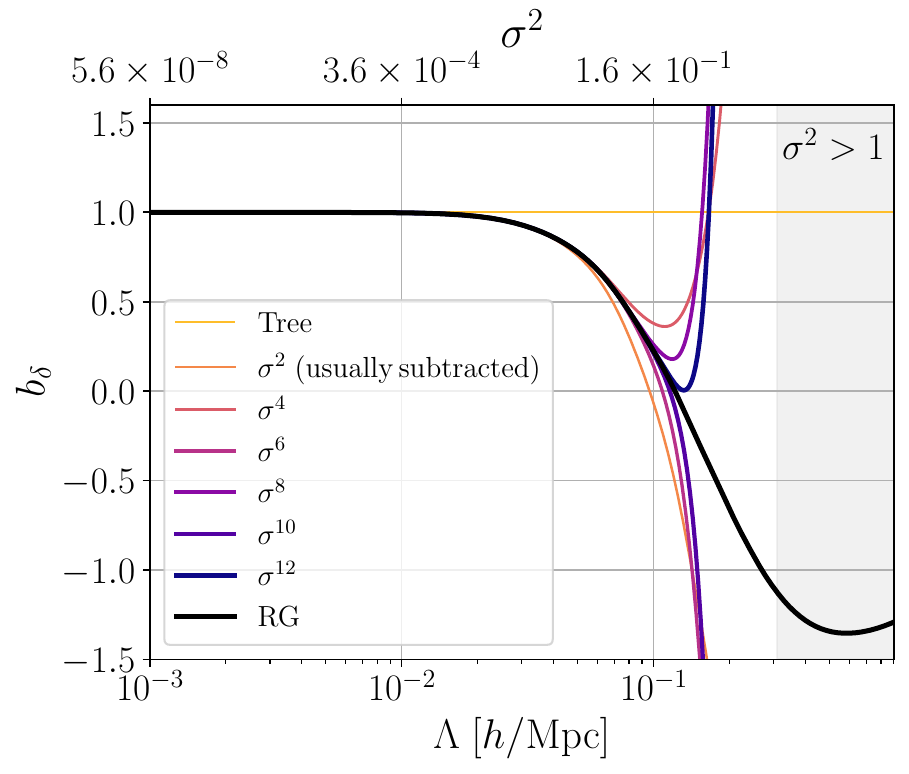}
    \caption{Solution of the RG equation for the linear bias $b_\d$ in black, given by the first line of \refeq{solODE}. We compare it to the series expansion in powers of $\s^2$ given by the second line of \refeq{solODE}. The usual $\s^2$ subtraction \refeq{Z1L} at one-loop order corresponds to keeping only the zeroth and first terms. }
    \label{fig:Taylor}
\end{figure}

\begin{figure}[t]
    \centering
    \includegraphics[width = 0.85\textwidth]{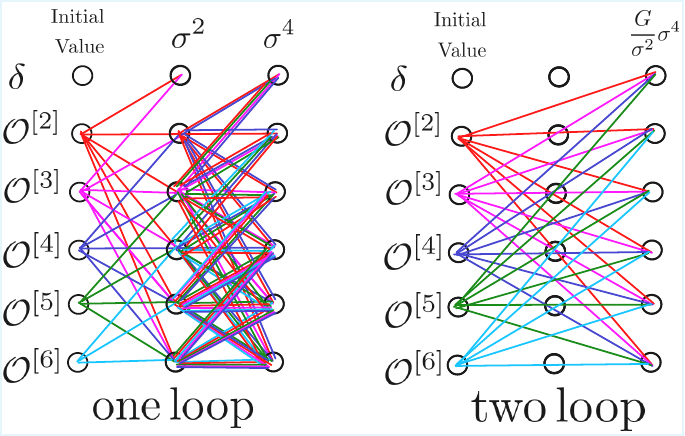}
    \caption{  Diagram showing how biases $b_a$ of operators ${\cal O}_a^{[n]}$ at order $n=1,2,\dots,6$ (row $1$ to $6$) source other biases $b_b$ due to operator mixing induced by renormalization. The columns show the order in powers of $\sigma^2$ for which a bias $b_a$ at the initial scale $\s^2_\ast=0$ can source $b_b$ at $\s^2>0$. The left diagram corresponds to the one-loop RG [solved iteratively in first ($\s^2$) and second ($\s^4$) iteration], and the right one to the (first iteration of the) two-loop contribution. At one-loop, an initial bias corresponding to an operator ${\cal O}_a^{[n]}$ can source all operators with $n'\geq n$ as well as those with $n'=n-1$ and $n'=n-2$, as shown by the coloured lines. An exception is the linear bias, related to momentum conservation, which cannot source any other (leading-gradient) bias. At two-loop, a single iteration of the RG yields already a factor $(G/\s^2)\s^4$ of order $\s^4$, with $G$ defined in \refeq{Gdef}. Operators ${\cal O}_a^{[n]}$ can source all operators with $n'\geq n$ as well as those with $n'=n-1,\dots,n-4$ at two-loop. In this work we derive all outgoing lines parting from fourth and fifth-order operators, all incoming lines to third-order operators (that lead the RG for those) at one-loop and all two-loop lines that go to $\d$. }
    \label{fig:Diag}
\end{figure}

In this part we discuss the relevance of considering the running of the bias coefficients of third-order operators, as \refeq{3rdrunning}, but omit fourth- and fifth order terms as well as two-loop contributions for now.  The latter approximation implies that we can change variables from $\Lambda$ to $\s^2\equiv \s_\Lambda^2$, 
\be \label{eq:dbds2}
\frac{d b_a}{d\s^2} = -\bar{s}_{ac}^{\rm 1L} b_c  \,,
\ee
where both indices $a,c$ range over all $7$ bias operators up to third order here, such that $\bar{s}_{ac}^{\rm 1L} = s_{ca}^{\rm 1L}$ is a $7\times 7$ matrix [see \refeq{sbardef}].
Eq.\,\eqref{eq:dbds2} admits solutions in terms of the matrix exponential of $\bar{s}^{\rm 1L}_{ac}$ with initial condition $b_a^\ast = b_a(\s^{2}_\ast)$ at some $\s^{2}_\ast$ (e.g. $\s^2_\ast=0$ when starting the evolution in the IR, in which case $b_a^\ast$ are the renormalized bias coefficients $b_{[a]}$ from Sec.\,\ref{sec:biasrenorm}): 
\begin{equation}\label{eq:solODE}
\boxed{
\begin{aligned}
b_a(\s^2) &= \left[e^{-\bar{s}^{\rm 1L}\times(\sigma^2-\sigma^2_\ast)}\right]_{ac}b_c^\ast \\
        &= b_a^\ast - (\sigma^2-\sigma^2_\ast)\bar{s}_{ac}^{\rm 1L} b_c^\ast + \frac{1}{2}(\sigma^2-\sigma^2_\ast)^2\bar{s}_{ab}^{\rm 1L}\bar{s}^{\rm 1L}_{bc} b_c^\ast - \frac{1}{6}(\sigma^2-\sigma^2_\ast)^3\bar{s}^{\rm 1L}_{ab}\bar{s}_{bd}^{\rm 1L}\bar{s}^{\rm 1L}_{dc} b_c^\ast \\
        & \quad + O\left((\s^2-\s^2_\ast)^4\right)\,. 
\end{aligned}
}
\end{equation}
Note that the difference of renormalized bias at one-loop order from \refeq{Z1L} (being the subtraction that is usually considered in this context) and of bias at smoothing scale $\Lambda$  can be obtained as the first term in the Taylor expansion of this RG solution for $\s^2_\ast=0$. The coupled set of ODE's determined by the RG equations, therefore, provide corrections in higher powers of $\s^2$ to the usual renormalization procedure. We display the solution expanded at different orders in $\s^2$ in \reffig{Taylor}. 
This illustrates that the RG resums higher powers of $\s^2$, suggesting that it accounts for higher-order hard-loops. We return to this point in \refsec{resum}. 
We also provide a visual illustration of the RG solutions in \reffig{Diag}. The left panel indicates how the operators connect to each other at one loop and at each order in powers of $\s^2$. Note that each operator of order $n$ sources all operators of order at least $n-2$ at every $\s^2$ step, as a consequence of the one-loop being related to the single-hard limit of the operator kernels.  

The solution of the one-loop RG Eq.~\eqref{eq:solODE}
can also be written in terms of the eigenmodes of the anomalous dimension matrix,
\ba \label{eq:solODEeigen}
b_a(\s^2) 
        &= p_{ai} e^{\lambda_i(\sigma^2-\sigma^2_\ast)} c_i\,,
\ea
with eigenvalues $\lambda_i$ of the matrix  $-\bar s^\text{1L}_{ab}$, associated eigenvectors  $p_{ai}$, and coefficients $c_i=(p^{-1})_{ic}b_c^\ast$ fixed by the choice of initial conditions $b_c^\ast$. Here $p^{-1}$ is the matrix inverse of $p$. 
The RG equation can also be diagonalized, in the operator basis
\be
  {\cal O}_i^\text{diag} \equiv {\cal O}_ap_{ai}\,,
\ee
for which the one-loop RG takes the decoupled form (no summation over $i$ here)
\be
  \frac{db_i^\text{diag}}{d\sigma^2}=\lambda_ib_i^\text{diag}\,, 
\ee
where $b_i^\text{diag}\equiv (p^{-1})_{ia}b_a$.

As an explicit (known from \cite{Rubira:2023vzw}) example, when restricting to first- and second-order operators $(\tr[\Pi^{[1]}],\tr[\Pi^{[1]}]^2,\tr[(\Pi^{[1]})^2])=(\delta,\delta^2,\tr[(\Pi^{[1]})^2])$, the eigenvalue set  is
\be \label{eq:eigen2}
\{\lambda_1, \lambda_2,\lambda_3\}
\ \equiv \ \{0,0,-\frac{8126}{2205}\} \ \simeq \
\{0,0,-3.69\}\,.
\ee
The two zero eigenvalues are associated to the non-renormalization of $\delta$ and ${\cal G}_2=\delta^2-\tr[(\Pi^{[1]})^2]$ at leading-gradient order, with eigenvectors $p_{a1}=(1,0,0)$ and $p_{a2}=(0,1,-1)$, consistent with~\refeq{Galileonconstraints}, while the eigenvector for the non-zero eigenvalue is $p_{a3}=(1, 127/3570, 656/595)$. This gives ${\cal O}_1^\text{diag}=\delta$, ${\cal O}_2^\text{diag}={\cal G}_2$, and ${\cal O}_3^\text{diag}=\delta+127/3570\,\delta^2+656/595\tr[(\Pi^{[1]})^2]$ with diagonalized one-loop RG equations (up to second order) for the associated bias coefficients $b_1^\text{diag}=b_{\delta}-210/239\,(b_{\delta^2}+b_{\tiny \tr[(\Pi^{[1]})^2]})$,
$b_2^\text{diag}=3936/4063\,b_{\delta^2}-127/4063\,b_{\tiny \tr[(\Pi^{[1]})^2]}$ and $b_3^\text{diag}=210/239\,(b_{\delta^2}+b_{\tiny \tr[(\Pi^{[1]})^2]})$,
\be
  \frac{db_1^\text{diag}}{d\sigma^2}=0,\quad
  \frac{db_2^\text{diag}}{d\sigma^2}=0,\quad
  \frac{db_3^\text{diag}}{d\sigma^2}=-\frac{8126}{2205}b_3^\text{diag}\,.\quad
\ee
In the usual RG classification\footnote{We use the terms marginal, relevant and irrelevant here in the original sense, i.e. with respect to the behaviour under RG evolution (constant, growing in the IR, or growing in the UV, respectively). We stress that all leading-gradient bias operators have identical mass dimension (being dimensionless), and thus the bias RG evolution can be seen as the analog of logarithmic scale evolution in QFT. This should not be confused with the common usage of the terms marginal, relevant and irrelevant in the context of effective theories in particle physics, where those terms are sometimes used to refer to operators of mass dimension four, less than four, or more than four, respectively. Instead, in the sense used here, the QCD coupling constant would be classified as relevant while the QED coupling would be irrelevant.} this means that (at leading gradient order) ${\cal O}_1^\text{diag}=\delta$ and ${\cal O}_2^\text{diag}={\cal G}_2$ are {\it marginal} operators, and  ${\cal O}_3^\text{diag}$ is {\it relevant}. While marginal operators do not change with scale, the relevant operators decay in the UV and grow in the IR.
The RG also has a set of fixed points, for $b_3^\text{diag}=0$ and arbitrary values of $b_1^\text{diag}$ and $b_2^\text{diag}$. In the original basis, the fixed points lie in the plane for which
\be \label{eq:planeb2}
b_{{\rm tr}[ ( \Pi^{[1]} )^2 ]} = - b_{\delta^2}\,.
\ee
Those fixed points are {\it stable}, since the system only has a negative eigenvalue.\footnote{In Sec.~2.4 of \cite{Rubira:2023vzw} it was mentioned that the system had an unstable fixed point, but this refers to running towards decreasing values of $\s^2$. Here we consider the running with increasing $\s^2$.} 
This means that even when starting with arbitrary renormalized biases $b_a^\ast$ in the IR (i.e. at $\s^2_\ast=0$), the RG solution will converge towards the two-dimensional subspace defined by Eq.\,\eqref{eq:planeb2} when running towards the UV ( i.e. larger values of $\s^2$). Since this subspace corresponds to $b_3^\text{diag}=0$, only the marginal bias operators contribute in the UV, being $\delta$ and ${\cal G}_2$.

Let us now discuss what happens when including the running of bias operators up to third order.
The system including the third-order operators has eigenvalues (rounded to three digits for the latter four)
\be \label{eq:eigen3}
\{0,0,0,-12.6,-3.44,-2.01,0.220\}\,,
\ee
with the zeros associated to eigenvectors that correspond to the non-renormalizable bias operators $\d$, ${\G_2}$ and ${\G_3}$, in line with \refsec{oneLoopRG} [more precisely, \refeq{sda} and \refeq{Galileonconstraints} provide a basis for the kernel of the matrix $\bar{s}_{ab}^\text{1L}$]. Those three operators thus belong to the ${\cal O}_i^\text{diag}$ basis, and are marginal. In addition, there are now three relevant operators
associated to the eigenmodes of the negative eigenvalues, and,
as a new feature, there is also a positive eigenvalue, and thus an {\it irrelevant} operator.
Formally, the system has fixed points for which, in the diagonal basis, only the bias coefficients of the marginal operators $\d$, ${\G_2}$ and ${\G_3}$ are non-zero.
However, due to the positive eigenvalue, the solution does not converge towards this subspace when running from the IR ($\s^2_\ast=0$)
towards the UV. Instead, when starting the RG evolution in the IR with arbitrary initial conditions, the operators that contribute in the UV are the marginal ones \emph{and} the irrelevant operator associated to the positive eigenvalue.

\begin{figure}[t]
    \centering
    \includegraphics[width = 0.49\textwidth]{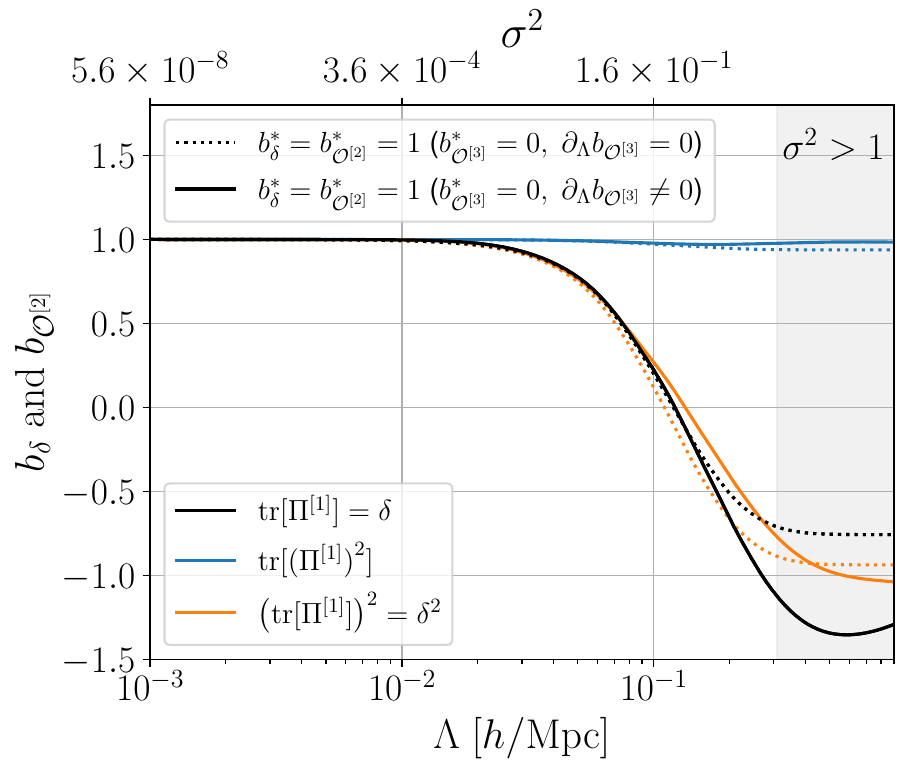}
    \includegraphics[width = 0.49\textwidth]{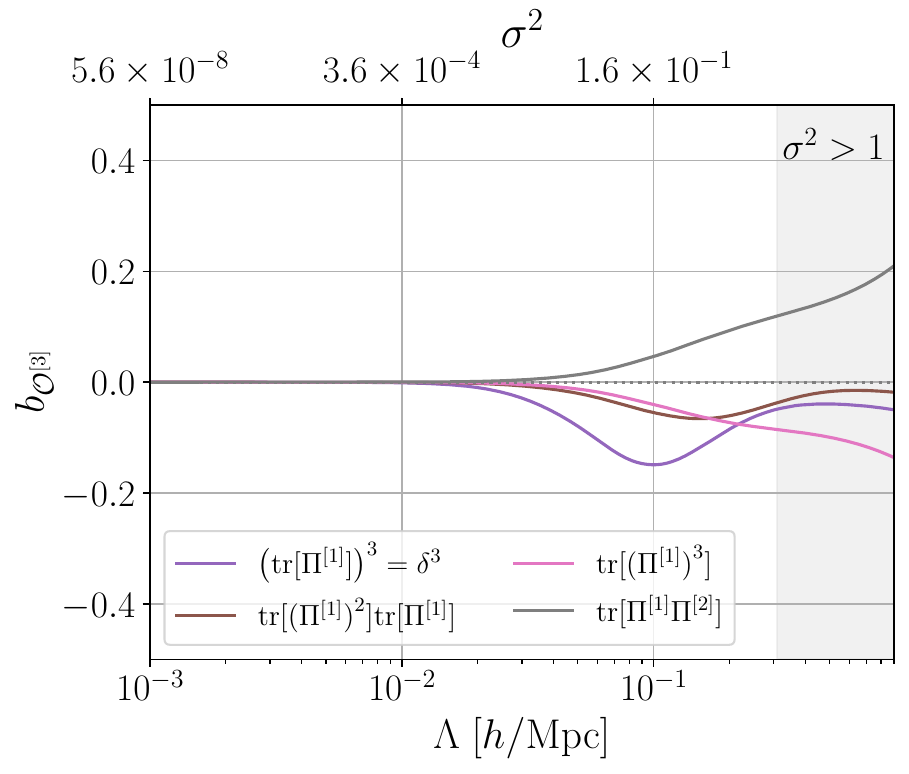}
    \caption{Impact of the running of third-order operators. The one-loop running of the linear, second (left) and third-order (right) bias parameters. For the dotted case, we completely remove the third-order terms. For the solid line, we let them run, but fix their initial conditions at zero, such that they are sourced by second-order operators via \refeq{3rdrunning}. We neglect fourth and fifth-order coefficients and two-loop running in this figure, and fix the initial conditions at $b^\ast = b(\Lambda^\ast)$ at $\Lambda^\ast = 10^{-3} h/$Mpc, with first and second-order parameters fixed at 1 and third-order parameters fixed at zero for illustration. }
    \label{fig:1Lrunning}
\end{figure}

We display in \reffig{1Lrunning} the running of linear, second and third-order bias parameters. 
We fix the initial conditions as $b^\ast = b(\Lambda^\ast)$ at $\Lambda^\ast = 10^{-3} h/$Mpc and use a $\L$CDM spectrum to match $\L$ to $\s^2$ as mentioned above. 
The colors represent different operators. For the dotted lines, we completely remove the third-order terms, while for the solid lines the third-order terms are zero at the initial conditions but can be sourced by second-order operators via \refeq{3rdrunning}.
We start by analyzing the running of the third-order bias coefficients. When fixing their initial conditions at zero and letting the second-order terms source the third-order parameters (solid) those terms are very slowly sourced, according to \reffig{Diag} at order $\s^2$. For the first- and second-order parameters, we find significant differences starting at $\s^2 \sim 10^{-1}$. The back-reaction of third-order terms into first and second-order operators is at order $\s^4$, according to \reffig{Diag}.  Moreover, while the terms converged to a fixed point when neglecting the third-order parameters running (dotted), this is not the case after including it. As mentioned before, this is due to the presence of a positive growing eigenmode in the ODE solution.  

Another central point regards the closure of the RGE system. This was addressed in Sec.~2.4 of \cite{Rubira:2023vzw}, which truncated the system at second-order and considered an exponential ansatz $b_{a} = b_{a}^* e^{-s (\s^2 - \s^2_\ast)}$ for the third-order bias coefficients, with some constant $s$. Note that this ansatz does not include the backreaction of second-order operators into the third-order parameters. By missing the backreaction contributions, this ansatz would miss, for instance, the growing mode \refeq{eigen3}. \reffig{Diag} suggests a different way to understand the truncation of the perturbative series: for a given operator, at a given order in $\s^2$, one can trace back all the relevant contributions for that term at that order, e.g., the contributions to $\d$ at order $\s^4$ only requires the computation of order five operators (there is no light-blue line entering that operator at that order). We explore in \refsec{resum} [see discussion around \refeq{1overlog}] a different way to interpret the perturbative parameters of the RG, in which we find that the series in terms of $\s^2$ can be further split into a perturbation in the amplitude of the primordial fluctuations and another perturbation in terms of large logarithms. 

\subsection{Including fourth and fifth-order operators in the initial conditions} 

Here we discuss the impact of taking fourth and fifth order operators in the RG Eqs.\,\eqref{eq:2ndrunning} and \eqref{eq:3rdrunning} into account, but still neglect two-loop running. 
Fourth-order biases induce a running in the second and third-order bias coefficients at one-loop order, see $c\in{\cal O}^{[4]}$ terms in Eqs.\,\eqref{eq:2ndrunning} and \eqref{eq:3rdrunning}, and fifth-order biases source a running of third-order bias, see $c\in{\cal O}^{[5]}$ terms in Eq.\,\eqref{eq:3rdrunning}.

\begin{figure}[t]
    \centering
    \includegraphics[width = 0.49\textwidth]{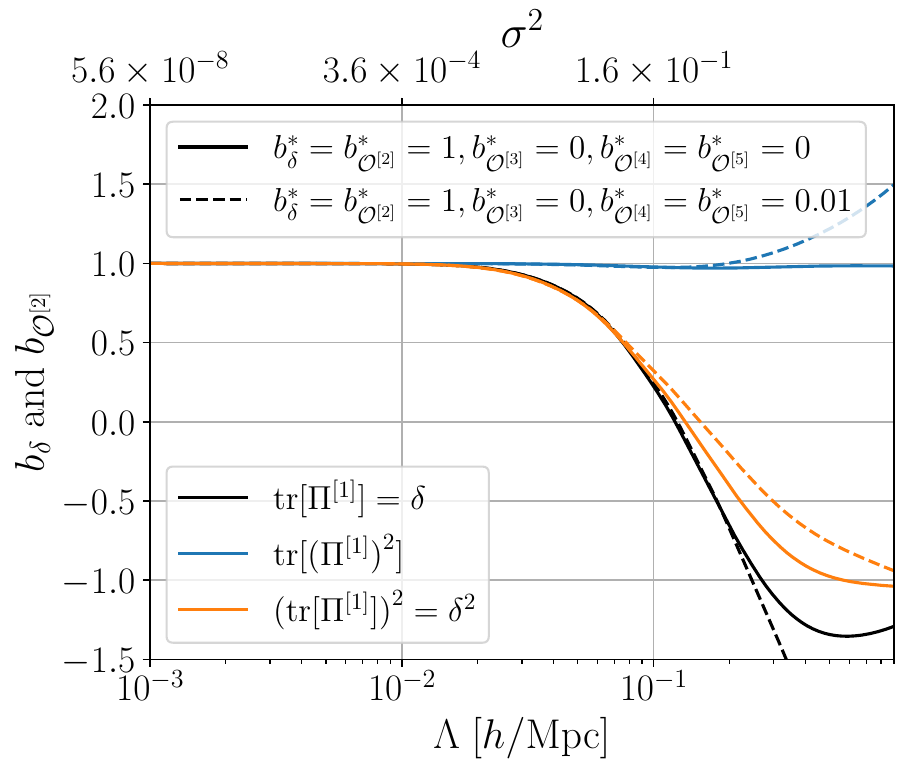}
    \includegraphics[width = 0.49\textwidth]{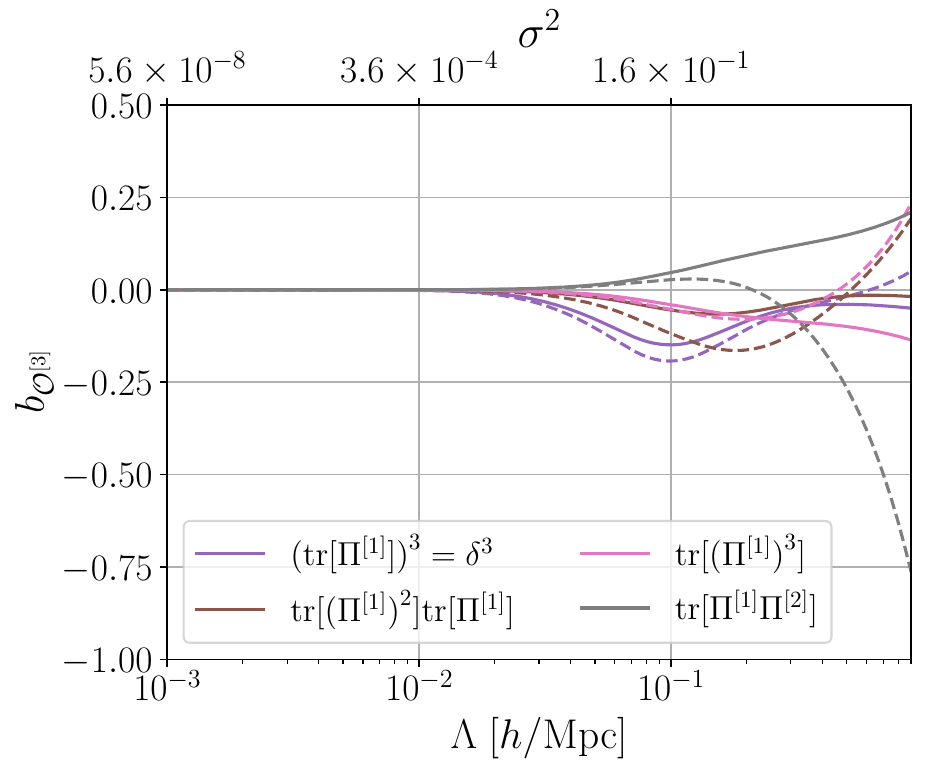}
    \caption{Running of first, second (left) and third-order (right) bias parameters. The solid lines indicate the same case as in \reffig{1Lrunning}. We include as dashed lines the cases with initial fourth and fifth-order operators. }
    \label{fig:1Lrunning_include5th}
\end{figure}

We display in \reffig{1Lrunning_include5th} the running of the linear, quadratic and cubic operators. The solid lines present a similar case as considered in \reffig{1Lrunning}, in which all parameters up to third-order are considered and run. The dashed lines indicate the running of the same operators, but now including fourth and fifth-order parameters fixed at a small value, 0.01, for illustration. The inclusion of those terms, even if their initial values are small, have a major impact on second and third-order coefficients, starting already at order $\s^2$. The linear bias is indirectly affected by the running of second and third-order terms, only differing by the solid line at order $\s^4$, according to \reffig{Diag}.  
We interpret this finding as a consequence of higher-order operators being very UV-sensitive, such that they strongly source other lower-order operators. This means that in practice, strong ($\L$-dependent) degeneracies between the bias coefficients are expected to occur, that must be carefully accounted for, both for field-level and $N$-point function analyses.  For instance, the degeneracies between two bias parameters when performing parameter inference may have a large $\L$-dependent component, since the operator definitions also depend on the choice of $\L$.

\subsection{The two-loop RGE} \label{sec:twoloopsolution}

\begin{figure}[t]
    \centering
    \includegraphics[width = 0.70\textwidth]{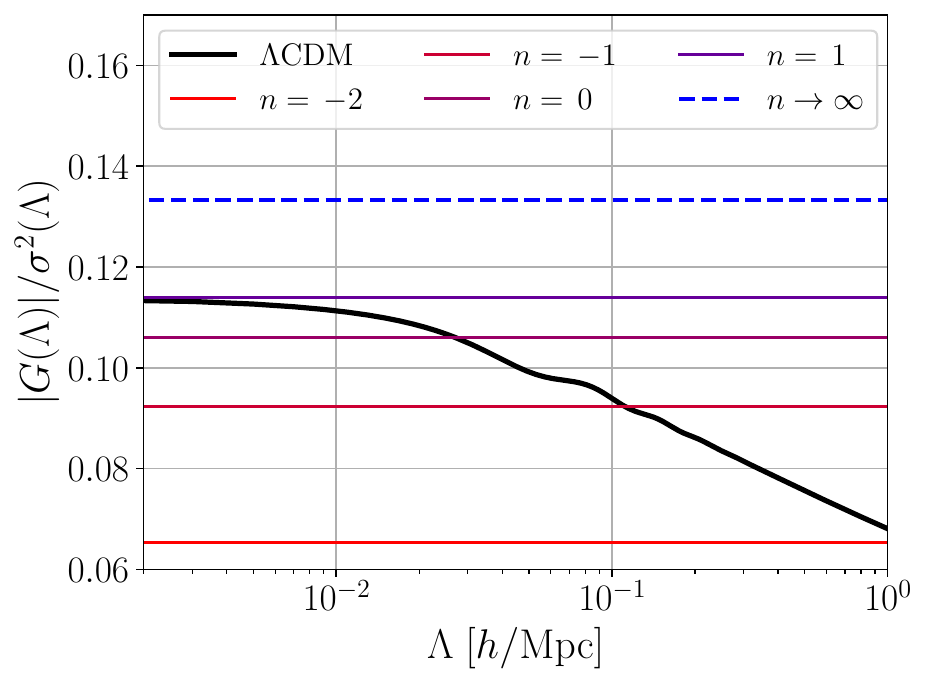}
    \caption{Integral $G(\L)$ from \refeq{Gdef} that enters the two-loop running, normalized by the variance $\s_\L^2$. We show power-law cases for various $n$, for which the ratio is a constant, and $\Lambda$CDM in black. In the asymptotic limit of a blue spectrum, $n \to \infty$, shown as dashed, the ratio approaches $-2/15 \approx -0.13$.  
    }
    \label{fig:2Lrunning}
\end{figure}

We now discuss the relevance of the two-loop running in the linear bias $b_\delta$, given by the last term of \refeq{drun}.
All bias operators up to fifth order contribute, controlled by the single quantity
\ba \label{eq:Gdef}
G(\L) &\equiv \int_0^\Lambda dq \frac{q^2P^{\text{lin}}(q)}{2\pi^2} g(q/\Lambda) \,,
\ea
containing the function $g(r)$ defined in \refeq{doublehardfuncg}.
At two-loop the dependence of the RG on the linear power spectrum $P^{\text{lin}}(q)$ is thus completely encapsulated in only one single additional function $G(\Lambda)$ along with $\s^2_\Lambda$. 

To build some intuition for those quantities, we start with the simpler case of a
power-law spectrum 
\be
k_*^3 P^{\text{lin}}(k) = A_s\bigg( \frac{k}{k_*} \bigg) ^n \,,
\ee
with reference scale $k_*$ and amplitude $A_s$. Its variance is given by (for $n>-3$)
\be
\s^2_\Lambda = \frac{1}{2 \pi ^2 (n+3)}A_s \bigg(\frac{\Lambda}{k_*}\bigg) ^{n+3}\,.
\ee
Furthermore, $G(\L)$ can be computed analytically for the power-law case, e.g.
\ba
G(\L) &= -\frac{2 (24 \log 2-13)}{315 \pi ^2}A_s\bigg(\frac{\Lambda}{k_*}\bigg) ^{2}
= -\frac{8 (24 \log 2-13)}{315 }\s^2_\Lambda \quad \textrm{for} \quad n = -1\,,\nn\\
G(\L) &=-\frac{2 (1+24 \log 2)}{3465 \pi ^2}A_s\bigg(\frac{\Lambda}{k_*}\bigg) ^{6}
= -\frac{8 (1+24 \log 2)}{1155 \pi ^2}\s^2_\Lambda \quad \textrm{for} \quad n = 3\,.
\ea
The ratio between $G$ and $\sigma^2$ is given by an $n$-dependent constant which we display in  \reffig{2Lrunning} for various $n$. It increases (in modulus) with the spectral index $n$.  This is because for larger $n$ the integral $G(\Lambda)$ is dominated by larger values of $q/\Lambda$, and since the weight function $g(q/\Lambda)$ increases in modulus with $q/\Lambda$ over the integration domain $q\leq\Lambda$  [see \reffig{gfunc} and \refeq{doublehardfuncg}].

We can also obtain some bounds that are independent of the linear input spectrum.
Since $g(r)$ is bounded by $-2/15\leq g <0$ [see \reffig{gfunc} and \refeq{doublehardfuncg}], we have
\be \label{eq:Gbound}
-\frac{2}{15}\leq \frac{G(\Lambda)}{\s^2_\Lambda}<0\,.
\ee
One can see that this bound is saturated in \reffig{2Lrunning} when $n\to \infty$, since in that case the spectrum is very blue and most of the support for the $G(\Lambda)$ integral is coming from wavenumbers near the cutoff, where $g(q/\Lambda)$ approaches $-2/15\simeq -0.13$. Therefore, the two-loop contribution becomes more relevant the more blue the spectrum is. 

\begin{figure}[t]
    \centering
    \includegraphics[width = 0.48\textwidth]{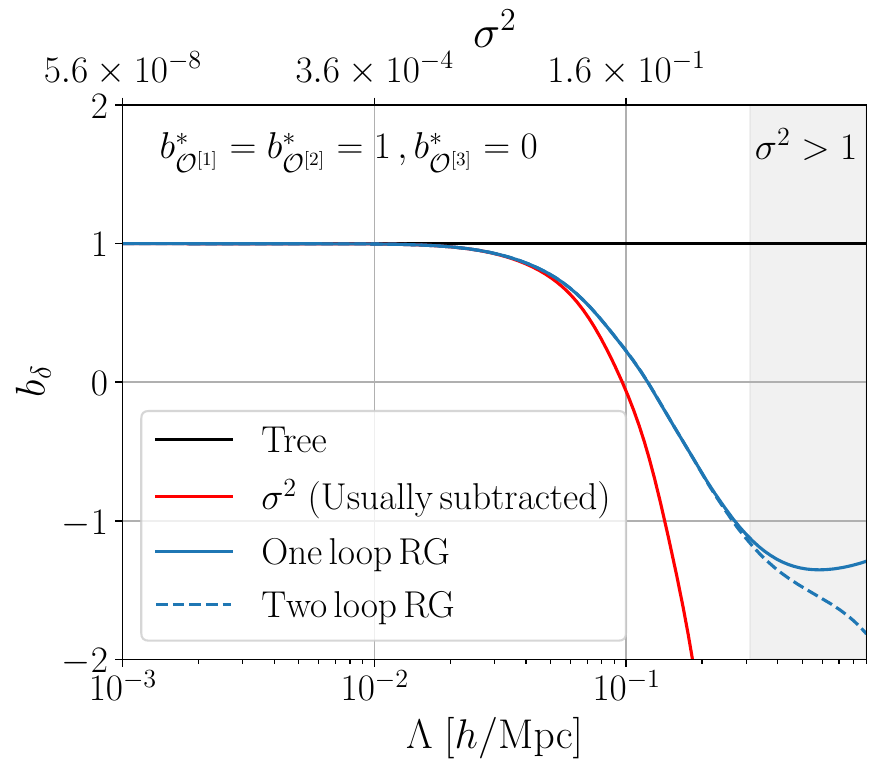}
    \includegraphics[width = 0.48\textwidth]{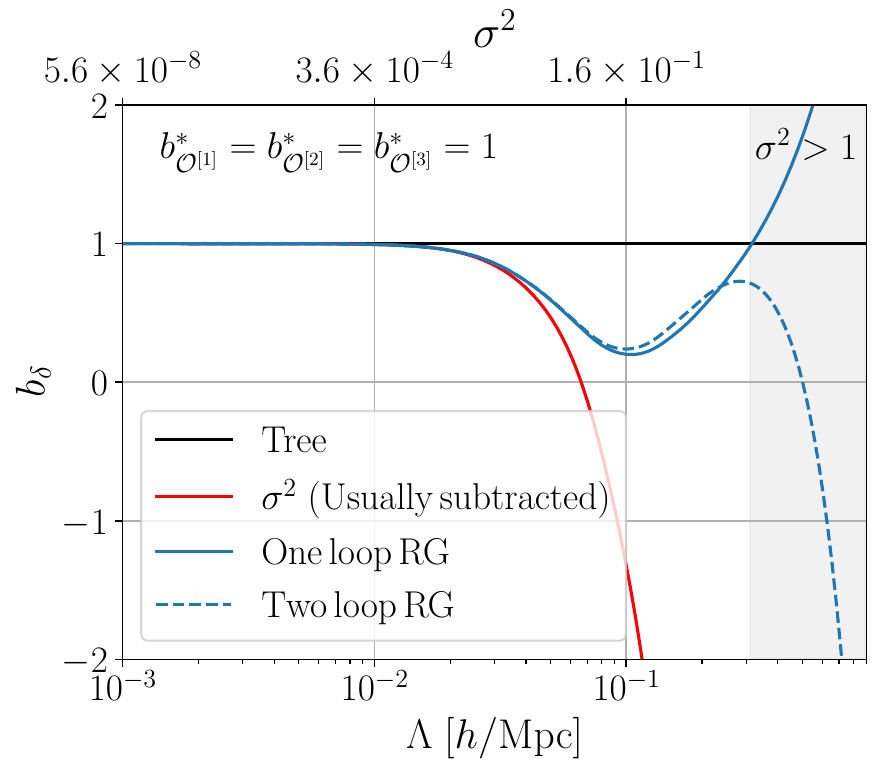}
    \caption{The difference between the tree-level ($\L$-independent), one-loop and two-loop running for the linear bias parameter. We also display the one-loop subtracted part of the renormalized bias (labelled as ``usually subtracted''), given by \refeq{Z1L} and that is also the first term in the expansion of the ODE solution \refeq{solODE}. In the left, we take as initial conditions $b^\ast = b(\Lambda^\ast) = 1$ with $\Lambda^\ast = 10^{-3} h/$Mpc, considering $b^\ast_{{\cal O}^{[1]}}  = b^\ast_{{\cal O}^{[2]}}  = 1$. In the right, we also consider non-zero initial conditions for the third-order operators $b^\ast_{{\cal O}^{[3]}}  = 1$. }
    \label{fig:1Lrunningover2L}
\end{figure}

In  \reffig{2Lrunning}, we also show the ratio $G(\Lambda)/\sigma^2$ obtained for a $\Lambda$CDM spectrum (black line), which features a $\Lambda$-dependence. As expected, it asymptotes towards the power-law case with $n\simeq 1$ for small $\Lambda$, and corresponds to the power-law value for increasingly negative $n$ when $\Lambda\gg 0.02h/$Mpc.
While $|G(\Lambda)|/\sigma_\Lambda^2$ is parametrically of order unity, the finding that its numerical value is small implies that the solutions of the RG equations with or without taking the two-loop $\beta$-function into account remain close to each other on weakly non-linear scales (see \reffig{1Lrunningover2L}). As we discuss in Sec.~\ref{sec:resum}, the smallness of this ratio is actually a necessary condition for claiming a systematic improvement when using the RG-evolved bias over the usual renormalized bias, and we can identify a regime where the ratio from \refeq{Gbound} is parametrically suppressed. 

We plot the constant tree-level contribution, the one-loop renormalized contribution from \cite{McDonald:2006mx,Assassi:2014fva} defined in \refsec{biasrenorm}, as well as the one and two-loop running in \reffig{1Lrunningover2L}. As commented below \refeq{solODE}, the one-loop renormalized bias is the first Taylor term of the RG solution. We consider only the linear and quadratic bias as non-zero for the initial conditions of the left panel and we also include third-order operators in the right panel. We also focus on the perturbative regime $\s^2<1$.
These results match the expectations of perturbation theory, that for low $\Lambda$ they all agree and each loop contribution starts to become relevant as $\L$ increases: the one-loop calculations deviate from the tree-level, then the RG solutions deviate from the one-loop renormalized result and finally the two-loop RG becomes slightly relevant. 
The inclusion of third-order operators in the initial condition (right panel) leads to an enhanced importance of the loop contributions, since the one-loop running starts to deviate from the renormalized result earlier in $\L$; also the two-loop becomes more relevant at lower $\L$.

\newpage
\section{Systematic resummation of higher loops via the RGE}\label{sec:resum}

As is well-known in QFT, the solution of RG equations corresponds to a partial resummation of higher-loop effects. The RG is useful in this context since {\it (i)} it resums the dominant parts of higher loops (in a precise sense reviewed below, related to the presence of two widely separated energy scales, and the appearance of `large logarithms') and {\it (ii)} solving the RG is much simpler than computing all those higher loops. While the second item applies also to the bias RG, as demonstrated in the last Section, we discuss in this Section to what extent the first one is met. In \refsec{QFT_resum} we briefly review the precise meaning of the resummation of `large logarithms' with the help of RG techniques in the QFT context, emphasizing that the resulting `RG-improved' perturbation theory allows for a systematic re-organization of the perturbative expansion under specific conditions. In \refsec{LSS_resum}  we transfer these techniques to the bias RG and identify the analog of the `large logarithm'.

\subsection{The analogy with QFT} \label{sec:QFT_resum}

To begin, we briefly remind the rationale for using RG evolution in the context of QFT, taking the evolution of the fine-structure constant $\alpha(\mu)$ with renormalization scale $\mu$ in quantum electrodynamics as a well-known example. The RG equation (computed in dimensional regularization within the modified minimal subtraction scheme $\overline{\text{MS}}$) takes the form
\be\label{eq:RGQED}
  \frac{d\alpha}{d\ln\mu} = \beta_\text{1L}\alpha^2 + \beta_\text{2L}\alpha^3 + O(\alpha^4)\,,
\ee
with $\beta_\text{1L}=2/(3\pi)$ and $\beta_\text{2L}=1/(4\pi^2)$ denoting the one- and two-loop $\beta$-functions, in analogy to the first and second lines in Eq.\,\eqref{eq:drun}, respectively. Keeping only the former yields the seminal solution
\be\label{eq:1LRGQED}
  \alpha(\mu)\big|_\text{LL} = \frac{\alpha}{1-\beta_\text{1L}\alpha\ln(\mu/\mu_*)} 
  = \alpha\left[1+\beta_\text{1L}\alpha\ln(\mu/\mu_*)-\beta_\text{1L}^2\alpha^2\ln^2(\mu/\mu_*)+\dots\right]\,,
\ee
where $\alpha\equiv\alpha(\mu_\ast)\simeq 1/137$ is the value inferred from low-energy experiments (at an IR scale $\mu_\ast$ related e.g.~to the electron mass), and $\alpha(\mu)$ the coupling strength relevant for predictions of high-energy collisions (e.g.~at an $e^-e^+$ collider with UV scale $\mu$ being chosen at the center of mass energy)\footnote{Even though the  scale $\mu$ introduced in dimensional regularization is distinct from the smoothing (or cutoff) scale $\L$, both are a priori unphysical quantities and observables such as cross-sections $\sigma$ in QFT or N-point functions of galaxies in LSS do not depend on it.
Nevertheless, in practice, such parameters are often highly useful for understanding and predicting the dependence of observables on widely separated physical scales, and developing suitable resummation techniques. In the LSS context a conceptually analogous situation of this kind is known from IR resummation, featuring an artificial separation scale $k_s$, the dependence on which would vanish exactly when adding IR-resummed loop corrections to all orders (see e.g. Sec.~4.3 in~\cite{Garny:2022fsh}).}. A tree-level computation using $\alpha(\mu)$ is then equivalent to resumming contributions from a series of loop computations performed at the IR scale, as given by the Taylor expansion of Eq.\,\eqref{eq:1LRGQED} in powers of $\alpha$. The terms of order $\alpha^\ell$ in the square bracket would result from an $\ell$-loop computation in this case. Nevertheless, those terms are not the complete $\ell$-loop, but only the part that involves one large logarithm $\ln(\mu/\mu_\ast)$ for each of the $\ell$-loop corrections, being the `leading logarithms' (LL) of the form $\alpha^\ell\ln^\ell(\mu/\mu_\ast)$. Solving the RGE and doing a tree-level computation with the running coupling $\alpha(\mu)$ as input is thus equivalent to doing a series of loop computations in terms of the IR coupling $\alpha$, and retaining only the LL part of each loop. The one-loop RG solution thus corresponds to LL resummation.

This scheme can be systematically improved by including also $\beta_\text{2L}$ when solving the RGE from Eq.\,\eqref{eq:RGQED}. In this case the running coupling corresponds to resumming also `next-to leading logarithms' (NLL) of the form $\alpha^\ell\ln^{\ell-1}(\mu/\mu_\ast)$ for all $\ell\geq 2$. Thus solving the RGE including the two-loop $\beta$-function and using the resulting $\alpha(\mu)$ as input for a one-loop computation is equivalent to a series of loop computations in terms of the IR coupling $\alpha$ and resumming LL and NLL contributions at all loop orders. To illustrate this point, consider the Taylor series expansion of the solution to the RG including both $\beta_\text{1L}$ and $\beta_\text{2L}$,
\be\label{eq:2LRGQED}
  \alpha(\mu)\big|_\text{NLL} 
  = \alpha\left[1+\beta_\text{1L}\alpha\ln(\mu/\mu_*)-\beta_\text{1L}^2\alpha^2\ln^2(\mu/\mu_*)
  +\beta_\text{2L}\alpha^2\ln(\mu/\mu_*) + O(\alpha^3) \right]\,.
\ee
When doing perturbation theory in the IR coupling $\alpha$, the two-loop corresponds to terms of order $\alpha^2$. In the RG approach, they are generated partially by the one-loop RG, being the `one-loop squared' term $\propto \beta_\text{1L}^2$ which yields the LL contribution, and the `two-loop' RG contribution $\propto\beta_\text{2L}$ which corresponds to the NLL. Provided $|\ln(\mu/\mu_\ast)|\gg 1$, the LL approximation thus already captures the dominant part, while the NLL approximation yields a correction that is relatively suppressed as $\beta_\text{1L}^2\ln(\mu/\mu_*)/\beta_\text{2L}$. 

\begin{figure}[t]
    \centering
    \includegraphics[width = 0.98\textwidth]{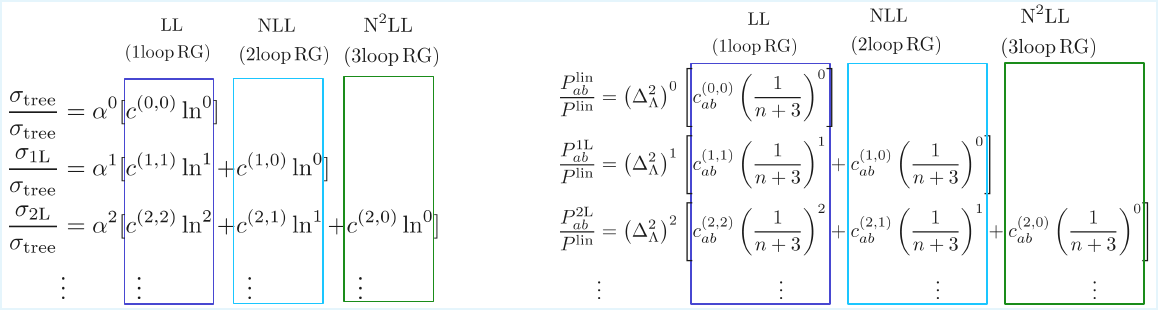}
    \caption{Schematic illustration of how the RG systematically resums higher-loop terms. In the left, the QFT analogy of a loop expansion of a generic cross-section $\sigma=\sigma_\text{tree}+\sigma_{1{\rm L}}+\dots$ is shown, in terms of powers of the fine-structure constant $\alpha$ (rows) and the hierarchy of powers of the `large logarithm' $\ln(\mu/\mu_\ast)$ at every loop order, see \refeq{sigmaLLoop}. The columns correspond to the infinite towers of terms that are resummed via solving RG equations, known as RG-improved perturbation theory, see \refeq{QFTRGimprovedPT},
    systematically organized in leading logarithmic (LL), next-to-leading logarithmic (NLL), next-to-next-to-leading logarithmic (N$^2$LL) approximation, etc.
    In the right, we illustrate the analogous expansion of the tracer power spectrum $P_{ab}(k)=P^\text{lin}_{ab}+P^{\text{1L}}_{ab}+\dots$, see \refeq{Pabloop}. The role of $\alpha$ is taken by the dimensionless input power spectrum $\Delta^2$, and the `large logarithm' corresponds to the ratio $\s_\L^2/\Delta_\L^2$, see Eq.\,\eqref{eq:mapsLCDM}, being for example given by $1/(n+3)$ for a power law spectrum with spectral index $n$ close to the critical value $-3$, and by $\ln(\L/k_\text{eq})/3$ for a $\L$CDM spectrum when $\L\gg k_\text{eq}$. Each column shows the tower of terms that are resummed by solving RG equations, allowing for a systematic computation within RG-improved perturbation theory, see \refeq{PabRGimprovedPT}. For simplicity we omit in each row terms with positive powers of $n+3$ since they vanish for $1/(n+3)\gg 1$. 
    }
    \label{fig:RGresumdiag}

\end{figure}

Thus, using the running coupling obtained from solving RG equations is a useful concept in QFT due to the existence of {\it two} small expansion parameters: {\it (a)} the small coupling constant $\alpha(\mu)\ll 1$, and {\it (b)} the inverse of the large logarithm $1/|\ln(\mu/\mu_\ast)|\ll 1$. The latter is an emergent scale, related to the logarithmic sensitivity of loop integrals to the UV and IR external scales in dimensional regularization (such as e.g.~the center of mass energy (UV) versus the electron mass (IR) in the example above). The LL approximation (tree-level computation with LL running coupling) corresponds to  resumming all terms that scale as $[\alpha\ln(\mu/\mu_\ast)]^\ell$ for all loop orders $\ell=1,2,3,\dots$. The NLL approximation (one-loop computation with NLL running coupling) corresponds to resumming $\alpha\times[\alpha\ln(\mu/\mu_\ast)]^\ell$ terms. This can be systematically extended to NNLL, and so on, and is known as RG-improved perturbation theory. 
Schematically an observable (e.g. a cross-section $\sigma$) has an expansion within RG-improved perturbation theory (relative to its tree-level value) 
\be\label{eq:QFTRGimprovedPT}
\frac{\sigma}{\sigma_\text{tree}} = f_\text{LL}(x)
+\alpha\times f_\text{NLL}(x)
+\alpha^2\times f_\text{NNLL}(x) + \dots\,,
\ee
with $x\equiv \alpha\times \ln(\mu/\mu_\ast)$ and functions $f_{\text{N}^n\text{LL}}(x)=\sum_{m\geq 0} x^m c^{(n+m,m)}$ that capture the resummation of large logarithms, for some coefficients $c^{(n+m,m)}$. Importantly, in \refeq{QFTRGimprovedPT} it is understood that we have chosen $\mu$ to be of order the center-of-mass energy $\sqrt{s}$ of the cross-section $\sigma$. We continue to make this assumption in the remainder of this subsection. The LL approximation corresponds to $f_\text{LL}(x)$, the NLL approximation to $f_\text{LL}(x)+\alpha f_\text{NLL}(x)$, etc.
For comparison, the naive loop expansion (i.e. at the IR scale) of the form $\sigma=\sigma_\text{tree}+\sigma_{\text{1L}}+\dots$ corresponds to a strict Taylor series in $\alpha$ (i.e. irrespective of the power of logarithms), such that the $\ell$-loop correction $\sigma_{\ell\text{L}}$ reads
\be\label{eq:sigmaLLoop}
  \frac{\sigma_{\ell\text{L}}}{\sigma_\text{tree}}
  = \alpha^\ell\Bigg[c^{(\ell,\ell)}\ln^\ell(\mu/\mu_\ast)+c^{(\ell,\ell-1)}\ln^{\ell-1}(\mu/\mu_\ast)+\dots\Bigg]\,.
\ee
The virtue of the RG is that it allows one to fix the entire $x$-dependence of the functions $f(x)$. More precisely, for every fixed value of $n$, the entire tower of coefficients $c^{(n+m,m)}$ for all $m=1,2,3,\dots$ are fixed by the $(n+1)$-loop $\beta$-functions entering the RGEs, with the only free parameter being $c^{(n,0)}$ (requiring to perform an $n$-loop computation at the IR scale to determine it for the cross-section of a given process). For example, if $\sigma_\text{tree}(s)=\alpha^2/s$ (with center of mass energy squared $s$) depends quadratically on $\alpha$, the one-loop RG yields 
\be\label{eq:fLLQFT}
  f_\text{LL}(x) = \left(\frac{\alpha(\mu)}{\alpha}\right)^2=\frac{1}{(1-\beta_\text{1L}x)^2}\,,
\ee
and thus all $c^{(m,m)}=(m+1)(\beta_\text{1L})^mc^{(0,0)}$ for $m>0$ are fixed in terms of $\beta_\text{1L}$, while $c^{(0,0)}=1$ is fixed by the tree-level result. 
Thus, in this example, LL resummation just amounts to replacing $\alpha$ by the running coupling $\alpha(\mu)$ in the tree-level cross-section, in line with the discussion above. We illustrate the resummation at each  $\text{N}^n\text{LL}$ order in the left panel of \reffig{RGresumdiag}. The right panel will be discussed in \refsec{LSS_resum}.

In summary, the RG approach is useful when choosing the arbitrary scale $\mu$ of order the UV scale (related to the momenta of particles in the process, e.g.~the center of mass energy $\sqrt{s}$), and $\mu_\ast$ of order of the IR scale (e.g.~the electron mass) at which the value of $\alpha\simeq 1/137$ is well-known. If the logarithm becomes so large that the product $x=\alpha\ln(\mu/\mu_\ast)\sim O(1)$ is of order unity (while $\alpha\ll 1$ still ensures perturbativity of the theory), using the RG-improved framework becomes mandatory, while the naive perturbative expansion (strict expansion in $\alpha$) would start to break down.

\subsection{Resummation in LSS} \label{sec:LSS_resum}

Let us now discuss how to translate these properties to the bias RG, i.e. the resummation of ~$\L$-dependent loop corrections to the power spectra and bias coefficients. First, we realize that in the example above $\alpha$ takes two roles, being the perturbative expansion parameter as well as the quantity for which RG evolution is considered, but a generalization is straightforward (as occurs also for renormalization of quantities apart from the coupling constant itself in QFT, such as e.g.~the electron mass). For the bias RG the latter role is played by the bias coefficients $b_a(\Lambda)$, while the  expansion `parameter' is the linear density contrast\footnote{More precisely, rescaling $\delta_L({\bf q})\mapsto \sqrt{\alpha}\delta_L({\bf q})$ with a formal book-keeping parameter $\alpha$ the usual perturbative loop expansion of power spectra $P(k)$ corresponds to a Taylor series in $\alpha$. We use $\Delta_\L^2$ as a proxy for this book-keeping parameter in the following. We note that $\Delta_\L^2$ is the literal expansion parameter for the power-law case in the limit $k\ll\L$ considered below, such as in Eq.\,\eqref{eq:PabRGimprovedPT}.} $\delta_L({\bf q})$, related to $\Delta_q^2=q^3P^{\text{lin}}(q)/(2\pi^2)=d\sigma_q^2/d\ln q$.
The two-loop $\beta$-function in the second line of Eq.\,\eqref{eq:drun} is of second order in this quantity, one more than the one-loop piece in the first line. This is analogous to~Eq.\,\eqref{eq:RGQED}. Second, the bias RG is a matrix equation for the set of all bias coefficients, instead of a single ODE Eq.\,\eqref{eq:RGQED}. This situation also often occurs in QFT, and does not change the parametric structure.

The main open question is if there exists a quantity that plays the role of the `large logarithm' within the bias RG. If yes, this emergent quantity would allow for a systematic formulation of RG-improved perturbation theory analogous to the LL, NLL, NNLL (etc.) resummation discussed above. This means that using the bias obtained from the solution of the RG equations absorbs dominant UV contributions of higher loop effects, analogously to using the running coupling in QFT. Thus, identifying a parameter akin to $\ln(\mu/\mu_\ast)$ is a crucial justification for the usefulness of the RG approach. 

Our result for the two-loop $\beta$-function allows us to perform a first non-trivial check. Namely, as in Eq.\,\eqref{eq:2LRGQED}, the analog of the (inverse of the) `large logarithm' is provided by the ratio of the `one loop squared' to the `two-loop' RG contribution when expanding the RG solution to second order in $\Delta_\L^2=d\sigma_\L^2/d\ln \L$ (for initial scale at $\Lambda_\ast=0$) 
\ba \label{eq:solODE2}
b_\delta(\L) 
        &= b_\delta^\ast - \sigma_\L^2\bar{s}_{\delta c}^{\rm 1L} b_c^\ast + \bar{s}_{\delta b}^{\rm 1L}\bar{s}^{\rm 1L}_{bc} b_c^\ast I_{\text{(1L)}^2}(\L)
         - \bar s^\text{2L}_{\delta c} b_c^\ast I_{\text{2L}}(\L) 
         \,\, + O[(\s_\L^2)^3]\,, 
\ea
where the first and second term are the bias defined at the IR scale and the one-loop correction, respectively, while the third term is the `one-loop squared' and the fourth the `two-loop RG' contribution, respectively, analogously to Eq.\,\eqref{eq:2LRGQED}. Here we introduced the notation $\bar s^\text{2L}_{\delta c}\equiv 30\tilde d_c^{(5)}$ 
for the numerical coefficients of the two-loop $\beta$-function, and 
\bea\label{eq:I1LsqI2L}
  I_{\text{(1L)}^2}(\L) &=& \int_0^\Lambda \frac{d\L'}{\L'} \frac{d\sigma^2_{\Lambda'}}{d\ln\Lambda'} \int_0^{\Lambda'} \frac{d\L''}{\L''} \frac{d\sigma^2_{\Lambda''}}{d\ln\Lambda''} = \frac{(\s^2_\L)^2}{2}\,,\nn\\
  I_{\text{2L}}(\L) &=& \int_0^\Lambda \frac{d\L'}{\L'} \frac{d\sigma^2_{\Lambda'}}{d\ln\Lambda'} \int_0^{\Lambda'} \frac{d\L''}{\L''} \frac{d\sigma^2_{\Lambda''}}{d\ln\Lambda''} g(\L''/\Lambda')\,,
\eea
with the function $g(r)$ from Eq.\,\eqref{eq:doublehardfuncg} controlling the two-loop RG. The analog of the emergent small parameter $1/|\ln(\mu/\mu_\ast)|$ in the bias RG is thus related to the ratio $I_{\text{2L}}(\L)/I_{\text{(1L)}^2}(\L)$,
\be \label{eq:1overlog}
  \frac{1}{\ln(\mu/\mu_\ast)} \mapsto \frac{I_{\text{2L}}(\L)}{I_{\text{(1L)}^2}(\L)}\,.
\ee
This correspondence is to be understood in the sense that the role of the two small expansion parameters $(a)$ $\alpha\ll 1$ and $(b)$ $1/|\ln(\mu/\mu_\ast)|\ll 1$ is played by $(a)$ $\Delta_q^2$ and $(b)$ a quantity that parametrically scales as the ratio stated above (up to some $O(1)$ numerical prefactors). Resummation via RG techniques and the framework of RG-improved perturbation theory becomes strictly valid only when the $(b)$ scale features a dependence on some input parameters that allows it to become arbitrarily small in a certain well-defined limit. It turns out that this parametric suppression arises when the linear power spectrum is a power law with spectral index close to $n = -3$, as we will see below. Later, we discuss the suppression for a $\L$CDM spectrum as well.

\subsubsection*{Power-law spectrum}

\begin{figure}[t]
    \centering
    \includegraphics[width = 0.7\textwidth]{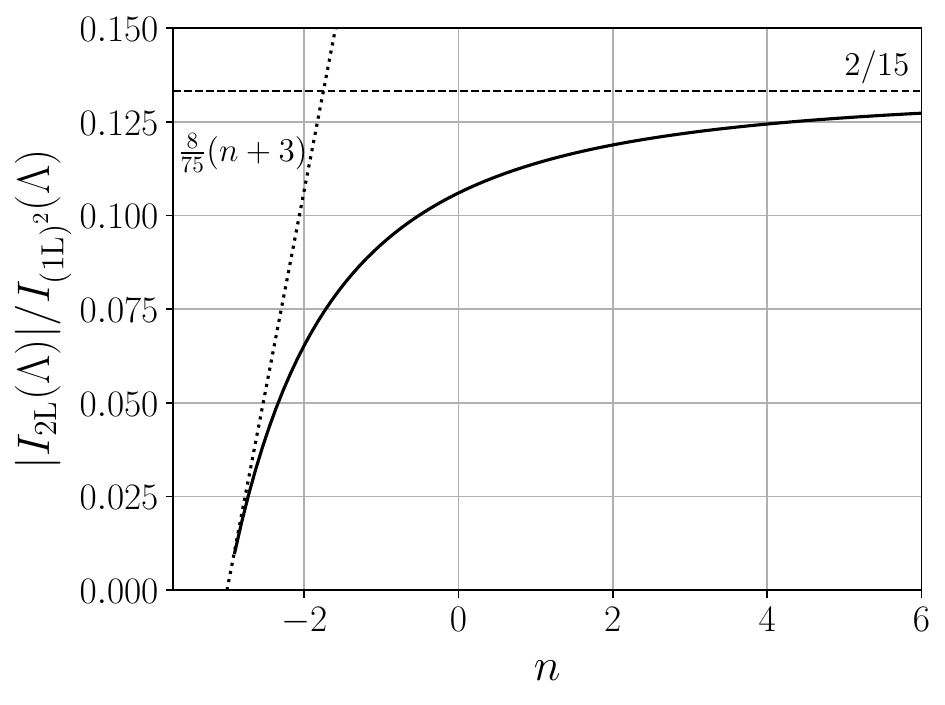}
    \caption{
    Ratio between the two-loop contributions to bias renormalization generated from two-loop RG $\propto I_{\text{2L}}$ and the iterated `one-loop squared' term from the one-loop RG $\propto I_{\text{(1L)}^2}$, versus the spectral index $n$ for power law cosmologies. This ratio is the analog of $1/|\ln(\mu/\mu_\ast)|$ in QFT, and a small value signals that the bias RG yields a systematic improvement over naive perturbation theory. The dotted line $8(n+3)/75$ is the limit for $n\to -3$, see Eq.\,\eqref{eq:supfaclimit}, and the dashed horizontal line at $2/15$ for $n\to\infty$. 
    }
    \label{fig:2Lrunning_I}
\end{figure}

Let us first consider the case of a power law spectrum $P^{\text{lin}}(k)\propto  k^n$. We have that the ratio $I_{\text{2L}}(\L)/I_{\text{(1L)}^2}(\L)$ is then a $\Lambda$-independent constant, and equal to $G(\Lambda)/\sigma_\L^2$ from Eq.\,\eqref{eq:Gdef}, for example
\ba\label{eq:supfac}
\bigg| \frac{I_{\text{2L}}(\L)}{I_{\text{(1L)}^2}(\L)} \bigg|&= \frac{8 (24 \log (2)-13)}{315} \simeq 0.092 \quad \textrm{for} \quad n = -1\,,\\
\bigg|\frac{I_{\text{2L}}(\L)}{I_{\text{(1L)}^2}(\L)}\bigg| &= \frac{8 (1+24 \log (2))}{1155} \simeq 0.122 \,\quad \textrm{for} \quad n = 3\,. 
\ea
While the numerical value is somewhat small (due to the smallness of $g(r)$ across its domain), it is not controlled by a parametrically small scale and thus, for a generic power law, no analog of the `large logarithm' emerges. For $n\to\infty$, the inner integral in Eq.\,\eqref{eq:I1LsqI2L} becomes dominated by the upper boundary and since $|g(1)|=2/15$ the ratio also approaches this value for a very blue power spectrum. On the other hand, for $n\to -3$ the variance $\sigma_\L^2\propto 1/(n+3)$ becomes parametrically large. This is because $\Delta_\L^2=d\sigma^2_\Lambda/d\ln\L=A_s(\Lambda/k_\ast)^{n+3}/(2\pi^2)$ becomes $\Lambda$-independent and thus the entire integration range between the IR and UV scale contributes to $\s_\Lambda^2$. 
Thus, $I_{\text{(1L)}^2}(\L) \propto 1/(n+3)^2$ for $n\to -3$ from above. On the other hand, we find $I_{\text{2L}}(\L)\propto 1/(n+3)$,
such that 
\ba\label{eq:supfaclimit}
\bigg| \frac{I_{\text{2L}}(\L)}{I_{\text{(1L)}^2}(\L)} \bigg|\ =&\ \frac{8}{75}(n+3)+O\left((n+3)^2\right)  \quad \textrm{for} \qquad \frac{1}{n+3}\gg 1. 
\ea
Thus, for a power-law spectrum with spectral index $n\to -3$ from above, the role of the `large logarithm' $|\ln(\mu/\mu_\ast)|\gg 1$ is taken by the large parameter $1/(n+3)\gg 1$. We show the ratio between the two-loop and one-loop squared contributions as a function of the spectral index in \reffig{2Lrunning_I}. The asymptotic limits $n\to -3$ and $n\to \infty$ are shown in dotted and dashed lines, respectively.

What is the origin of this `emergent' scale? We need to compare the two lines in Eq.\,\eqref{eq:I1LsqI2L}. While the integrals over $d\s^2/d\ln\Lambda'$ as well as the one over $d\s^2/d\ln\Lambda''$ in $I_{{(1L)}^2}$ pick up contributions from the entire interval between the IR and UV limits (corresponding to a factor $1/(n+3)$ for each such integral and therefore being proportional to $1/(n+3)^2$), the integral weighted by $g(r)$ entering $I_{\text{2L}}$ is still dominated by the region close to the boundary, due to the property $g(0)=0$, and $g(r)\propto r^2$ for small $r\to 0$. 
Hence, the $g$ function acts like a `filter' and parametrically picks up the diagonal region in the $\L'\times\L''$ plane, such that $I_{\text{2L}}(\L)\propto 1/(n+3)$.
This is not at all a coincidence, but required by the structure of the two-loop $\beta$-function Eq.\,\eqref{eq:RG2L}, containing the difference $s^\text{2L}(r)-s^\text{2L}(0)$ which suppresses the integrand for $r=\Lambda''/\L'\ll 1$.
In turn, this property is deeply related to the structure of the RG, meaning that, parametrically, only the double-hard region $r\sim O(1)$ of the two-loop contributes to the two-loop $\beta$-function, while the single-hard region with $r\ll 1$ does not, and is instead accounted for by the `one-loop squared'. Thus, the smallness of the ratio in Eq.\,\eqref{eq:supfaclimit} when $1/(n+3)\gg 1$ can be considered as a systematic suppression, akin to the suppression of NLL to LL contributions in QFT. This is related to the scale-invariance of $\Delta_q^2=A_s(q/k_\ast)^{n+3}/(2\pi^2)$ for $n\to -3$, and thus the sensitivity of loop integrals over the entire range from IR to UV scales. 

The  arguments from above suggest a generalization. In particular, the structure of loop integrals related to deterministic bias renormalization at leading order in gradient expansion is of the form (at $\ell$-loop order)
\be\label{eq:lLoopstructure}
  \int_0^\Lambda \frac{d\L_1}{\L_1} \frac{d\sigma^2_{\Lambda_1}}{d\ln\Lambda_1} \int_0^{\Lambda_1} \frac{d\L_2}{\L_2} \frac{d\sigma^2_{\Lambda_2}}{d\ln\Lambda_2}\cdots \int_0^{\Lambda_{\ell-1}} \frac{d\L_\ell}{\L_\ell} \frac{d\sigma^2_{\Lambda_\ell}}{d\ln\Lambda_\ell}\, g^{(i)}_\ell(\L_2/\L_1,\dots,\L_\ell/\L_1)\,,
\ee
in analogy to Eq.\,\eqref{eq:I1LsqI2L}. Here $g^{(i)}_\ell$ stand for a set of (dimensionless) weight functions (labeled with some index $i$) obtained from the bias kernels expanded for $k\ll\L$ and averaged over angles. For dimensional reasons, they can only depend on ratios of loop wavenumbers. For example, as discussed above, at two-loop order and for up to fifth order bias operators only two weight functions occur, corresponding to $g_2^{(1)}(r)=1$ (related to the `one-loop squared' RG contribution) and $g_2^{(2)}(r)=g(r)$ (for the `two-loop RG' contribution). Let us now discuss the parametric scaling in the limit $n\to -3$. Consider first the simplest case for which $g^{(i)}_\ell=1$ is constant (for some $i$). Then Eq.\,\eqref{eq:lLoopstructure} simply yields $(\s_\L^2)^\ell/\ell!=(\Delta_\L^2)^\ell\times (1/(n+3))^\ell/\ell!$, i.e.~one factor of $1/(n+3)$ for each loop. In addition each loop trivially contributes a factor of $A_s$, and since the result depends on scale only via $\L$ in the limit $k\ll\L$ we obtain a factor $(\Delta_\L^2)^\ell$. Note that this matches precisely the parametric structure of the terms that are generated by solutions of the one-loop RG, see Eq.~\eqref{eq:solODE}.

How does this change when considering other weight functions? The latter argument still applies, i.e.~we generically obtain a dependence on $(\Delta_\L^2)^\ell$. However, the scaling with $1/(n+3)$ depends on the weight function. For $n\to -3$ and a constant weight function, each loop integral receives equal contributions per logarithmic interval in $d\Lambda_k/\Lambda_k$ for all $k=1,\dots,\ell$, giving rise to one enhancement factor $1/(n+3)$ per loop. For  $g^{(i)}_\ell(r_2,\dots,r_\ell)$ we have $0\leq r_k\leq 1$ within the integration domain. If the weight function is suppressed if one or more of ratios tend to zero, such as e.g. occurs for $g(r)\propto r^2$ for $r\to 0$, this reduces the integration region that yields relevant contributions and effectively `filters' out some of the parts of the full integration domain. If for example $g^{(i)}_\ell(r_2,\dots,r_\ell)\propto r_\ell^p$ for $r_\ell\to 0$ with some $p>0$\footnote{The bias kernels entering in $g^{(i)}_\ell(r_2,\dots,r_\ell)$ have a dependence on loop wavenumbers ${\bf p}_j$ of the form `${\bf p}_j,-{\bf p}_j$', and the relevant limit is related to the case in which one (or several) of them (say with indices $j=1,\dots, m$ for some $m<\ell$) become much larger than the rest ($j'=m+1,\dots,\ell$), i.e. the wavenumber arguments entering the bias kernels obey the hierarchy $k\ll \Lambda_{j'}\equiv |{\bf p}_{j'}| \ll \Lambda_j\equiv |{\bf p}_j|$. While we do not present a proof here, general decoupling arguments~\cite{Peebles:1980yev} for non-linearities due to gravitational forces suggest that  small-scale perturbations ($\Lambda_j$) generate an impact on larger scales (here $\Lambda_{j'}$) that can be multipole-expanded and thus do not lead to divergences, meaning that $p\geq 0$. This is also in line with similar arguments for the matter field~\cite{1986ApJ...311....6G}.} the last integral in Eq.\,\eqref{eq:lLoopstructure} becomes dominated by the region that is (parametrically) close to the upper boundary even when $n\to -3$ (i.e.~even when $d\sigma^2_{\Lambda_\ell}/d\ln\Lambda_\ell\propto \L_\ell^{n+3}$ becomes scale-independent). Thus this last integral remains finite for $n\to -3$ and no $1/(n+3)$ factor arises. Even if the weight function is flat over the entire domain of the remaining $\ell-1$ integrations, the result scales only as $1/(n+3)^{\ell-1}$ in this case. Alternatively, if the weight function is further suppressed also when $r_{\ell-1}$ or more ratios become small, the resulting integral scales with correspondingly fewer powers of $1/(n+3)$. This argument can also be related to estimating the parametric scaling based on the method of regions~\cite{Beneke:1997zp}. For example, a weight function that is suppressed if any of the $r_k$ become small corresponds to `filtering' out the parametric $\ell$-hard region in which all loop wavenumbers tend to infinity uniformly for large $\L$, i.e. parametrically with ratios of order unity. In this case only the last $\L_1$ integration in Eq.\,\eqref{eq:lLoopstructure} is enhanced for $n\to -3$, and  thus a weight function that is unsuppressed only within the all-hard region yields a result that scales with a single power $1/(n+3)$. This can be generalized to weights that are suppressed when only a subset of the $r_k$ tend to zero, and suggests a systematic power counting in terms of powers of $1/(n+3)$, with a maximum of $1/(n+3)^\ell$ at $\ell$-loop order. We proceed in the following by assuming this pattern to hold, noting the close analogy to the counting of poles in dimensional regularization within QFT.

This means that for power law spectra with $n$ close to $-3$, the perturbative expansion can be organized as a power series in {\it two} small quantities: $(a)$ the usual expansion in the linear density field,
and $(b)$ the deviation of $n$ from $-3$, $n+3$. They are the analog of $(a)$ the small coupling constant $\alpha$ and $(b)$ the inverse of the large logarithm $1/\ln(\mu/\mu_\ast)$ in QFT, i.e.
\be \label{eq:maps}
\boxed{
  \ln(\mu/\mu_\ast)\mapsto \frac{1}{n+3}\,.
}
\ee
The existence of these two distinct parameters (related to amplitude $A_s$ and spectral index $n$ of the linear spectrum) implies that the bias RG can be used to systematically resum higher-loop UV-sensitive contributions by solving the RG equations for the bias coefficients, analogously as in QFT, provided that $\Delta_q^2\ll 1$ and $1/(n+3)\gg 1$. 

\medskip

Concretely, we can apply RG-improved perturbation theory to the cross spectrum of unrenormalized (i.e.~$\L$-dependent) bias operators ${\cal O}_a$ and ${\cal O}_b$ in the limit $k\ll \L$. Focusing for simplicity on the contributions related to deterministic bias renormalization  (i.e. after subtracting the stochastic contribution) denoted by $P_{ab}(k)$
yields
\be\label{eq:PabRGimprovedPT}
\frac{P_{ab}(k)}{P^\text{lin}(k)}\bigg|_{k\ll\L}
= f_{ab}^\text{LL}(x)+\Delta_\L^2\times f_{ab}^\text{NLL}(x)
+(\Delta_\L^2)^2\times f_{ab}^\text{NNLL}(x)+\dots \,,
\ee
where $x\equiv \Delta_\L^2\times 1/(n+3) = \s_\L^2$ and $f_{ab}^{\text{N}^n\text{LL}}(x)=\sum_{m\geq 0} x^m c_{ab}^{(n+m,m)}$ with some coefficients $c^{(n+m,m)}$, in complete analogy to Eq.\,\eqref{eq:QFTRGimprovedPT}. As in the QFT example, the $f(x)$ functions capture a resummation of `large logarithms', being in the present context all powers of $x=\s_\L^2$ while treating $\Delta_\L^2\ll 1$ as small, and $1/(n+3)\gg 1$ as large, such that parametrically $x\sim O(1)$. The LL approximation is $f_{ab}^\text{LL}(x)$, the NLL approximation is $f_{ab}^\text{LL}(x)+\Delta_\L^2\times f_{ab}^\text{NLL}(x)$, etc., as illustrated in the right part of Fig.~\ref{fig:RGresumdiag}. Comparison of \refeq{PabRGimprovedPT} with the naive loop expansion $P_{ab}=P_{ab}^\text{lin}+P_{ab}^{1\text{L}}+\dots$, being a strict Taylor series in the single quantity $\Delta_\L^2$ [irrespective of the value of $1/(n+3)$] gives for the $\ell$-loop part\footnote{Note that all integrals are IR finite for $n>-3$, and bounded from above by $\Lambda$. Technically, for $n\to -3$, the loops become IR divergent and the spectral index $n$ plays a role similar to the dimensional regularization parameter $\epsilon=(4-d)/2$ in QFT, except that small but finite $n+3$ is a physically viable value, while $\epsilon$ is an artificial regularization parameter.} 
\be\label{eq:Pabloop}
  \frac{ P^{\ell\text{L}}_{ab}(k)}{P^\text{lin}(k)}\bigg|_{k\ll\L} = (\Delta_\Lambda^2)^\ell\times\Bigg[c_{ab}^{(\ell,\ell)}\left(\frac{1}{n+3}\right)^{\ell}
  + c_{ab}^{(\ell,\ell-1)}\left(\frac{1}{n+3}\right)^{\ell-1} +\dots\Bigg]\,,
\ee
in analogy to Eq.\,\eqref{eq:sigmaLLoop} and in line with the discussion around Eq.\,\eqref{eq:lLoopstructure}. 

As in the QFT case, the virtue of the RG is that it allows one to compute the complete $x$-dependence of the function $f_{ab}^\text{LL}(x)$ in terms of the one-loop $\beta$-function $s_{ab}^\text{1L}$ (as well as the linear tracer power spectrum), the function $f_{ab}^\text{NLL}(x)$ in terms of the one- and two-loop $\beta$-function (as well as the one-loop tracer power spectrum), and so on. Specifically, this means all coefficients $c_{ab}^{(\ell,\ell)}$ are fixed in terms of the one-loop $\beta$-function and the tree-level spectrum $c_{ab}^{(0,0)}$, all coefficients $c_{ab}^{(\ell,\ell-1)}$ are fixed in terms of the one- and two-loop $\beta$-function and the one-loop spectrum related to $c_{ab}^{(1,0)}$, and so on. This means that the bias RG can be used to predict and resum the dominant UV-sensitive parts of higher loops at all loop orders in the exact sense described above, provided the spectral index is close to $-3$. 
We illustrate the resummation at each  $\text{N}^n\text{LL}$ order  in the right panel of \reffig{RGresumdiag}.
We discuss next how this happens in practice for the LL case.

Concretely, we can use that (by definition) the complete tracer power spectrum $P_{gg}(k)=P_{ab}(k,\L)b_a(\L)b_b(\L)+\text{cst.}$ has to be independent of $\Lambda$, and use the LL solution of the one-loop bias RG Eq.\,\eqref{eq:RG1L}, following \refeq{solODE}, which reads (denoted by the LL superscript in analogy to the solution of the one-loop RG in the QFT example above) 
\be\label{eq:BLL}
  b_a^\text{LL}(\Lambda)=[e^{-\bar s^\text{1L}\sigma_\Lambda^2}]_{ab}b_{[b]}\,,
\ee
with initial condition at $\L_\ast=0$ and initial values given by the renormalized bias coefficients $b_{[b]}$. We observe that its dependence on $\L$ is entirely encapsulated in $\s_\L^2=x$, and is thus a function of the variable $x$ only. This matches precisely with the $\L$-dependence of the LL term in Eq.\,\eqref{eq:PabRGimprovedPT}. Note that we can formally treat $x$ and $\Delta^2_\L$ as two independent expansion parameters (corresponding to the two expansion parameters $(a)$ and $(b)$ discussed above, related to $A_s$ and $1/(n+3)$, respectively). Thus the $\L$-dependence has to cancel among terms that scale with some given powers of each of these two parameters. The LL terms correspond to the zeroth power in $\Delta_\L^2$, but arbitrary powers of $x$. Using Eq.\,\eqref{eq:BLL} and Eq.\,\eqref{eq:PabRGimprovedPT}, we see that $\L$-independence of $P_{gg}$ thus implies that\footnote{Note that the same result could have been obtained from $0=(d/d\Lambda)[P_{ab}b_ab_b+\text{cst.}]$ and using the bias RG for $b_a$ and $b_b$ to infer an RG for $P_{ab}$, and solving it with one-loop $\beta$-functions to obtain the LL approximation. 
} (in analogy to Eq.\,\eqref{eq:fLLQFT} in the QFT example)
\be\label{eq:PabLL}
f_{ab}^\text{LL}(x) = [e^{+ s^\text{1L}x}]_{a\delta}[e^{+\bar s^\text{1L}x}]_{\delta b}\,,
\ee
where we used that $s_{ab}^\text{1L}$ is the transpose of $\bar s_{ab}^\text{1L}$ and that $c_{ab}^{(0,0)}$ is fixed by the tree-level spectrum, being unity for $a=b=\delta$ and zero otherwise, i.e. $c_{ab}^{(0,0)} = \delta^K_{a\delta}\delta^K_{\delta b}$. Thus, we showed that the LL resummed power spectrum is completely determined by the one-loop RG coefficients $s^\text{1L}$.
Taylor expanding \refeq{PabLL} in powers of $x=\s_\L^2$ yields explicit expressions for the $c_{ab}^{(\ell,\ell)}$ at all loop orders $\ell=1,2,3,\dots$ in terms of the one-loop $\beta$-function coefficient matrix $s^\text{1L}_{ab}$,
\bea
  c_{ab}^{(1,1)} &=& s^\text{1L}_{a\delta}\delta^K_{\delta b}+\delta^K_{a\delta}\bar s^\text{1L}_{\delta b}\,,\nn\\
  c_{ab}^{(2,2)} &=& s^\text{1L}_{a\delta}\bar s^\text{1L}_{\delta b}+\frac12 s^\text{1L}_{ac}s^\text{1L}_{c\delta}\delta^K_{\delta b}+\frac12 \delta^K_{a\delta}\bar s^\text{1L}_{\delta c}\bar s^\text{1L}_{cb}\,,
\eea
and so on, where $\delta^K$ is the Kronecker symbol.
The argument above could be extended by considering the two-loop RG along with the one-loop tracer power spectrum, from which the NLL coefficients $c_{ab}^{(\ell,\ell-1)}$ for all $\ell\geq 2$ can be derived.

We can check that those coefficients indeed agree with the explicit result for the one- and two-loop power spectrum in the limit $k\ll\L$. At one-loop those are given by the first line in Eq.\,\eqref{eq:singlehardrencorrespondence}, and noting $P^\text{lin}_{ab}=\delta^K_{a\delta}\delta^K_{b\delta}P^\text{lin}$ precisely match with the RG result for $c_{ab}^{(1,1)}$ from above.
At two-loop we use Eq.\,\eqref{eq:doublehardrencorrespondence}. The LL part of this double-hard limit, i.e. the part that is enhanced as $1/(n+3)^2$ for $n\to -3$, can be extracted by writing $s^\text{2L}(q/p)$ as $s^\text{2L}(q/p)-s^\text{2L}(0)+s^\text{2L}(0)$ and noting that the difference between the first two terms is proportional to $g(r)$ from \refeq{doublehardfuncg}. Hence, following the discussion around \refeq{lLoopstructure} this piece yields only an NLL contribution $\propto 1/(n+3)$. This is not surprising since this difference is what enters the two-loop $\beta$-function, see Eq.\,\eqref{eq:RG2L}, which controls the NLL terms. The LL part of the two-loop power spectrum can thus be obtained by replacing $s^\text{2L}(q/p)\to s^\text{2L}(0)$ in Eq.\,\eqref{eq:doublehardrencorrespondence}. After using Eq.\,\eqref{eq:commutationoflimitsrelation}, we find agreement with $c_{ab}^{(2,2)}$. Validating the prediction for $c_{ab}^{(3,3)}$ (being easily obtained from Taylor expanding Eq.\,\eqref{eq:PabLL} to third order in $x=\s_\L^2$) would require to compute the triple-hard limit of the three-loop contribution to the tracer power spectrum, which is beyond the scope of this work. 

Therefore, this systematic treatment crucially relies on the presence of an additional expansion parameter besides the usual density variance, that plays the role of the `large logarithm' within QFT. For a power-law spectrum, we identified this quantity as $1/(n+3)$. While usual perturbation theory breaks down when $n\to-3$ from above, RG-resummed results such as Eq.\,\eqref{eq:PabLL} remain valid even when the ratio $\s_\L^2=\Delta_\L^2/(n+3)$ is of order unity, as long as $\Delta_\L^2\ll 1$.  

Finally, we note that the parameter $x=\s_\L^2$ controlling the resummation for $n\simeq -3$ can also be rewritten as
\be\label{eq:sLtolog}
  \s_\L^2=\frac{\Delta_\L^2}{n+3}=\frac{1}{n+3}\frac{A_s}{2\pi^2}\left(\frac{\Lambda}{k_\ast}\right)^{n+3} = \frac{A_s}{2\pi^2}\times\left[\frac{1}{n+3}+\ln\left(\frac{\L}{k_\ast}\right)+O(n+3)\right]\,,
\ee
while the second expansion parameter is $\Delta_\L^2=A_s/(2\pi^2)+O(n+3)$. This means RG-improved perturbation theory, being the resummation of powers of $x$ while treating $\Delta_\L^2$ as small, can also formally be related to a logarithmic sensitivity to the UV scale $\L$, making the analogy to QFT even more direct. 
Nevertheless, for a power-law Universe, the IR and UV scales $k_\ast$ and $\Lambda$ are somewhat artificial, and we proceed to discuss the $\L$CDM case for which those scales can be given a physical interpretation.

\subsubsection*{$\L$CDM spectrum}
Here we discuss the generalization of our previous findings to $\L$CDM cosmologies. To build some intuition and understand the parametric dependence we start by considering scales (far) above the equality scale $k_\text{eq}\simeq 0.02\,h/$Mpc, but within the realm of perturbation theory, i.e. assuming a large separation to the non-linear scale $k_\text{nl}(z)\gg k_\text{eq}$. In order to obtain analytical insights, we momentarily consider the sub-horizon (sH) asymptotic form of the linear power spectrum above the equality scale (approximating the primordial scalar spectral index as $n_s\simeq 1$),
\be\label{eq:EisensteinHulargek}
  P^\text{lin}(k)\big|_\text{sH} = A_s k^{-3}\ln^2(k/k_\ast)\,,
\ee
with $k_\ast$ of order $k_\text{eq}$ and a normalization factor $A_s$. Since loop integrals would be IR divergent when literally extrapolating this ansatz to $k\to 0$, and since it applies for $\L$CDM cosmology only above the equality scale, we impose an IR cutoff at $k_\ast$ in addition to the UV cutoff at $\L$. For $\L\gg k_\ast$ we then have $\Delta_\L^2|_\text{sH}=A_s\ln^2(\L/k_\ast)/(2\pi^2)$ and thus
$(\s_\L^2/\Delta_\L^2)|_\text{sH}=(1/3)\times \ln(\L/k_\ast)$. We recall that in the power law case this ratio (being $1/(n+3)$) played the role of the `large logarithm', and from Eq.\,\eqref{eq:lLoopstructure} expect this identification to carry over to generic loop integrals for deterministic bias renormalization at leading gradient order. Thus, for a linear spectrum given by Eq.\,\eqref{eq:EisensteinHulargek},  $1/(n+3)\mapsto \ln(\L/k_\ast)/3$ compared to the power law case. 
In comparison to the QFT example, the translation to the LSS context is then taken to be
\be \label{eq:mapsLCDM}
\boxed{
   \ln(\mu/\mu_\ast)\mapsto \frac{\s_\L^2}{\Delta_\L^2} = \int_0^\L \frac{d\L'}{\L'}\left(\frac{\L'}{\L}\right)^3\frac{P^\text{lin}(\L')}{P^\text{lin}(\L)}\sim\frac13\ln(\L/k_\text{eq})\,,
}
\ee
where the last expression is approached in the asymptotic limit $\L\gg k_\text{eq}$.
RG-improvement systematically resums dominant $\Lambda$-dependent parts of higher loops if $\s_\L^2/\Delta_\L^2\gg 1$ while perturbativity requires (as usual) $\Delta_\L^2\ll 1$. More precisely, parametrically, for the cross-power spectrum of bias operators (in the limit $k\ll\L$) the RG resums all  UV contributions to higher loops that scale with arbitrary powers in $x=\s_\L^2$ while simultaneously expanding in powers of $\Delta_\L^2\ll 1$, with the zeroth order in $\Delta_\L^2$ being the LL resummed result, the first order the NLL result, etc.

\begin{figure}[t]
    \centering
    \includegraphics[width = 0.7\textwidth]{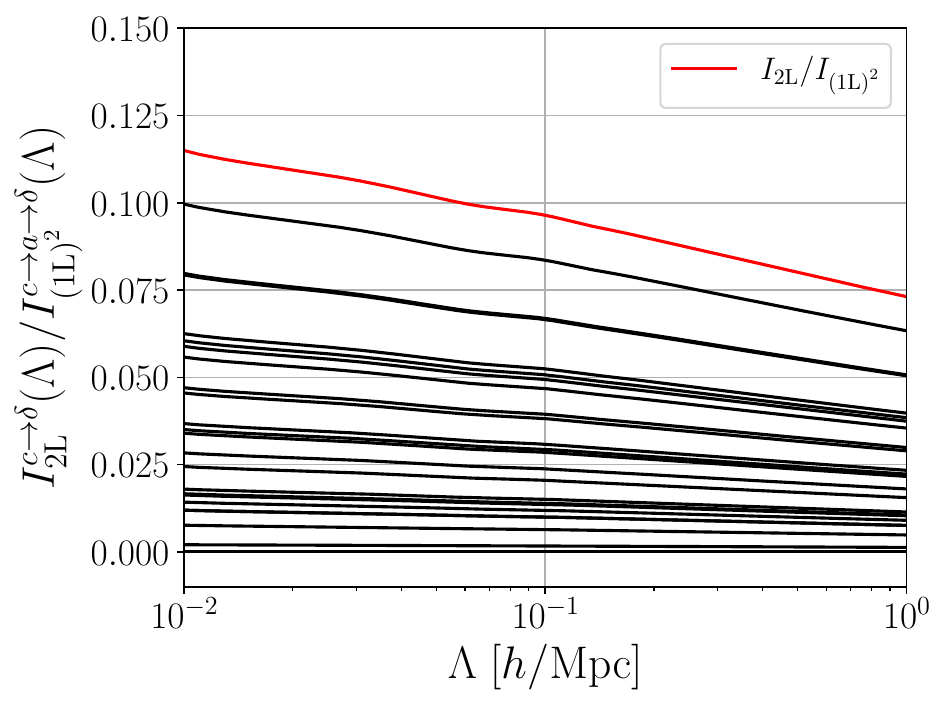}
    \caption{
    In black, we plot the ratio of the 'two-loop RG' and 'one-loop squared' contributions from Eq.\,\eqref{eq:Iratio} to the renormalization of $b_\d$ for a $\L$CDM spectrum. Each of the $29$ lines represents a different bias operator ${\cal O}_c$ sourcing the running of $b_\delta$ (note that some lines overlap). A small value of this ratio indicates that the bias RG systematically resums dominant UV-sensitive contributions from higher loops. Here we also include numerical factors from the one- and two-loop $\beta$-functions as in Eq.\,\eqref{eq:Iratio}. Since the two-loop $\beta$-function is independent of the operator (up to an overall factor), all lines are multiples of each other. In red, we also show the ratio of the contributions \refeq{I1LsqI2L} without including any numerical prefactors. 
    }
    \label{fig:2Lrunning_lcdm}
\end{figure}

As a first non-trivial check of this scheme, we return to the discussion around Eq.\,\eqref{eq:solODE2} and check whether 
the ratio $|I_\text{2L}(\L)/I_{(\text{1L})^2}(\L)|$ is small. Let us start again with the analytical ansatz from Eq.\,\eqref{eq:EisensteinHulargek}. In this case we find for $G(\L)$ (which enters the two-loop $\beta$-function, see Eq.\,\eqref{eq:Gdef})
\be
G(\L)|_\text{sH}= -\frac{8}{75}\frac{A_s}{2\pi^2}\ln^2(\L/k_\ast) +O\big(\ln(\L/k_\ast)\big),
\ee 
such that $|G(\L)/\s_\L^2|_\text{sH}= 8/25\times 1/|\ln(\L/k_\ast)|+O( 1/|\ln^2(\L/k_\ast)|)$ is small when the log becomes large. Using this in Eq.\,\eqref{eq:I1LsqI2L} yields
\ba\label{eq:supfaclimitEH}
\bigg| \frac{I_{\text{2L}}(\L)}{I_{\text{(1L)}^2}(\L)} \bigg|_\text{sH}\ =&\ \frac{48}{125}\frac{1}{\ln(\L/k_\ast)}+O\left(\frac{1}{\ln^2(\L/k_\ast)}\right)  \quad \textrm{for} \qquad \L\gg k_\ast\simeq k_\text{eq}\,. 
\ea
This shows that, as expected, the LL contribution (`one-loop squared') indeed dominates over the NLL (from the two-loop RG) by one power of the logarithm, making the ratio small and supporting the rationale for using RG-improvement.
The reason for why $G|_\text{sH}\propto \ln^2(\L/k_\ast)$ while $\s_\L^2|_\text{sH}\propto \ln^3(\L/k_\ast)$ (and thus $I_\text{2L}|_\text{sH}\propto \ln^5(\L/k_\ast)$ while $I_{\text{(1L)}^2}|_\text{sH}\propto \ln^6(\L/k_\ast)$) is again the suppression of the function $g(r)$ for $r\ll 1$, in analogy to the discussion below Eq.\,\eqref{eq:supfaclimit}, being deeply rooted in the very way how the RG approach works.

Let us finally consider the ratio $I_{\text{2L}}(\L)/I_{\text{(1L)}^2}(\L)$ obtained numerically from a full $\L$CDM input power spectrum, shown in Fig.\,\ref{fig:2Lrunning_lcdm} as a red line.
We further include the numerical prefactors from the one- and two-loop $\beta$-functions 
\bea
\label{eq:Iratio} I_{\text{(1L)}^2}^{c\to a\to \d}(\L) &=& \bar s_{\d a}^{\rm 1L}\bar s_{ac}^{\rm 1L}I_{\text{(1L)}^2}(\L)\,,\\
 I_{\text{2L}}^{c\to\d}(\L) &=& \bar s^\text{2L}_{\delta c} I_{\text{2L}}(\L)\,, \nonumber
\eea
for all bias operators $c$ up to fifth order, and $a$ summed over all operators (with non-zero contributions up to third order).
The analog of the ratio of NLL to LL terms in Eq.\,\eqref{eq:2LRGQED} is thus $I_{\text{2L}}^{c\to\d}(\L)/I_{\text{(1L)}^2}^{c\to a\to \d}(\L)$,
shown in Fig.\,\ref{fig:2Lrunning_lcdm}. The meaning of this ratio is the following: a value of e.g. $0.05$ means that $95\%$ of the two-loop bias renormalization is determined by solving the one-loop RG equations, while only $5\%$ arises from the two-loop RG. The smallness of this quantity is thus a proof that solving the one-loop bias RG and using the resulting running bias coefficients $b_a(\Lambda)$ captures already the dominant $\Lambda$-dependent part of higher loops, specifically of the two-loop, for $\L$CDM cosmology. 

While these findings are promising, we note that applications in its present form are limited by the setup of considering RG evolution in terms of the (a priori unphysical) smoothing scale $\L$, being restricted to the UV range $\L\gg k$.
Ideally, applications should involve two widely separated physical scales, across which RG evolution forms the bridge, similarly as the center of mass energy and electron mass in the QFT example. An example in this direction could be the squeezed limit of the bispectrum, for which the RG setup would have to be adapted. Alternatively, the bias RG in its present form can be of interest in field-level analyses, which by construction keeps the cutoff scale $\L$ explicit. In that context, in field-level analysis the smoothing scale $\L$ of the linear density field appears as an explicit parameter that can be chosen appropriately (see e.g.~Fig.~1 of \cite{Schmidt:2020viy} for the dependence of field-level results on $\Lambda$). For example, our results could be useful to translate (unrenormalized) bias $b_a(\L)$ measured by cross correlating the tracer field with field-level bias operators constructed from a smoothed density field~\cite{Schmittfull:2014tca,Lazeyras:2017hxw,Abidi:2018eyd} to renormalized bias $b_{[a]}$ determined e.g.~from response function techniques~\cite{Lazeyras:2015lgp, Baldauf:2015vio, Li:2015jsz}. For instance, one can use the RG equations to match a field-level measurement of a bias parameter at a fixed cutoff, running towards $\Lambda_\ast \to 0$ limit and directly compare that to the renormalized bias typically used for parameters extraction in $N$-point function analysis. Mapping between the values obtained from these two methods is usually performed by accounting for the leading UV-sensitive one-loop corrections, which can be improved by using instead RG evolution of bias parameters from $\L$ to $\L_\ast=0$. We plan to explore that connection in a future work. Another possible application of the RG to field-level analysis involves optimizing the cross-correlations among bias coefficients. Since the RG effectively acts as a rotation in bias-parameter space, one can use it to move across different values of $\L$ so as to minimize the cross-correlation between bias parameters, thereby improving parameter sampling.

\newpage
\section{Conclusions}\label{sec:conclusions}

Extending perturbative calculations for biased tracers in LSS to higher-order is of interest for extracting additional information on cosmology from galaxy surveys. Specifically, computing the  next-to-next-to-leading-order (two-loop) power spectrum and bispectrum demands the inclusion of fifth-order operators. In this work, we have re-derived the full list of fifth-order operators in the bias expansion in \refsec{fifthorder}, finding consistency with previous independent calculations \cite{Donath:2023sav,Ansari:2025nsf,Schmidt:2020tao}, as well as additional degeneracies of four fourth- and eight fifth-order operators that enter the two-loop power spectrum through the $P^{(24)}$ and $P^{(15)}$ diagrams, confirming the general arguments and related findings for the one-loop bispectrum at fourth order  of~\cite{egge_oneloopbisp}. 

Consistent operator renormalization is a vital aspect of the large-scale bias expansion. In that context, we have extended the renormalization prescription for the bias operators from \cite{Assassi:2014fva} to fifth-order operators and two-loops considering renormalization conditions that match the renormalized bias values to the IR limit of $N$-point functions (see \refsec{biasrenorm}). 
We show in \refsec{oneloop_renorm} that the one-loop renormalization can be written in terms of the \textit{single-hard} limit of the bias kernels, calculated in \refsec{singledoublelimits}.
Extending the approach to two-loop corrections requires the calculation of the \textit{double-hard} limit of (at least fifth-order expanded) operator kernels. Remarkably, the  dependence of the double-hard contribution of all fifth-order kernels on the ratio of loop wavenumbers $p/q$ is \emph{universal}, i.e. it is specified in terms of a single function $g(p/q)$ given in \refeq{doublehardfuncg}. We provide very simple expressions for the renormalized deterministic bias in terms of the single- and double-hard limit of the operator kernels in \refeq{Oren1L2L} and, after incorporating stochasticity at leading order in the gradient expansion, the two-loop power spectrum in \refeq{pren1L2L}. 

A complementary approach to the renormalized bias parameters is to consider their running with the smoothing scale $\L$ based on RG techniques applied to the LSS context \cite{Rubira:2023vzw,Rubira:2024tea,Nikolis:2024kbx}. Using the (double-)hard fifth-order kernel limits from \refsec{singledoublelimits}, we perform a complete derivation of the running of the linear bias parameter (at leading order in gradient expansion) at one and two-loop order and second- and third-order bias parameters at one-loop, with two-loop $\beta$-function (`anomalous dimension') given in \refeq{RG2L}. 
Alternatively, we show in \refsec{connection_RG_renorm} that the anomalous dimension matrix $\gamma_{ab}$ for the bias coefficients in the RG approach  can be expressed in terms of the renormalization constants $Z_{ab}$ from \refeq{biasren} in the approach of \cite{Assassi:2014fva} at all loop orders, and when expanding both expressions order by order in loops, we rederive our earlier results. 
We discuss solutions of the bias RG in Sec.\,\ref{sec:solutions}, pointing out the relevance of including third- and higher-order biases for the running of $b_\delta$, and identify a growing-in-$\Lambda$ eigenmode of the one-loop RG as cause. 

Finally, we discuss in how far the framework of RG-improved perturbation theory in QFT can be applied to the cross-power spectrum of bias operators  using the bias RG. It relies on a double expansion, usually in terms of a small coupling constant and a `large logarithm', and corresponds to a systematic re-organization of the perturbative expansion allowing for a resummation of all-order terms enhanced by large logarithms based on solving RG equations. We generalize this concept to the cross-power spectrum of bias operators, see Eq.\,\eqref{eq:PabRGimprovedPT}, and identify the quantity that corresponds to the `large logarithm' in the context of LSS in Eq.\,\eqref{eq:mapsLCDM}. Furthermore, we explicitly derive  the cross-power spectrum of bias operators  in `leading logarithmic' approximation using the one-loop RG in Eq.\,\eqref{eq:PabLL}. Finally, we demonstrate the relative suppression of `next-to-leading logarithmic' contributions related to the two-loop RG as compared to leading logarithmic terms, being a first non-trivial check, and providing evidence that the concept of RG-improvement could be systematically applied to the tracer power spectrum for $\L$CDM cosmologies.

In conclusion, this work offers significant theoretical advances in understanding the renormalization of bias operators and their application to the two-loop power spectrum. In addition, we have extended the RG approach to include higher-order and two-loop contributions.
Moreover, our results suggest that applying RG techniques to LSS observables may enable information gain by resumming dominant UV-sensitive parts of higher loops, using the systematic framework of RG-improved perturbation theory.  Altogether, these results further strengthen the connections between QFT and cosmology and prepare the stage for applications in galaxy clustering analyses. 

\section*{Acknowledgements}
The authors acknowledge Alexander Eggemeier, Fabian Schmidt, Rom\'an Scoccimarro and Felix Yu for useful conversations, and Fabian Schmidt for insightful comments on the draft. This publication is part of the project `A rising tide: Galaxy intrinsic alignments as a new probe of cosmology and galaxy evolution' (with project number VI.Vidi.203.011) of the Talent programme Vidi which is (partly) financed by the Dutch Research Council (NWO). For the purpose of open access, a CC BY public copyright license is applied to any Author Accepted Manuscript version arising from this submission. We further acknowledge support by the Excellence Cluster ORIGINS, which is funded by the Deutsche Forschungsgemeinschaft (DFG, German Research
Foundation) under Germany's Excellence Strategy - EXC-2094 - 390783311. H.R and  Z.V. acknowledges the support of the Kavli Foundation. 
\newpage
\appendix

\section{Kernel recursive relations} \label{app:recursivekernel}

Following the results presented in \cite{Vlah:2019}, we can write any of the `building block' operators $\Pi_{ij}^{[n]}$ occurring in the bias expansion in Sec.~\ref{sec:fifthorder} as a sum over $m$-th order contributions, where $m \geq n$:
\eq{
\label{eq:Pi^n_ij}
\Pi^{[n]}_{ij}(\vec k) = \sum_{m=n}^\infty (2\pi)^3 \dirac(\vec k - \vec q_{1 m})
D^m \pi^{(n, m)}_{ij} (\vec q_1, \ldots , \vec q_m) \df_L(\vec q_1) \ldots  \df_L(\vec q_m)\, .
}
For $\Pi^{[1]}_{ij}$ we simply have 
\eeq{
\pi^{(1, m)}_{ij} (\vec q_1, \ldots , \vec q_m) =  
\frac{(\vec q_{1 m})_i(\vec q_{1 m})_j}{ q^2_{1 m}} F^{(m)} (\vec q_1, \ldots , \vec q_m)\, .
}
For the higher $n$ orders, one can derive a recursive relation for the kernels of $\Pi^{[n]}_{ij}$, ending up with
\eq{
\pi^{(n,m)}_{ij} &(\vec q_1, \ldots , \vec q_m) = \frac{ m - n + 1 }{(n-1)!} \pi^{(n-1,m)}_{ij} (\vec q_1, \ldots , \vec q_m) \\
&\hspace{2cm} -  \frac{1}{(n-1)!} \sum_{\ell = n-1}^{m-1}
\frac{ \vec q_{1\ell}. \vec q_{\ell+1,m} }{ q_{\ell+1,m}^2}
\pi^{(n-1,\ell)}_{ij} ( \vec q_1, \ldots , \vec q_\ell)  G^{(m - \ell)} (\vec q_{\ell+1}, \ldots , \vec q_m)\, . \non
}
In the above expression, we always have $m \geq n$ and $n\geq2$. Note that for the case when $n=m$, only one term $\ell=n-1$ of the second part contributes.
To get the symmetrized kernels, we can sum over all the wavenumber permutations 
\eeq{
\pi^{(n,m)}_{ij, {\rm sim} } (\vec q_1, \ldots , \vec q_m) = \frac{1}{m!} \sum_{\pi \rm-all} \pi^{(n,m)}_{ij} (\vec q_{\#_1}, \ldots , \vec q_{\#_m})\, . 
}
Combining the last two relations, 
\eq{
\pi^{(n,m)}_{ij, {\rm sim}} &(\vec q_1, \ldots , \vec q_m) = 
\frac{ m - n + 1 }{(n-1)!} \pi^{(n-1,m)}_{ij,{\rm sim}} (\vec q_1, \ldots , \vec q_m) \\
&\hspace{-0.7cm} -  \frac{1}{(n-1)!} \sum_{\ell = n-1}^{m-1} \frac{l! (m-l)!}{m!} \sum_{\pi \rm-cross} 
\frac{ \vec q_{1\ell}. \vec q_{\ell+1,m} }{ q_{\ell+1,m}^2}
\pi^{(n-1,\ell)}_{ij,{\rm sim}} ( \vec q_1, \ldots , \vec q_\ell)  G^{(m - \ell)}_{\rm sim} (\vec q_{\ell+1}, \ldots , \vec q_m) , \non
}
where, in the last term, only the distinct unordered sets of $(q_1, \ldots q_\ell)$ wavemodes need to be considered in the $\pi-{\rm cross}$ sum.
Further on, in the main text, we dispense with explicit symmetrization labels in $\Pi_{ij}^{[n]}$. 

\section{Deterministic contributions} \label{app:det}
\newgeometry{left=1cm,bottom=1cm}

\be\label{eq:doublehardcoeff}
\begin{array}{c|ccc}
 a  & {c^{(3)}_{a\d}}  & {d^{(5)}_a} & {\tilde d^{(5)}_a}  \\ \\
 \hline 
 \\
 {\rm tr}\big[ \Pi^{[1]} \big] & 0 & 0 & 0  \\ \\
 \hline 
 \\
 {\rm tr}\big[ \big( \Pi^{[1]} \big)^2 \big] & \frac{68}{63} & \frac{862}{1575} & \frac{376}{6615}\\
 \Big( {\rm tr}\big[ \Pi^{[1]} \big] \Big)^2 & \frac{68}{63}  & \frac{862}{1575} & \frac{376}{6615} \\ \\
 \hline
\\
 \Big( {\rm tr}\big[ \Pi^{[1]} \big] \Big)^3 & 1 & \frac{70739}{33075} & \frac{4}{105}  \\
 {\rm tr}\big[ \big( \Pi^{[1]} \big)^2 \big] {\rm tr}\big[ \Pi^{[1]} \big] & \frac{5}{9}  & \frac{2917}{2205} & \frac{716}{1323}  \\
 {\rm tr}\big[ \big( \Pi^{[1]} \big)^3 \big] & \frac{1}{3} & \frac{30263}{33075} & \frac{1748}{2205} \\
 {\rm tr}\big[\Pi^{[1]} \Pi^{[2]} \big] & \frac{41}{63}  & \frac{134957}{99225} & \frac{148}{441} \\ \\
 \hline 
\\
 \color{gray}{\Big( {\rm tr}\big[ \Pi^{[1]} \big] \Big)^4} & 0 & \frac{272}{105} & 0  \\
 \color{gray}{{\rm tr}\big[ \big( \Pi^{[1]} \big)^3 \big] {\rm tr}\big[ \Pi^{[1]} \big]} & 0 & \frac{82}{105} & \frac{5}{21} \\
 \color{gray}{{\rm tr}\big[ \big( \Pi^{[1]} \big)^2 \big] \big({\rm tr}\big[ \Pi^{[1]} \big]\big)^2} & 0 & \frac{6352}{4725} & \frac{4}{21} \\
 \color{gray}{\Big( {\rm tr}\big[ \big( \Pi^{[1]} \big)^2 \big] \Big)^2} & 0 & \frac{592}{675} & \frac{8}{63}  \\
 {\rm tr}\big[\Pi^{[1]} \Pi^{[1]} \Pi^{[2]} \big] & 0 & \frac{16112}{19845} & \frac{3706}{6615}  \\
 {\rm tr}\big[ \Pi^{[1]} \big] {\rm tr}\big[\Pi^{[1]} \Pi^{[2]} \big] & 0 & \frac{18166}{14175} & \frac{373}{735} \\
 {\rm tr}\big[\Pi^{[1]} \Pi^{[3]} \big] & 0 & \frac{12814}{11025} & \frac{401}{2205} \\
 {\rm tr}\big[\Pi^{[2]} \Pi^{[2]} \big] & 0 & \frac{27784}{19845} & \frac{232}{315} \\ \\
 \hline
\\
 \color{gray}{\Big( {\rm tr}\big[ \Pi^{[1]} \big] \Big)^5} & 0 & 1 & 0  \\
 \color{gray}{{\rm tr}\big[ \big( \Pi^{[1]} \big)^3 \big] \left({\rm tr}\big[ \Pi^{[1]} \big]\right)^2} & 0 & \frac{11}{45} & 0  \\
 \color{gray}{{\rm tr}\big[ \big( \Pi^{[1]} \big)^2 \big] \left({\rm tr}\big[ \Pi^{[1]} \big]\right)^3} & 0 & \frac{7}{15} & 0 \\
 \color{gray}{{\rm tr}\big[ \big( \Pi^{[1]} \big)^3 \big]{\rm tr}\big[ \big( \Pi^{[1]} \big)^2 \big]} & 0 & \frac{31}{225} & 0  \\
 \color{gray}{{\rm tr}\big[ \Pi^{[1]} \big]\Big( {\rm tr}\big[ \big( \Pi^{[1]} \big)^2 \big] \Big)^2} & 0 & \frac{163}{675} & 0  \\
 \color{gray}{\big({\rm tr}\big[ \Pi^{[1]}\big] \big)^2 {\rm tr}\big[\Pi^{[1]} \Pi^{[2]} \big]} & 0 & \frac{47}{105} & \frac{2}{21}  \\
 \color{gray}{{\rm tr}\big[ \Pi^{[1]}  \Pi^{[1]} \big] {\rm tr}\big[\Pi^{[1]} \Pi^{[2]}\big]} & 0 & \frac{173}{675} & \frac{2}{63}  \\
 \color{gray}{{\rm tr}\big[ \Pi^{[1]} \big] {\rm tr}\big[\Pi^{[1]} \Pi^{[1]} \Pi^{[2]}\big]} & 0 & \frac{391}{1575} & \frac{5}{63} \\
 {\rm tr}\big[\Pi^{[1]} \Pi^{[2]} \Pi^{[2]} \big] & 0 & \frac{197}{735} & \frac{82}{441} 
\\
 {\rm tr}\big[ \Pi^{[1]} \big] {\rm tr}\big[\Pi^{[2]} \Pi^{[2]} \big] & 0 & \frac{5057}{11025} & \frac{110}{441}
\\
 {\rm tr}\big[\Pi^{[1]} \Pi^{[1]} \Pi^{[3]} \big] & 0 & \frac{643}{2835} & \frac{104}{945} 
\\
 {\rm tr}\big[ \Pi^{[1]} \big] {\rm tr}\big[\Pi^{[1]} \Pi^{[3]} \big] & 0 & \frac{5477}{14175} & \frac{4}{35} 
\\
 {\rm tr}\big[\Pi^{[2]} \Pi^{[3]} \big] & 0 & \frac{8629}{19845} & \frac{491}{2205}\\
 {\rm tr}\big[\Pi^{[1]} \Pi^{[4]} \big] & 0 & \frac{33403}{198450} & \frac{74}{6615} \\
\end{array}
\ee
\restoregeometry
\be\label{eq:singlehardcoeffK4}
\begin{array}{c|ccc}
   c^{(4)}_{ab} &&b& \\ \\ \hline
  a & {\rm tr}\big[ \Pi^{[1]} \big] & {\rm tr}\big[ \big( \Pi^{[1]} \big)^2 \big] & \Big( {\rm tr}\big[ \Pi^{[1]} \big] \Big)^2 \\ \\
 \hline
 \\
 {\rm tr}\big[ \Pi^{[1]} \big] & 0 & 0 & 0 \\
 {\rm tr}\big[ \big( \Pi^{[1]} \big)^2 \big] & \frac{34}{63} & \frac{127}{6615} & \frac{1312}{2205} \\
 \Big( {\rm tr}\big[ \Pi^{[1]} \big] \Big)^2 & \frac{34}{63} & \frac{127}{6615} & \frac{1312}{2205} \\
 \\
 \hline
 \\
 \Big( {\rm tr}\big[ \Pi^{[1]} \big] \Big)^3 & \frac{1}{2} & 0 & \frac{34}{21} \\
 {\rm tr}\big[ \big( \Pi^{[1]} \big)^2 \big] {\rm tr}\big[ \Pi^{[1]} \big] & \frac{5}{18} & \frac{58}{315} & \frac{88}{105} \\
 {\rm tr}\big[ \big( \Pi^{[1]} \big)^3 \big] & \frac{1}{6} & \frac{29}{105} & \frac{47}{105} \\
 {\rm tr}\big[\Pi^{[1]} \Pi^{[2]} \big] & \frac{41}{126} & \frac{38}{189} & \frac{134}{147} \\
 \\
 \hline
 \\
 \color{gray}{\Big( {\rm tr}\big[ \Pi^{[1]} \big] \Big)^4} & 0 & 0 & 1 \\
 \color{gray}{{\rm tr}\big[ \big( \Pi^{[1]} \big)^3 \big] {\rm tr}\big[ \Pi^{[1]} \big]} & 0 & \frac{1}{6} & \frac{1}{6} \\
 \color{gray}{{\rm tr}\big[ \big( \Pi^{[1]} \big)^2 \big] \big({\rm tr}\big[ \Pi^{[1]} \big]\big)^2} & 0 & \frac{1}{6} & \frac{7}{18} \\
 \color{gray}{\Big( {\rm tr}\big[ \big( \Pi^{[1]} \big)^2 \big] \Big)^2} & 0 & \frac{19}{45} & \frac{2}{45} \\
 {\rm tr}\big[\Pi^{[1]} \Pi^{[1]} \Pi^{[2]} \big] & 0 & \frac{103}{630} & \frac{17}{90} \\
 {\rm tr}\big[ \Pi^{[1]} \big] {\rm tr}\big[\Pi^{[1]} \Pi^{[2]} \big] & 0 & \frac{2}{21} & \frac{3}{7} \\
 {\rm tr}\big[\Pi^{[1]} \Pi^{[3]} \big] & 0 & \frac{134}{945} & \frac{17}{45} \\
 {\rm tr}\big[\Pi^{[2]} \Pi^{[2]} \big] & 0 & \frac{233}{2205} & \frac{1084}{2205} \\
\end{array}
\ee
\newgeometry{left=1cm,bottom=1cm}
\be\label{eq:singlehardcoeffK5}
\begin{array}{c|ccccccc}
   c^{(5)}_{ab} &b&& \\ \\ \hline
  a& {\rm tr}\big[ \Pi^{[1]} \big] & {\rm tr}\big[ \big( \Pi^{[1]} \big)^2 \big] & \Big( {\rm tr}\big[ \Pi^{[1]} \big] \Big)^2 & \Big( {\rm tr}\big[ \Pi^{[1]} \big] \Big)^3 & {\rm tr}\big[ \big( \Pi^{[1]} \big)^2 \big] {\rm tr}\big[ \Pi^{[1]} \big] & {\rm tr}\big[ \big( \Pi^{[1]} \big)^3 \big] & {\rm tr}\big[\Pi^{[1]} \Pi^{[2]} \big] \\ \\
  \hline
  \\
 {\rm tr}\big[ \Pi^{[1]} \big] & 0 & 0 & 0 & 0 & 0 & 0 & 0 \\ \\
 \hline
 \\
 {\rm tr}\big[ \big( \Pi^{[1]} \big)^2 \big] & \frac{34}{105} & \frac{127}{11025} & \frac{1312}{3675} & \frac{70832}{509355} & \frac{2928}{94325} & \frac{228808}{12733875} &
-\frac{1322}{67375} \\
 \Big( {\rm tr}\big[ \Pi^{[1]} \big] \Big)^2 & \frac{34}{105} & \frac{127}{11025} & \frac{1312}{3675} & \frac{70832}{509355} & \frac{2928}{94325} & \frac{228808}{12733875}
& -\frac{1322}{67375} \\ \\
\hline 
\\
 \Big( {\rm tr}\big[ \Pi^{[1]} \big] \Big)^3 & \frac{3}{10} & 0 & \frac{34}{35} & \frac{1312}{1225} & \frac{127}{3675} & 0 & 0 \\
 {\rm tr}\big[ \big( \Pi^{[1]} \big)^2 \big] {\rm tr}\big[ \Pi^{[1]} \big] & \frac{1}{6} & \frac{58}{525} & \frac{88}{175} & \frac{877}{1715} & \frac{23129}{77175} & \frac{892}{8575} & -\frac{5}{63}
\\
 {\rm tr}\big[ \big( \Pi^{[1]} \big)^3 \big] & \frac{1}{10} & \frac{29}{175} & \frac{47}{175} & \frac{3971}{17150} & \frac{2224}{5145} & \frac{1338}{8575} & -\frac{5}{42}
\\
 {\rm tr}\big[\Pi^{[1]} \Pi^{[2]} \big] & \frac{41}{210} & \frac{38}{315} & \frac{134}{245} & \frac{2489609}{5093550} & \frac{30616}{94325} & \frac{1631026}{12733875} &
-\frac{19231}{173250} \\ \\
\hline 
\\
 \color{gray}{\Big( {\rm tr}\big[ \Pi^{[1]} \big] \Big)^4} & 0 & 0 & \frac{3}{5} & \frac{68}{35} & 0 & 0 & 0 \\
 \color{gray}{{\rm tr}\big[ \big( \Pi^{[1]} \big)^3 \big] {\rm tr}\big[ \Pi^{[1]} \big]} & 0 & \frac{1}{10} & \frac{1}{10} & \frac{47}{175} & \frac{76}{175} & \frac{29}{175} & 0 \\
 \color{gray}{{\rm tr}\big[ \big( \Pi^{[1]} \big)^2 \big] \big({\rm tr}\big[ \Pi^{[1]} \big]\big)^2} & 0 & \frac{1}{10} & \frac{7}{30} & \frac{358}{525} & \frac{286}{525} & 0 & 0 \\
 \color{gray}{\Big( {\rm tr}\big[ \big( \Pi^{[1]} \big)^2 \big] \Big)^2} & 0 & \frac{19}{75} & \frac{2}{75} & \frac{16}{245} & \frac{3092}{3675} & \frac{464}{3675} & 0 \\
 {\rm tr}\big[\Pi^{[1]} \Pi^{[1]} \Pi^{[2]} \big] & 0 & \frac{103}{1050} & \frac{17}{150} & \frac{64334}{231525} & \frac{10862}{25725} & \frac{40009}{231525} & \frac{11}{525}
\\
 {\rm tr}\big[ \Pi^{[1]} \big] {\rm tr}\big[\Pi^{[1]} \Pi^{[2]} \big] & 0 & \frac{2}{35} & \frac{9}{35} & \frac{162353}{231525} & \frac{5416}{15435} & \frac{103736}{1157625} & \frac{197}{7875}
\\
 {\rm tr}\big[\Pi^{[1]} \Pi^{[3]} \big] & 0 & \frac{134}{1575} & \frac{17}{75} & \frac{940493}{1697850} & \frac{30455}{67914} & \frac{605653}{4244625} & -\frac{3307}{86625}
\\
 {\rm tr}\big[\Pi^{[2]} \Pi^{[2]} \big] & 0 & \frac{233}{3675} & \frac{1084}{3675} & \frac{174028}{231525} & \frac{1012}{3087} & \frac{4822}{46305} & \frac{758}{11025} \\ \\
 \hline
\\
 \color{gray}{\Big( {\rm tr}\big[ \Pi^{[1]} \big] \Big)^5} & 0 & 0 & 0 & 1 & 0 & 0 & 0 \\
 \color{gray}{{\rm tr}\big[ \big( \Pi^{[1]} \big)^3 \big] \left({\rm tr}\big[ \Pi^{[1]} \big]\right)^2} & 0 & 0 & 0 & \frac{1}{10} & \frac{1}{5} & \frac{1}{10} & 0 \\
 \color{gray}{{\rm tr}\big[ \big( \Pi^{[1]} \big)^2 \big] \left({\rm tr}\big[ \Pi^{[1]} \big]\right)^3} & 0 & 0 & 0 & \frac{3}{10} & \frac{3}{10} & 0 & 0 \\
 \color{gray}{{\rm tr}\big[ \big( \Pi^{[1]} \big)^3 \big]{\rm tr}\big[ \big( \Pi^{[1]} \big)^2 \big]} & 0 & 0 & 0 & 0 & \frac{7}{50} & \frac{9}{50} & 0 \\
 \color{gray}{{\rm tr}\big[ \Pi^{[1]} \big]\Big( {\rm tr}\big[ \big( \Pi^{[1]} \big)^2 \big] \Big)^2} & 0 & 0 & 0 & \frac{2}{75} & \frac{29}{75} & 0 & 0 \\
 \color{gray}{\big({\rm tr}\big[ \Pi^{[1]}\big] \big)^2 {\rm tr}\big[\Pi^{[1]} \Pi^{[2]} \big]} & 0 & 0 & 0 & \frac{67}{210} & \frac{4}{35} & 0 & \frac{1}{10} \\
 \color{gray}{{\rm tr}\big[ \Pi^{[1]}  \Pi^{[1]} \big] {\rm tr}\big[\Pi^{[1]} \Pi^{[2]}\big]} & 0 & 0 & 0 & \frac{19}{735} & \frac{1819}{7350} & \frac{116}{3675} & \frac{19}{150} \\
 \color{gray}{{\rm tr}\big[ \Pi^{[1]} \big] {\rm tr}\big[\Pi^{[1]} \Pi^{[1]} \Pi^{[2]}\big]} & 0 & 0 & 0 & \frac{17}{150} & \frac{143}{1050} & \frac{1}{21} & \frac{1}{15} \\
 {\rm tr}\big[\Pi^{[1]} \Pi^{[2]} \Pi^{[2]} \big] & 0 & 0 & 0 & \frac{211}{1715} & \frac{1037}{10290} & \frac{1493}{51450} & \frac{64}{525} \\
 {\rm tr}\big[ \Pi^{[1]} \big] {\rm tr}\big[\Pi^{[2]} \Pi^{[2]} \big] & 0 & 0 & 0 & \frac{428}{1225} & \frac{313}{3675} & 0 & \frac{2}{21} \\
 {\rm tr}\big[\Pi^{[1]} \Pi^{[1]} \Pi^{[3]} \big] & 0 & 0 & 0 & \frac{6407}{66150} & \frac{1189}{7350} & \frac{5371}{66150} & \frac{1}{50} \\
 {\rm tr}\big[ \Pi^{[1]} \big] {\rm tr}\big[\Pi^{[1]} \Pi^{[3]} \big] & 0 & 0 & 0 & \frac{2579}{9450} & \frac{10}{63} & \frac{136}{3375} & \frac{41}{2250} \\
 {\rm tr}\big[\Pi^{[2]} \Pi^{[3]} \big] & 0 & 0 & 0 & \frac{71047}{231525} & \frac{9997}{77175} & \frac{2411}{92610} & \frac{229}{3150} \\
 {\rm tr}\big[\Pi^{[1]} \Pi^{[4]} \big] & 0 & 0 & 0 & \frac{261944}{2546775} & \frac{98489}{1131900} & \frac{1332383}{50935500} & \frac{373}{28875} \\
\end{array}
\ee
\restoregeometry

\section{Stochastic contributions}\label{sec:stoch}

\be\label{eq:doublehardcoeffnoise}
\begin{array}{c|ccccc} 
 a & {n^{(2)}_a}   & {n^{(3)}_a} & {\tilde n^{(3)}_a} & {n^{(4)}_a} & {\tilde n^{(4)}_a}   \\ \\
 \hline
 \\
 {\rm tr}\big[ \Pi^{[1]} \big] & 0  & 0 & 0 & 0 & 0  \\ \\
 \hline
 \\
 {\rm tr}\big[ \big( \Pi^{[1]} \big)^2 \big] & 1  & \frac{13}{21} & \frac{8}{21} & \frac{4063}{6615} & \frac{4}{63} \\
 \Big( {\rm tr}\big[ \Pi^{[1]} \big] \Big)^2 & 1   & \frac{13}{21} & \frac{8}{21} & \frac{4063}{6615} & \frac{4}{63} \\ \\
 \hline
\\
 \Big( {\rm tr}\big[ \Pi^{[1]} \big] \Big)^3 & 0 & 1 & 0 & \frac{34}{21} & 0   \\
 {\rm tr}\big[ \big( \Pi^{[1]} \big)^2 \big] {\rm tr}\big[ \Pi^{[1]} \big] & 0   & \frac{1}{3} & \frac{2}{3} & \frac{46}{45} & \frac{10}{21}   \\
 {\rm tr}\big[ \big( \Pi^{[1]} \big)^3 \big] & 0  & 0 & 1 & \frac{76}{105} & \frac{5}{7}  \\
 {\rm tr}\big[\Pi^{[1]} \Pi^{[2]} \big] & 0   & \frac{10}{21} & \frac{11}{21} & \frac{1472}{1323} & \frac{19}{63}  \\ \\
 \hline
\\
 \color{gray}{\Big( {\rm tr}\big[ \Pi^{[1]} \big] \Big)^4}  & 0 & 0 & 0 & 1 & 0   \\
 \color{gray}{{\rm tr}\big[ \big( \Pi^{[1]} \big)^3 \big] {\rm tr}\big[ \Pi^{[1]} \big]}  & 0 & 0 & 0 & \frac{1}{3} & 0 \\
 \color{gray}{{\rm tr}\big[ \big( \Pi^{[1]} \big)^2 \big] \big({\rm tr}\big[ \Pi^{[1]} \big]\big)^2} & 0 & 0 & 0 & \frac{5}{9} & 0  \\
 \color{gray}{\Big( {\rm tr}\big[ \big( \Pi^{[1]} \big)^2 \big] \Big)^2} & 0 & 0 & 0 & \frac{7}{15} & 0   \\
 {\rm tr}\big[\Pi^{[1]} \Pi^{[1]} \Pi^{[2]} \big] & 0 & 0 & 0 & \frac{37}{105} & \frac{5}{21}   \\
 {\rm tr}\big[ \Pi^{[1]} \big] {\rm tr}\big[\Pi^{[1]} \Pi^{[2]} \big] & 0 & 0 & 0 & \frac{11}{21} & \frac{5}{21} \\
 {\rm tr}\big[\Pi^{[1]} \Pi^{[3]} \big] & 0 & 0 & 0 & \frac{491}{945} & \frac{2}{63}  \\
 {\rm tr}\big[\Pi^{[2]} \Pi^{[2]} \big] & 0 & 0 & 0 & \frac{439}{735} & \frac{10}{21} \\
\\
\end{array}
\ee
\newgeometry{left=1cm,bottom=1cm}

\be\label{eq:singlehardcoeffnoisetwoloop}
\begin{array}{c|ccccccc}
   n_{ab} &b&& \\ \\ \hline
  a& {\rm tr}\big[ \Pi^{[1]} \big] & {\rm tr}\big[ \big( \Pi^{[1]} \big)^2 \big] & \Big( {\rm tr}\big[ \Pi^{[1]} \big] \Big)^2 & \Big( {\rm tr}\big[ \Pi^{[1]} \big] \Big)^3 & {\rm tr}\big[ \big( \Pi^{[1]} \big)^2 \big] {\rm tr}\big[ \Pi^{[1]} \big] & {\rm tr}\big[ \big( \Pi^{[1]} \big)^3 \big] & {\rm tr}\big[\Pi^{[1]} \Pi^{[2]} \big] \\ \\
  \hline 
  \\
 {\rm tr}\big[ \Pi^{[1]} \big] & 0 & 0 & 0 & 0 & 0 & 0 & 0 \\ \\
 \hline 
 \\
 {\rm tr}\big[ \big( \Pi^{[1]} \big)^2 \big] & 0 & \frac{55106}{2205} & \frac{55106}{2205} & \frac{230}{7} & \frac{302}{15} & \frac{482}{35} & \frac{49414}{2205} \\
 \Big( {\rm tr}\big[ \Pi^{[1]} \big] \Big)^2 & 0 & \frac{55106}{2205} & \frac{55106}{2205} & \frac{230}{7} & \frac{302}{15} & \frac{482}{35} & \frac{49414}{2205} \\ \\
 \hline 
 \\
 \Big( {\rm tr}\big[ \Pi^{[1]} \big] \Big)^3 & 0 & \frac{230}{7} & \frac{230}{7} & 18 & 10 & 6 & \frac{82}{7} \\
 {\rm tr}\big[ \big( \Pi^{[1]} \big)^2 \big] {\rm tr}\big[ \Pi^{[1]} \big] & 0 & \frac{302}{15} & \frac{302}{15} & 10 & \frac{94}{15} & \frac{22}{5} & \frac{106}{15} \\
 {\rm tr}\big[ \big( \Pi^{[1]} \big)^3 \big] & 0 & \frac{482}{35} & \frac{482}{35} & 6 & \frac{22}{5} & \frac{18}{5} & \frac{166}{35} \\
 {\rm tr}\big[\Pi^{[1]} \Pi^{[2]} \big] & 0 & \frac{49414}{2205} & \frac{49414}{2205} & \frac{82}{7} & \frac{106}{15} & \frac{166}{35} & \frac{5926}{735} \\ \\
 \hline 
 \\
 \color{gray}{\Big( {\rm tr}\big[ \Pi^{[1]} \big] \Big)^4} & 0 & 12 & 12 & 0 & 0 & 0 & 0 \\
 \color{gray}{{\rm tr}\big[ \big( \Pi^{[1]} \big)^3 \big] {\rm tr}\big[ \Pi^{[1]} \big]} & 0 & 4 & 4 & 0 & 0 & 0 & 0 \\
 \color{gray}{{\rm tr}\big[ \big( \Pi^{[1]} \big)^2 \big] \big({\rm tr}\big[ \Pi^{[1]} \big]\big)^2} & 0 & \frac{20}{3} & \frac{20}{3} & 0 & 0 & 0 & 0 \\
 \color{gray}{\Big( {\rm tr}\big[ \big( \Pi^{[1]} \big)^2 \big] \Big)^2} & 0 & \frac{28}{5} & \frac{28}{5} & 0 & 0 & 0 & 0 \\
 {\rm tr}\big[\Pi^{[1]} \Pi^{[1]} \Pi^{[2]} \big] & 0 & \frac{148}{35} & \frac{148}{35} & 0 & 0 & 0 & 0 \\
 {\rm tr}\big[ \Pi^{[1]} \big] {\rm tr}\big[\Pi^{[1]} \Pi^{[2]} \big] & 0 & \frac{44}{7} & \frac{44}{7} & 0 & 0 & 0 & 0 \\
 {\rm tr}\big[\Pi^{[1]} \Pi^{[3]} \big] & 0 & \frac{1964}{315} & \frac{1964}{315} & 0 & 0 & 0 & 0\\
 {\rm tr}\big[\Pi^{[2]} \Pi^{[2]} \big] & 0 & \frac{1756}{245} & \frac{1756}{245} & 0 & 0 & 0 & 0 \\
\end{array}
\ee
\restoregeometry

\section{Commutation of limits for stochastic terms}\label{app:hierarchynoise}

Similarly as for the deterministic terms, we can compare the single-hard limit to the limiting case for which the two loop wavenumbers entering the double-hard limit are hierarchical, i.e. of very different size. More precisely, for the double-hard limit of $P^{(24)}$ in~\refeq{doublehardnoise}  the first summand (proportional to the coefficients $n_b^{(4)}$) is actually independent of the ratio of $p$ and $q$ and has already the same structure as the single-hard result. The second summand (proportional to the coefficients $\tilde n_b^{(4)}$) contains the function $g(p/q)$, which vanishes in the hierarchical limits, $g(0)=g(\infty)=0$. The double-hard limit of $P^{(33),II}$ contains a sum of three terms. The first summand contains the integral $\int_{pq}P(p)P(q)P(|{\bm p}+{\bm q}|)$. There are three possible hierarchical limits, $p,|{\bm p}+{\bm q}|\gg q$, $q,|{\bm p}+{\bm q}|\gg p$, $p,q\gg |{\bm p}+{\bm q}|$. In each case the integral approaches the same limit (after relabelling loop wavenumbers) $\int_pP(p)^2\int_qP(q)$. Thus we obtain again a result of the same form as the single-hard limit. The same is true for the second and third summand, realizing that the angle-average can be done analytically in the hierarchical limit, and that for e.g. the limit $p,|{\bm p}+{\bm q}|\gg q$ one has $f({\bm p},{\bm q})\to ({\bm p}\cdot{\bm q})^2/(p^2q^2)$, such that $f({\bm p},{\bm q})_\text{av}\to 1/3$ and $[f({\bm p},{\bm q})^2]_\text{av}\to 1/5$. The same average is approached in the other two hierarchical limits, giving again an overall factor of three. This means the limit of hierarchical loop wavenumbers for the double-hard contribution of stochastic terms yields
\bea
  \lefteqn{P^\text{2L}_{ab}(k)\Big|^\text{double-hard}_\text{stochastic}\Bigg|^\text{hierarchical\ loop\ wavenumbers}}\nn\\
  &=& \Bigg(12n_a^{(2)}n_b^{(4)}+12n_a^{(4)}n_b^{(2)}+3\times6\times\Big(n^{(3)}_an^{(3)}_b+(n^{(3)}_a\tilde n^{(3)}_b+\tilde n^{(3)}_an^{(3)}_b)/3\nn\\
  && {} +\tilde n^{(3)}_a\tilde n^{(3)}_b/5\Big)\Bigg)\left(\int_pP(p)^2\right)\times\left(\int_q P(q)\right)\,.
\eea
We find that the sum of coefficients in the round bracket precisely agrees with the $n_{ab}$ coefficients, 
\bea
  n_{ab} &=& 12n_a^{(2)}n_b^{(4)}+12n_a^{(4)}n_b^{(2)}+3\times6\times\Big(n^{(3)}_an^{(3)}_b+(n^{(3)}_a\tilde n^{(3)}_b+\tilde n^{(3)}_an^{(3)}_b)/3\nn\\
  && {} +\tilde n^{(3)}_a\tilde n^{(3)}_b/5\Big)\,,
\eea
such that
\be
  P^\text{2L}_{ab}(k)\Big|^\text{double-hard}_\text{stoch.}\Bigg|^\text{hierarchical\ loop\ wavenumbers} = P^\text{2L}_{ab}(k)\Big|^\text{single-hard}_\text{stoch.}\Bigg|^\text{hard}\,.
\ee
This implies again a commutation of limits of either taking a limit where first $p,q\to\infty$ with $p/q$ fixed and then taking the hierarchical limit (e.g. $q/p\to 0$), or alternatively first $p\to\infty$ with $q$ fixed and then $q\to\infty$, and is the analog of the result~\refeq{doublehardvssequentialsinglehard}.

\bibliographystyle{JHEP}
\bibliography{main}

\end{document}